\newcommand{\be}{\begin{equation}}
\newcommand{\ee}{\end{equation}}
\newcommand{\bea}{\begin{eqnarray}}
\newcommand{\eea}{\end{eqnarray}}
\newcommand{\nn}{\nonumber}
\newcommand{\Appendix}[1]%
    {\renewcommand{\thesection}{Appendix~\Alph{section}:}%
         \section{#1}}%
\newcommand*\patchAmsMathEnvironmentForLineno[1]{%
  \expandafter\let\csname old#1\expandafter\endcsname\csname #1\endcsname
  \expandafter\let\csname oldend#1\expandafter\endcsname\csname end#1\endcsname
  \renewenvironment{#1}%
     {\linenomath\csname old#1\endcsname}%
     {\csname oldend#1\endcsname\endlinenomath}}%
\newcommand*\patchBothAmsMathEnvironmentsForLineno[1]{%
  \patchAmsMathEnvironmentForLineno{#1}%
  \patchAmsMathEnvironmentForLineno{#1*}}%
\AtBeginDocument{%
\patchBothAmsMathEnvironmentsForLineno{equation}%
\patchBothAmsMathEnvironmentsForLineno{align}%
\patchBothAmsMathEnvironmentsForLineno{flalign}%
\patchBothAmsMathEnvironmentsForLineno{alignat}%
\patchBothAmsMathEnvironmentsForLineno{gather}%
\patchBothAmsMathEnvironmentsForLineno{multline}%
}

\catcode`@=11
\long\def\@makecaption#1#2{
   \vskip 10pt
   \setbox\@tempboxa\hbox{{\small\bf #1.} \ {\small #2}}
   \ifdim \wd\@tempboxa >\hsize       
   {\small\bf #1.} \ {\small #2}\par  
   \else                              
        \hbox to\hsize{\hfil\box\@tempboxa\hfil}
   \fi}
\catcode`@=12

\catcode`@=11
\def\secteqno{\@addtoreset{equation}{section}%
\def\theequation{\thesection.\arabic{equation}}}
\def\endsecteqno{\def\theequation{\@ifundefined{chapter}%
{\arabic{equation}}{\thechapter.\arabic{equation}}}}
\newcounter{subequation}
\def\thesubequation{\alph{subequation}}
\def\sneqnarray{\stepcounter{equation}\let\@currentlabel=\theequation
\setcounter{subequation}{1}
\def\@eqnnum{{\rm (\theequation\thesubequation)}}
\global\@eqcnt\z@\tabskip\@centering\let\\=\@eqncr\let\@@eqncr=\@@sneqncr
$$\halign to \displaywidth\bgroup\@eqnsel\hskip\@centering
 $\displaystyle\tabskip\z@{##}$&\global\@eqcnt\@ne
 \hskip 2\arraycolsep \hfil${##}$\hfil
 &\global\@eqcnt\tw@ \hskip 2\arraycolsep
$\displaystyle\tabskip\z@{##}$\hfil
tabskip\@centering&\llap{##}\tabskip\z@\cr}
\def\endsneqnarray{\@@sneqncr\egroup $$\global\@ignoretrue}
\def\@@sneqncr{\let\@tempa\relax
   \ifcase\@eqcnt \def\@tempa{& & &}\or \def\@tempa{& &}
   \else \def\@tempa{&}\fi
     \@tempa \if@eqnsw\@eqnnum\stepcounter{subequation}\fi
     \global\@eqnswtrue\global\@eqcnt\z@\cr}
\def\nobiblabels{\def\@lbibitem[##1]##2{\@bibitem{##2}}}
\catcode`@=12

\def\beq{\begin{equation}}
\def\eeq{\end{equation}}
\def\bea{\begin{eqnarray}}
\def\eea{\end{eqnarray}}
\def\nn{\nonumber}

\def\la{\lambda} \def\lap{\lambda^{\prime}} \def\pa{\partial} \def\de{\delta}  \def\dag{\dagger}
   \def\Oc{{\rm O}} \def\S{{\rm S}}
\def\bnabla{{\bm \nabla}}

\documentclass[a4paper, aps, 11pt, superscriptaddress, prd, showpacs, 
amsmath, amssymb, nofootinbib, showkeys]{revtex4-1}

\usepackage[german,english]{babel}
\usepackage{hyperref}
\usepackage{bm}
\usepackage{bbm}
\usepackage{braket}
\usepackage{slashed}
\usepackage{multirow}
\usepackage{epsfig}
\usepackage{epstopdf}
\usepackage{diagbox}
\usepackage[mathlines]{lineno}
\usepackage[utf8]{inputenc}

\allowdisplaybreaks

\begin{document}

\preprint{TUM-EFT 129/19}
\title{QCD spin effects in the heavy hybrid potentials and spectra}

\author{Nora Brambilla}
\email{nora.brambilla@ph.tum.de}
\affiliation{Physik-Department, Technische Universit\"at M\"unchen, \\ 
James-Franck-Str. 1, 85748 Garching, Germany}
\affiliation{Institute for Advanced Study, Technische Universit\"at M\"unchen, 
Lichtenbergstrasse 2a, 85748 Garching, Germany}
\author{Wai Kin Lai}
\email{wk.lai@tum.de }
\affiliation{Physik-Department, Technische Universit\"at M\"unchen, \\ 
James-Franck-Str. 1, 85748 Garching, Germany}
\author{Jorge Segovia}
\email{jsegovia@upo.es}
\affiliation{Departamento de Sistemas F\'isicos, Qu\'imicos y Naturales, \\
Universidad Pablo de Olavide, E-41013 Sevilla, Spain}
\author{Jaume Tarr\'us Castell\`a}
\email{jtarrus@ifae.es}
\affiliation{Grup de F\'\i sica Te\`orica, Dept. F\'\i sica and IFAE-BIST, 
Universitat Aut\`onoma de Barcelona, \\ E-08193 Bellaterra (Barcelona), Spain}

\date{\today}

\begin{abstract}
The spin-dependent operators for heavy quarkonium hybrids have been recently obtained in a nonrelativistic effective field theory approach up to next-to-leading order in the heavy-quark mass expansion. In the effective field theory for hybrids several operators not found in standard quarkonia appear, including an operator suppressed by only one power of the heavy-quark mass. We compute the matching coefficients for these operators in the short heavy-quark-antiquark distance regime, $r\ll  1/\Lambda_{\rm QCD}$, by matching weakly-coupled potential NRQCD to the effective field theory for hybrids. In this regime the perturbative and nonperturbative contributions to the matching coefficients factorize, and the latter can be expressed in terms of purely gluonic correlators whose form we explicitly calculate with the aid of the transformation properties of the gluon fields under discrete symmetries. We detail our previous comparison with direct lattice computations of the charmonium hybrid spectrum, from which the unknown nonperturbative contributions can be obtained, and extend it to data sets with different light-quark masses.
\end{abstract}


\maketitle

\section{Introduction}

One of the long-standing, unconfirmed, predictions of QCD is the existence of hadrons in which gluonic excitations play an analogous role as constituent quarks in traditional hadrons. This kind of states are divided into two classes depending on whether they contain quark degrees of freedom or not. In the case that the state is formed purely by gluonic excitations it is called a glueball, while when the state contains both quark and gluonic  degrees of freedom it is called a hybrid. The experimental identification of any of such states has been up until now unsuccessful. In the case of glueballs, this  can be understood as owing to the fact that the lowest-lying states, as predicted by lattice QCD calculations~\cite{Morningstar:1999rf,Chen:2005mg}, have quantum numbers coinciding with those of standard isosinglet mesons, and therefore a strong mixing is expected. Glueballs with exotic $J^{PC}$, such as $0^{+-},\,2^{-+}$, or $1^{-+}$, are expected to appear at rather large masses.

For hybrid states the experimental identification may be simpler since the interplay of quark and gluonic degrees of freedom enlarges the range of possible quantum numbers $J^{PC}$, including exotic ones among its lowest mass states. Nevertheless, if the quarks forming the hybrid state are light, the hybrids are still expected to appear at the same scale as conventional mesons, $\Lambda_{\rm  QCD}$,   leading again to the expected large mixings if the quantum numbers $J^{PC}$ of the hybrids are not explicitly exotic. On the other hand, hybrids containing heavy quarks, called heavy or quarkonium hybrids, develop a gap of order $\Lambda_{\rm QCD}$ with respect to the respective states containing only the heavy-quark component, i.e. the standard quarkonium states. Therefore, quarkonium hybrids are  expected to be the states including gluonic excitations that are easier to identify experimentally.

It is precisely in the quarkonium spectrum, close and above the open-flavor thresholds, that in the last decade tens of exotic heavy quarkonium-like states have been discovered in experiments at B-factories (BaBar, Belle, and CLEO), $\tau$-charm facilities (CLEO-c and BESIII) and hadron colliders (CDF, D0, LHCb, ATLAS, and CMS). These states are the so-called XYZ mesons. Several interpretations of the  XYZ mesons have been proposed. In these interpretations, XYZ mesons are bound states of a heavy-quark-antiquark pair with non-trivial light degrees of freedom. In the case that the light degrees of freedom are light quarks, different tetraquark pictures emerge depending on the spatial arrangement of the light quarks with respect to the heavy quarks. If the light degrees of freedom are gluonic, the picture that emerges is that of a quarkonium hybrid. So far there is no conclusion on which interpretation is the correct one, see Refs.~\cite{Brambilla:2014jmp,Olsen:2014qna,Esposito:2016noz,Guo:2017jvc,Brambilla:2019esw} for reviews of the experimental and theoretical status of the subject.

Quarkonium hybrids are characterized by the separation between the dynamical energy scales of the heavy quarks and the gluonic degrees of freedom. The gluon dynamics is nonperturbative and, therefore, happens at the scale $\Lambda_{\rm QCD}$, while the nonrelativistic heavy-quark-antiquark pair bind together in the background potential created by the gluonic excited state at a lower energy scale $mv^2\ll \Lambda_{\rm QCD}$, where $m$ is the  heavy-quark mass and $v$ their  relative velocity. This separation of energy scales is analogous to that of the electrons and nuclei in molecules, and has led to the observation that quarkonium hybrids can be treated in a framework inspired by the Born-Oppenheimer approximation~\cite{Griffiths:1983ah,Juge:1997nc,Braaten:2014qka,Braaten:2014ita,Braaten:2013boa,Meyer:2015eta,Lebed:2017min}. In recent papers \cite{Berwein:2015vca,Oncala:2017hop,Brambilla:2017uyf,TarrusCastella:2019rit} an effective field theory (EFT) formulation of the Born-Oppenheimer approximation, called the BOEFT, has been developed and used to compute the quarkonium hybrid spectrum. In this paper we will rely on the hierarchy described above, i.e. $\Lambda_{\rm QCD} \gg mv^2$ and work under the further assumption that $mv\gg \Lambda_{\rm QCD}$. The advantage of this assumption is that the nonperturbative dynamics can be factored out and its effects encoded in nonperturbative gluonic correlators, allowing for a clear theoretical analysis of the heavy hybrid spin contributions. In the case in which $mv \sim \Lambda_{\rm QCD}$, the potentials will be given
by generalized Wilson loops, however their spin structure will be the same.

The spin-dependent operators for the BOEFT have been presented in Ref.~\cite{Brambilla:2018pyn} up to $\mathcal{O}(1/m^2)$. The most interesting feature, also pointed out in Ref.~\cite{Soto:2017one}, is that quarkonium hybrids, unlike standard quarkonium, receive spin-dependent contributions already at order $1/m$. At order $1/m^2$ there are spin-dependent operators analogous to those appearing in the case  for standard quarkonium as well as three new operators that are unique to quarkonium hybrids. The matching coefficients of these operators, the spin-dependent potentials, are generically characterized as the sum of a perturbative contribution and a nonperturbative one. The perturbative contribution corresponds to the spin-dependent octet potentials and only appears in the operators analogous to those of standard quarkonium. The nonperturbative contributions can be written as a power series in $r^2$ with coefficients encoding the nonperturbative dynamics of the gluon fields. 
In this paper, we compute the spin-dependent potentials by matching weakly-coupled potential NRQCD (pNRQCD)~\cite{Pineda:1997bj,Brambilla:1999xf} to the BOEFT and obtain the detailed expressions for the nonperturbative matching coefficients in terms of gluonic correlators. To complete the computation, it is necessary to use discrete symmetries to reduce the pNRQCD two-point functions into the structures  matching the ones in the BOEFT. The values of the nonperturbative contributions are unknown, nevertheless our explicit formulas will allow a future direct lattice calculation of these objects. Alternatively, the nonperturbative matching coefficients can be obtained by comparing with lattice calculations of the charmonium hybrid spectrum and the values used to predict the spin-splittings in the bottomonium hybrid sector as shown in Ref.~\cite{Brambilla:2018pyn}. We provide in this paper a detailed description of the fitting procedure and enlarge the analysis to older lattice data with larger light-quark masses.

The paper is organized as follows: in Sec.~\ref{sec1} we review the discussion on the relevant scales for quarkonium hybrid systems and summarize weakly-coupled pNRQCD and the BOEFT for hybrid states. In Sec.~\ref{pot} we demonstrate the essential calculation steps and present the results for the matching of the spin-dependent potentials, and give explicit formulas for the gluonic correlators. In Sec.~\ref{sec4} we compute the mass shifts in the hybrid spectrum due to the spin-dependent potentials and compare them with the charmonium hybrid spectrum obtained from two different lattice QCD calculations at different light-quark masses and fit the values of the nonperturbative matching coefficients. We use these values to give a prediction for the spin-dependent mass shifts in the bottomonium sector. We give our summary and conclusion in Sec.~\ref{conc}. In Appendix~\ref{app_CPT}, using discrete symmetries, we obtain the relations between
the gluonic correlators that are needed  to complete the matching calculation of the spin-dependent potentials. A detailed overview of the matching of the spin-dependent terms of the two-point functions in pNRQCD and the BOEFT is given in Sec.~\ref{pot} and Appendix~\ref{app_matching}. Finally, in Appendix~\ref{app} and  \ref{app_elements} we work out the matrix elements of the spin-dependent operators.
\section{Scales and effective field theory description} \label{sec1}
In heavy quarkonium systems there are several well-separated scales typical of nonrelativistic bound states: the heavy-quark mass $m$ (hard scale), the relative momentum between the heavy quarks $m v\sim 1/r$ (soft scale), where  $v\ll 1$ is the relative velocity and $r$ the relative distance, and the heavy-quark binding energy $mv^2$  (ultrasoft scale). Additionally, we also encounter the scale of the QCD nonperturbative physics $\Lambda_{\rm QCD}$. 
  
Heavy quarkonium hybrids are bound states of a heavy-quark-antiquark pair with a gluonic excitation. In quarkonium hybrids an interesting scale separation pattern appears similar to the one of diatomic molecules bound by electromagnetic interactions. The heavy quarks play the role of the nuclei and the gluons (and the light quarks) play the role of the electrons. In a diatomic molecule the electrons are non-relativistic and their energy levels can be studied in the nuclei static limit due to the latter larger mass. These electronic energy levels are called electronic static energies and are of order $m_e\alpha^2$, with $m_e$ the electron mass and $\alpha$ the fine structure constant. The nuclei vibrational (bound) states occur around the minima of these electronic static energies and have energies smaller than $m_e\alpha^2$.

In quarkonium hybrids, the light degrees of freedom are relativistic with a typical energy and momentum of order $\Lambda_{\rm QCD}$. This implies that the typical size of a hybrid state is of the order of $1/\Lambda_{\rm QCD}$. The scaling of the typical distance of the heavy-quark-antiquark pair, $r\sim 1/(mv)$, depends on the details of the full inter-quark potential, which has a long-range nonperturbative part and a short-range Coulomb-like interaction. Therefore, it may happen that the heavy-quark-antiquark pair is more closely bound than the light degrees of freedom. This situation is interesting because the hybrid would present a hierarchy between the distance of the quark-antiquark pair and the typical size of the light degrees of freedom that does not exist in the case of diatomic molecules, where the electron cloud and the distance between the nuclei are of the same size. A consequence of this is that while the molecules are characterized by a cylindrical symmetry,  the symmetry group for hybrids at leading order in a (multipole) expansion in the distance of the heavy-quark-antiquark pair is a much stronger spherical symmetry. This modifies significantly the power counting of the EFT for hybrids with respect to the case of diatomic molecules, leading to new effects. In the following we consider this case with the interquark distance of order $r \ll 1/\Lambda_{\rm QCD}$. As in diatomic molecules, in order for a Born-Oppenheimer  picture to emerge it is crucial that the binding energy of the heavy particles, $mv^2$, is smaller than the energy scale of the light degrees of freedom. In summary, we will require the following hierarchy of energy scales to hold true: $m\gg m v\gg\Lambda_{\rm QCD} \gg m  v^2 $. We can then build an EFT to describe quarkonium hybrids by sequentially integrating out the scales above $mv^2$ \cite{Berwein:2015vca,Brambilla:2017uyf}. In this paper we focus our attention on the spin-dependent terms up to $\mathcal{O}(1/m^2)$. 
\subsection{Weakly-coupled pNRQCD}\label{s1}
The first step in constructing the quarkonium hybrid BOEFT is to integrate out the hard scale which produces the well-known NRQCD \cite{Caswell:1985ui,Bodwin:1994jh,Manohar:1997qy}. The next step is to integrate out the soft scale,  i.e., expand in small relative distances between the heavy quarks. In the short-distance regime, $r \ll 1/\Lambda_{\rm QCD}$, this step can be performed in perturbation theory and one arrives at pNRQCD \cite{Pineda:1997bj,Brambilla:1999xf,Brambilla:2004jw}, which is the starting point of our discussion. The weakly-coupled pNRQCD Lagrangian ignoring light quarks\footnote{In this work we will not consider light quarks, see Refs.~\cite{Brambilla:2017uyf,TarrusCastella:2019rit} for a discussion on their inclusion and the use of the BOEFT formalism for tetraquark states.} and including the gluon interaction operators from Ref.~\cite{Brambilla:2003nt} that will be needed for the present work is 
\begin{align}  L_{\rm pNRQCD}  =& \int  d^3R\Bigg\{\int d^3r  \,\Bigl(
{\rm                 Tr}\left[\S^{\dag}\left(i\partial_0-h_s\right)\S+
\Oc^{\dagger}\left(iD_0-h_o\right)\Oc\right]     \nn    \\     &+g{\rm
Tr}\left[\S^{\dag}\bm{r}\cdot\bm{E}\,\Oc+\Oc^{\dag}\bm{r}\cdot\bm{E}\,\S+\frac{1}{2}\Oc^{\dag}\bm{r}\cdot\{\bm{E},\Oc\}
-\frac{1}{8}\Oc^\dagger r^ir^j[\bm{D}^iE^j,\Oc]\right]\nonumber\\
&+\frac{g}{4m}{\rm{Tr}}\left[\Oc^{\dag}\bm{L}_{Q\bar
Q}\cdot[\bm{B},\Oc]\right] \nn\\
&    +\frac{gc_F}{m}{\rm     Tr}
\left[\S^{\dag}(\bm{S}_1-\bm{S}_2)\cdot\bm{B}\,\Oc+\Oc^{\dag}(\bm{S}_1-\bm{S}_2)\cdot\bm{B}\,\S
+\Oc^{\dag}\bm{S}_1\cdot\bm{B}\,\Oc-\Oc^{\dag}\bm{S}_2
\Oc\cdot\bm{B}\right] \nn\\
& +\frac{gc_s}{2m^2}{\rm
Tr}\left[\S^{\dag}(\bm{S}_1+\bm{S}_2)\cdot(\bm{E}\times\bm{p})\,\Oc+\Oc^{\dag}(\bm{S}_1+\bm{S}_2)\cdot(\bm{E}\times\bm{p})\,\S
\right.                                                          \nn\\
&\left.+\Oc^{\dag}\bm{S}_1\cdot(\bm{E}\times\bm{p})\,\Oc-\Oc^{\dag}\bm{S}_2\cdot(\bm{p}\,\Oc\times\bm{E})\right]\Bigr)-\frac{1}{4}
G_{\mu \nu}^a G^{\mu  \nu\,a}+\dots\Biggr\}
\,.
\label{pnrqcd1}
\end{align}  
$\S$ and $\Oc$ are the heavy-quark-antiquark singlet and octet fields respectively, normalized with respect to color as $\S=S \bm{1}_c/\sqrt{N_c}$ and $\Oc=O^a  T^a/\sqrt{T_F}$. They should be understood as functions of $t$, the relative coordinates $\bm{r}$, and the center of mass coordinate $\bm{R}$ of the heavy quarks. All the gluon fields in Eq.~\eqref{pnrqcd1} are multipole-expanded in $\bm{r}$ and therefore evaluated at $\bm{R}$ and~$t$: in particular the gluon field strength $G^{\mu\nu\,a}\equiv G^{\mu\nu\,a}(\bm{R},t)$, and the covariant derivatives  $iD_0O\equiv  i\partial_0O-g\left[A_0(\bm{R},t),O\right]$ and $i\bm{D}E^i\equiv i\boldsymbol{\nabla}_{\bm{R}}E^i(\bm{R},t)+g[\bm{A}(\bm{R},t),E^i(\bm{R},t)]$. The momentum and orbital angular momentum of the reduced mass of the heavy-quark-antiquark pair are respectively denoted by $\bm{p}=m\frac{d{\bm
 r}}{dt}=-i\boldsymbol{\nabla}_{\bm{r}}$ and $\bm{L}_{Q\bar Q}=\bm{r}\times\bm{p}$. The spin vectors of the heavy quark and heavy antiquark are $\bm{S}_1$ and $\bm{S}_2$ respectively. The terms with explicit factors of the chromoelectric field $\bm{E}$ and the chromomagnetic field $\bm{B}$ are obtained by matching NRQCD to weakly-coupled pNRQCD at tree level. The coefficients $c_F$ and $c_s$ are matching coefficients of NRQCD (see e.g. Ref.~\cite{Manohar:1997qy}), calculated in perturbation theory, as $\alpha_s$ is small at the scale $m$ that characterizes these coefficients. They are equal to $1$ at leading order in $\alpha_s$. The ellipsis denotes other spin-independent operators, operators higher order in the multipole expansion or $1/m$, and perturbative corrections of higher orders in $\alpha_s$.
The Hamiltonian densities $h_s$ and $h_o$ of the singlet and octet fields respectively read
\begin{align}
h_s=&-\frac{\bnabla^2_r}{m}+V_s(r)\,,\\
h_o=&-\frac{\bnabla^2_r}{m}+V_o(r)\,.
\end{align}
It is useful to organize $V_o(r)$\footnote{An analogous expansion can be written for $V_s(r)$, see \cite{Brambilla:2004jw}. We omit it here since we will not use it.} as an expansion in $1/m$ and separate the spin-independent (SI) and spin-dependent (SD) terms:
\begin{align}
V_o(r)&=V^{(0)}_o(r)+\frac{V^{(1)}_o(r)}{m}+\frac{V^{(2)}_o(r)}{m^2}+\dots\,, \label{ocpot}\\
V_o^{(2)}(r)&=V_{o\,SD}^{(2)}(r)+V_{o\,SI}^{(2)}(r)\,,\\
V_{o\,SD}^{(2)}(r)&=V_{o\,SL}(r)\bm{L}_{Q\bar{Q}}\cdot\bm{S}+V_{o\,S^2}(r)\bm{S}^2+V_{o\,S_{12}}(r){S}_{12}\,,
\end{align}
where $\bm{S}=\bm{S}_1+\bm{S}_2$ and ${S}_{12}=12(\bm{S}_1\cdot\hat{\bm{r}})(\bm{S}_2\cdot\hat{\bm{r}})-4\bm{S}_1\cdot\bm{S}_2$. 
The octet-field spin-dependent potentials can be found in Ref.~\cite{Pineda:1996nw} \footnote{A contribution to $V_{o\,S^2}$, proportional to the $f_8$'s, which originate in quark-antiquark annihilation diagrams, is missing in Ref.~\cite{Pineda:1996nw}. Setting $c_F=c_s=1$ and neglecting the contribution from the quark-antiquark annihilation diagrams in  Eqs.~(\ref{vols})-(\ref{vos12}) would recover the corresponding expressions in Ref.~\cite{Pineda:1996nw}.}. They are given from the tree-level matching of NRQCD to weakly-coupled pNRQCD by
\begin{align}
V_{o\,SL}(r)&=\left(C_F-\frac{C_A}{2}\right)\left(\frac{c_s}{2}+c_F\right)\frac{\alpha_s(\nu)}{r^3}\,, \label{vols}\\
V_{o\,S^2}(r)&=\left[\frac{4\pi}{3}\left(C_F-\frac{C_A}{2}\right)c^2_F\alpha_s(\nu)+T_F\left(f_8(^1S_0)-f_8(^3S_1)\right)\right]\de^3(\bm{r})\,,\label{vos2}\\
V_{o\,S_{12}}(r)&=\left(C_F-\frac{C_A}{2}\right)c_F^2\frac{\alpha_s(\nu)}{4r^3} \label{vos12}\,,
\end{align}
where $C_F=\left(\frac{N_c^2-1}{N_c}\right)T_F$ and $C_A=2N_cT_F$ are the Casimir factors for the fundamental and adjoint representations of the color gauge group $SU(N_c)$ respectively. We define $T_F$ by $\textrm{Tr}[T^aT^b]=T_F\delta^{ab}$, where $T^a$ are the color generators in the fundamental representation. The renormalization scale, $\nu$, is naturally of order $mv\sim1/r$. The matching coefficients $f_8$'s originate in heavy-quark-antiquark annihilation diagrams. To $\mathcal{O}(\alpha_s)$ they read~\cite{Bodwin:1994jh,Pineda:1998kj}
\begin{align}
f_8(^1S_0)=0\,,\quad f_8(^3S_1)=-\pi \alpha_s(m)\,.
\end{align}
At the accuracy of this work, we will use the tree-level expressions of $c_F$ and $c_s$, i.e. $c_F=c_s=1$, for the spin-dependent octet potentials
in Eqs.~(\ref{vols})-(\ref{vos12}).

\subsection{The BOEFT}\label{s2}
The final step is to build an EFT, which we call the BOEFT, that describes the heavy-quark-antiquark pair dynamics in the presence of a background gluonic excited state by integrating out the scale $\Lambda_{\rm QCD}$. First, we have to identify the degrees of freedom in the BOEFT. 

In the short-interquark-distance limit $r\rightarrow 0$ and the static limit $m\rightarrow \infty$, quarkonium hybrids reduce to gluelumps, which are color-singlet combinations of a local static octet color source coupled to a gluonic field. The gluonic excitations can be characterized by the so-called gluelump operators~\cite{Brambilla:1999xf,Berwein:2015vca}. The Hamiltonian for the gluons at leading order in the $1/m$- and multipole expansions, corresponding to the $G^{a}_{\mu\nu}G^{a\mu\nu}$ term in the Lagrangian in Eq.\eqref{pnrqcd1}, is given by
\begin{align}
H_{0} &= \int d^3\bm{R}\,\frac{1}{2} \left[\bm{E}^a\cdot\bm{E}^a+\bm{B}^a\cdot\bm{B}^a\right]\,.
\label{h-light}
\end{align} 
We define the gluelump operators, $G_{\kappa}^{ia}$, as the Hermitian color-octet operators that generate the eigenstates of $H_{0}$ in the presence of a local heavy-quark-antiquark octet source:
\begin{align}
H_0 G^{ia}_{\kappa}(\bm{R},t)|0\rangle &= \Lambda_{\kappa} G^{ia}_{\kappa}(\bm{R},t)|0\rangle\,,
\end{align}
where $a$ is the color index, $\kappa$ labels the quantum numbers $K^{PC}$ of the gluonic degrees of freedom and $i$ labels its spin components. 
The spectrum of the mass eigenvalues, $\Lambda_{\kappa}$, called the gluelump mass, has been computed on the lattice in Refs.~\cite{Foster:1998wu,Bali:2003jq,Marsh:2013xsa}.

At the next-to-leading order in the multipole expansion the system is no longer spherically symmetric
but acquires instead a  cylindrical symmetry \footnote{The symmetry group is changing from $O(3) \times C$ to $D_{\infty  h}$, with $P$ replaced by $CP$.} around the
heavy-quark-antiquark axis. Therefore it is convenient to work with a basis of states with good transformation properties under $D_{\infty  h}$. Such states can be constructed by projecting the gluelump operators on various directions with respect to the heavy-quark-antiquark axis:
\begin{align}
|\kappa,\,\lambda;\,\bm{r},\bm{R},t\rangle &=P^i_{\kappa\lambda} O^{a\,\dagger}\left(\bm{r},\bm{R},t\right) G_{\kappa}^{ia}(\bm{R},t)|0\rangle\,,
\label{eigen1}
\end{align}
where summations over indices $i$ and $a$ are implied. $P^i_{\kappa\lambda}$ is a projector that projects the gluelump operator to an eigenstate of $\bm{K}\cdot\hat{\bm{r}}$ with eigenvalue $\lambda$, where $\bm{K}$ is the angular momentum operator for the gluonic degrees of freedom and $\hat{\bm{r}}$ the unit vector along the heavy-quark-antiquark axis. It is therefore natural to define the degrees of freedom of the BOEFT as the operator
$\hat{\Psi}_{\kappa\lambda}(\bm{r},\,\bm{R},\,t)$ defined by
\begin{align}
P^{i\dag}_{\kappa\lambda} O^{a}\left(\bm{r},\bm{R},t\right) G_{\kappa}^{ia}(\bm{R},t)
&=Z_\kappa^{1/2}(\bm{r},\bm{R},\bm{p},\bm{P})\hat{\Psi}_{\kappa\la}(\bm{r},\bm{R},t)\,,\label{eq:relation}
\end{align}
where $\bm{P}$ is the momentum operator conjugate to $\bm{R}$, and $Z$ is a field renormalization
factor, normalized such that the following commutation relations hold:
\begin{align}
[O^a(\bm{r},\bm{R},t),O^{b\dagger}(\bm{r}',\bm{R}',t)]&=\delta^{ab}\mathbb{I}\delta^3(\bm{r}-\bm{r}')\delta^3(\bm{R}-\bm{R}')\,,\\
[\hat{\Psi}_{\kappa\la}(\bm{r},\bm{R},t),\hat{\Psi}^\dag_{\kappa'\la'}(\bm{r}',\bm{R}',t)]&=\delta_{\kappa\kappa'}\delta_{\la\la'}\mathbb{I}
\delta^3(\bm{r}-\bm{r}')\delta^3(\bm{R}-\bm{R}')\,,
\end{align}
where $\mathbb{I}$ is the identity matrix of the spin indices of the heavy quark and antiquark.
The BOEFT is obtained by integrating out modes of scale $\Lambda_{\rm QCD}$, i.e. the gluonic excitation. The Lagrangian of the BOEFT reads as
\begin{align}
L_{\rm BOEFT} &= \int d^3Rd^3r \, \sum_{\kappa} \sum_{\lambda\lambda^{\prime}}
\textrm{Tr}\left\{
\hat{\Psi}^{\dagger}_{\kappa\lambda}(\bm{r},\,\bm{R},\,t) \left[i\partial_t - V_{\kappa\lambda\lambda^{\prime}}(r)+ P^{i\dag}_{\kappa\lambda}\frac{\bnabla^2_r}{m}P^i_{\kappa\lambda^{\prime}}\right]\hat{\Psi}_{\kappa\lambda^{\prime}}(\bm{r},\,\bm{R},\,t)\right\}+\dots\,,
\label{bolag2}
\end{align}
where the trace is over spin indices of the heavy quark and antiquark, and the ellipsis stands for operators producing transitions to standard quarkonium states and transitions between hybrid states of different $\kappa$. The former are beyond the scope of this work\footnote{Transitions to standard quarkonium states are discussed in Ref.~\cite{Oncala:2017hop}.} and the latter are suppressed at least by $1/\Lambda_{\rm QCD}$ since the static energies for different $\kappa$ are separated by a gap $\sim\Lambda_{\rm QCD}$. The potential $V_{\kappa\lambda\lambda^{\prime}}$ can be organized into an expansion in $1/m$ and a sum of spin-dependent (SD) and independent (SI) parts:
\begin{align}
V_{\kappa\la\lap}(r)&=V^{(0)}_{\kappa\la}(r)\de_{\la\lap}+\frac{V^{(1)}_{\kappa\la\lap}(r)}{m}+\frac{V^{(2)}_{\kappa\la\lap}(r)}{m^2}+\dots\,, \label{eq:V_full}\\
V_{\kappa\la\lap}^{(1)}(r)&=V_{\kappa\la\lap\,SD}^{(1)}(r)+V_{{\kappa\la\lap}\,SI}^{(1)}(r)\,,\label{eq:V_1}\\
V_{\kappa\la\lap}^{(2)}(r)&=V_{\kappa\la\lap\,SD}^{(2)}(r)+V_{{\kappa\la\lap}\,SI}^{(2)}(r)\,.\label{eq:V_2}
\end{align}

\begin{figure}[ht]
\centerline{\includegraphics[width=.9\textwidth]{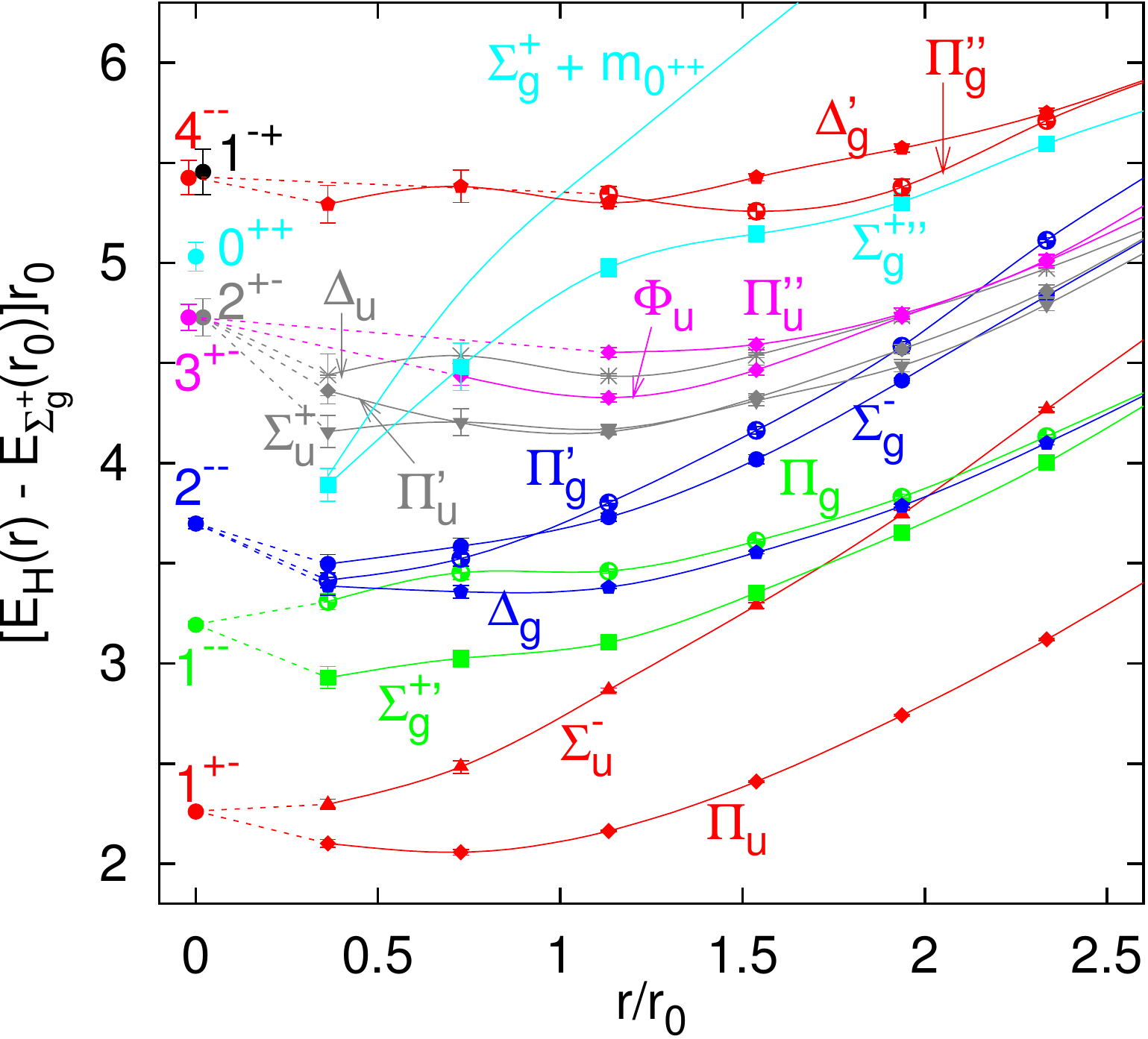}}
\caption{The lowest hybrid static energies~\cite{Juge:2002br} and gluelump masses~\cite{Foster:1998wu,Bali:2003jq} in units of $r_0\approx 0.5$~fm. The absolute values have been fixed such that the ground state $\Sigma_g^+$ static energy (not displayed) is zero at $r_0$. The data points at $r=0$, labeled with $\kappa=K^{PC}$, are the gluelump masses. The gluelump spectrum has been shifted by an arbitrary constant to adjust the $1^{+-}$ state with the $\Pi_u$ and $\Sigma^-_u$ potentials at short distances, with the dashed lines indicating the 
expected extrapolation to degeneracy at $r=0$. The behavior of the static energies at short distances becomes rather unreliable for some hybrids, especially the higher excited ones. This is largely due to the difficulty in lattice calculations to distinguish between states with the same quantum numbers, which mix. The figure is taken from~\cite{Bali:2003jq}.}
\label{jkmse}
\end{figure}

In Ref.~\cite{Berwein:2015vca} the static potential $V^{(0)}_{\kappa\lambda}(r)$ was matched to the quark-antiquark hybrid static energies computed on the lattice. In Fig.~\ref{jkmse} we show the QCD static energies computed using lattice NRQCD from Ref.~\cite{Juge:2002br}: they are plotted as a function of the quark-antiquark distance $r$ and only states with excited glue are presented. The standard quarkonium static energy, without gluonic excitations, would lie below in energy and is not shown. Recently, a new comprehensive lattice study of the hybrid static energies has appeared in Ref.~\cite{Capitani:2018rox}.

One of the major features of this spectrum is that in the short-distance region the static energies can be organized in quasi-degenerate multiplets corresponding to the gluelump spectrum. This is a direct consequence of the breaking of spherical symmetry into a cylindrical symmetry once the subleading contributions in the multipole expansion are included. Indeed, at leading order in the multipole expansion $V^{(0)}_{\kappa\lambda}(r)$ reads~\cite{Berwein:2015vca}
\begin{align}
V^{(0)}_{\kappa\lambda}(r)=\Lambda_{\kappa}+V^{(0)}_o(r)+\dots\,.\label{eq:static_V_small_r}
\end{align}
That is, the potential in the short-distance limit only depends on the quantum numbers of the gluelump $\kappa$ and not on its projection $\lambda$.

The lowest gluelump has quantum numbers $\kappa=1^{+-}$.
In Ref.~\cite{Berwein:2015vca} the matrix elements of $P^{i\dag}_{\kappa\lambda}\frac{\bnabla^2_r}{m}P^i_{\kappa\lambda^{\prime}}$ were obtained for $\kappa=1^{+-}$ and it was shown to contain off-diagonal terms in $\lambda$-$\lambda^{\prime}$ that lead to coupled Schr\"odinger equations. The
 Schr\"odinger equations were solved numerically and the spectrum and wave functions of hybrid states generated the static energies labeled by $\Sigma^-_u$ and $\Pi_u$ were obtained.

\section{Matching of the Spin-dependent potentials}\label{pot}
We present now the results of the matching for the spin-dependent potentials in Eqs.~(\ref{eq:V_1}) and (\ref{eq:V_2}) for the lowest-lying gluelump ($\kappa=1^{+-}$) in the short distance regime $1/r \gg \Lambda_{\rm{QCD}}$. We will first write down the formulation of the matching for general $\kappa$ and then focus on the case $\kappa=1^{+-}$, for which we will demonstrate the essential steps of the calculation and present the final results, and leave the more involved details of calculation in Appendix~\ref{app_matching}.

\begin{figure}[ht]
\caption{Matching of two-point function in the hybrid BOEFT on the left-hand side to weakly-coupled pNRQCD on the right-hand side. The diagrams are in coordinate space. The single, double, and curly lines represent the heavy-quark singlet, heavy-quark octet, and gluon fields respectively. The black dots stand for vertices from the Lagrangian in Eq.~\ref{pnrqcd1} and the gray circles represent the nonperturbative gluon dynamics.}
\centerline{\includegraphics[width=.99\textwidth]{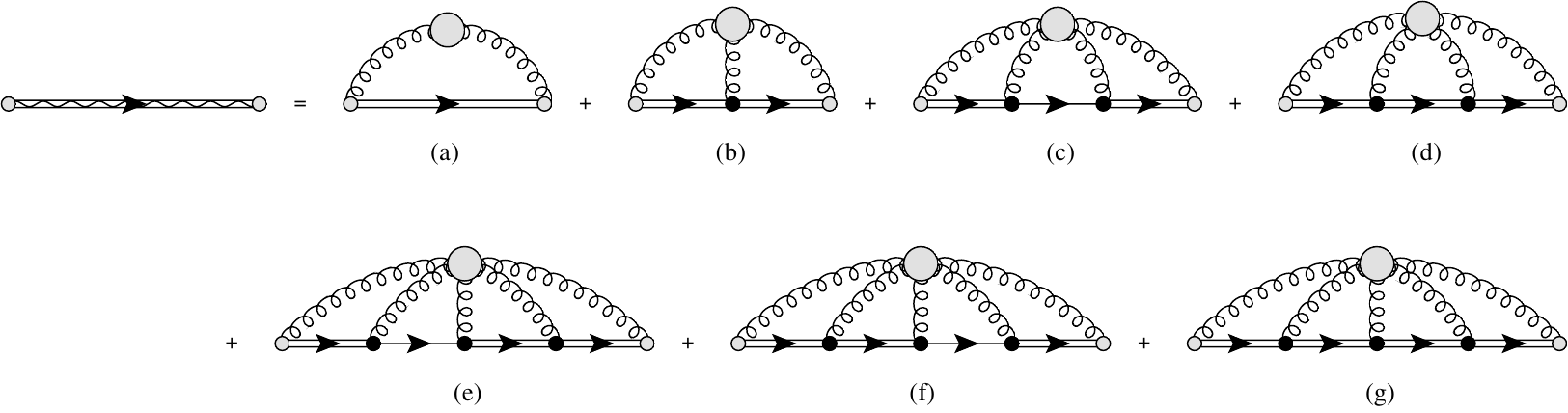}}
\label{match}
\end{figure}

The matching between weakly-coupled pNRQCD and the BOEFT at the scale $\Lambda_{\rm QCD}$ is performed by considering the following gauge-invariant two-point Green's function defined in terms of the fields in pNRQCD:
\begin{align}
&I_{\kappa\la\la'}(\bm{r},\bm{R},\bm{r}',\bm{R}')\nonumber\\
\equiv&\,\lim_{T\to\infty}
\langle 0|
P_{\kappa\la}^{i\dag} G^{ia\dag}_\kappa(\bm{R},T/2)O^a(\bm{r},\bm{R},T/2)
O^{b\dag}(\bm{r}',\bm{R}',-T/2)P_{\kappa\la'}^j G^{jb}_\kappa(\bm{R}',-T/2)|0\rangle\,,
\label{eq:I_pNRQCD_1}
\end{align}
where only the repeated color indices $a,\,b$ and spin indices $i,\,j$ are summed. In the BOEFT, with Eq.~(\ref{eq:relation}), the two-point Green's function is given by
\begin{align}
&I_{\kappa\la\la'}(\bm{r},\bm{R},\bm{r}',\bm{R}')\nonumber\\
=&\,\lim_{T\to\infty}
Z_\kappa^{1/2}(\bm{r},\bm{R},\bm{p},\bm{P})
\exp\left[-i\left( V_{\kappa\lambda\lambda^{\prime}}(r)- P^{i\dag}_{\kappa\lambda}\frac{\bnabla^2_r}{m}
P^i_{\kappa\lambda^{\prime}}\right)T \right]
Z_\kappa^{\dag 1/2}(\bm{r},\bm{R},\bm{p},\bm{P})\nonumber\\
&\,\quad\quad\times \mathbb{I}\delta^3(\bm{r}-\bm{r}')\delta^3(\bm{R}-\bm{R}')\,.\label{eq:I_hybrid}
\end{align}
The Green's function in pNRQCD (Eq.~(\ref{eq:I_pNRQCD_1})) has the form
\begin{align}
I_{\kappa\la\la'}(\bm{r},\bm{R},\bm{r}',\bm{R}')&=\lim_{T\to\infty}\exp\left\{-i\left[(h_o)_{\kappa\la\la'}+\Lambda_\kappa+\delta V_{\kappa\la\la'}\right]T\right\}
\mathbb{I}\delta^3(\bm{r}-\bm{r}')\delta^3(\bm{R}-\bm{R}')\,,\label{eq:I_pNRQCD}
\end{align}
where 
\begin{align}
(h_o)_{\kappa\la\la'}&\equiv P^{i\dag}_{\kappa\la}h_o P^{i}_{\kappa\la'}
=P^{i\dag}_{\kappa\lambda}\left(V_o-\frac{\bnabla^2_r}{m}\right)P^i_{\kappa\lambda^{\prime}}\,,
\end{align}
and the gluelump mass $\Lambda_\kappa$ is related to a gluonic correlator by
\begin{align}
e^{-i\Lambda_\kappa T}&= \langle 0|G^{ia}_\kappa(\bm{R},T/2)\phi^{ab}(T/2,-T/2)G^{ib}_\kappa(\bm{R}',-T/2)|0\rangle
\,,
\end{align}
with $\phi^{ab}(t,t')$ the adjoint static Wilson line defined by
\be
\phi(t,t')={\rm{P}}\exp\left[-ig\int^{t}_{t'}dt\, A_0^{adj}(\bm{R},t)\right]\,.\label{eq:Wilson}
\ee
In Eq.~(\ref{eq:I_pNRQCD}), $\delta V_{\kappa\la\la'}$ is obtained from the contributions to Eq.~(\ref{eq:I_pNRQCD_1}) from
insertions of singlet-octet- and octet-octet-gluon coupling operators from the Lagrangian in Eq.~(\ref{pnrqcd1}). 
Comparing Eqs.~(\ref{eq:I_pNRQCD}) and~(\ref{eq:I_hybrid}), we obtain $Z_\kappa(\bm{r},\bm{R},\bm{p},\bm{P})=1$ and
\be
V_{\kappa\la\la'}=P^{i\dag}_{\kappa\lambda}V_oP^i_{\kappa\lambda^{\prime}} +\Lambda_\kappa+\delta V_{\kappa\la\la'}\,,\label{eq:matching_potential}
\ee 
which reduces to Eq.~(\ref{eq:static_V_small_r}) at leading order in $1/m$ and the multipole expansion. 
The matching condition is schematically depicted in Fig.~\ref{match}. In Fig.~\ref{match}, the left-hand side and the right-hand side correspond to the two-point Green's function computed in the BOEFT and pNRQCD respectively. Diagram (a) gives the perturbative term 
$P^{i\dag}_{\kappa\lambda}V_oP^i_{\kappa\lambda^{\prime}}$ in Eq.~(\ref{eq:matching_potential}), which is inherited from the octet potential in Eq.~\eqref{ocpot}, as well as a nonperturbative term $\Lambda_\kappa$, the gluelump mass. Diagrams like (b), (c), (d), (e), (f), and (g), with black dots, which denote the singlet-octet- and octet-octet-gluon coupling operators in the pNRQCD Lagrangian Eq.~\eqref{pnrqcd1}, give another nonperturbative contribution $\delta V_{\kappa\la\la'}$. All diagrams in pNRQCD are computed in coordinate space, similar to what is done in Ref.~\cite{Brambilla:1999xf}. 
%
%

Now we focus on the case $\kappa=1^{+-}$. To simplify the notation we will drop the subscript $\kappa$ for the rest of the manuscript and it should be understood that we are always referring to $\kappa=1^{+-}$, and so $\lambda$ takes the values $0,\pm 1$. 
The spin-dependent potentials in Eqs.~(\ref{eq:V_1}) and (\ref{eq:V_2}) for $\kappa=1^{+-}$ read as follows:
\begin{align}
V_{\la\lap\,SD}^{(1)}(r)&=V_{SK}(r)\left(P^{i\dag}_{\la}\bm{K}^{ij}P^j_{\lap}\right)\cdot\bm{S}\nonumber\\
&\quad + V_{{SK}b}(r)\left[\left(\bm{r}\cdot \bm{P}^{\dag}_{\la}\right)\left(r^i\bm{K}^{ij}P^j_{\lap}\right)\cdot\bm{S}
-\left(r^i\bm{K}^{ij}P^{j\dag }_{\la}\right)\cdot\bm{S} \left(\bm{r}\cdot \bm{P}_{\lap}\right)\right]
\,,\label{sdm}\\
V_{\la\lap\,SD}^{(2)}(r)&=V_{SLa}(r)\left(P^{i\dag}_{\la}\bm{L}_{Q\bar{Q}}P^i_{\lap}\right)\cdot\bm{S}+V_{SLb}(r)P^{i\dag}_{\la}\left(L_{Q\bar{Q}}^iS^j+S^iL_{Q\bar{Q}}^j\right)P^{j}_{\lap}\nonumber\\
&\quad + V_{SLc}(r)\left[\left(\bm{r}\cdot \bm{P}^{\dag}_{\la}\right)\left(\bm{p}\times\bm{S}\right)\cdot \bm{P}_{\lap}
+\bm{P}^{\dag }_{\la}\cdot\left(\bm{p}\times\bm{S}\right)\left(\bm{r}\cdot \bm{P}_{\lap}\right)\right] \nonumber\\
&\quad +V_{S^2}(r)\bm{S}^2\de_{\la\lap}+V_{S_{12}a}(r)S_{12}\de_{\la\lap}+V_{S_{12}b}(r)P^{i\dag}_{\la}P^j_{\lap}\left(S^i_1S^j_2+S^i_2S^j_1\right)
\,,\label{sdm2}
\end{align}
where $\left(K^{ij}\right)^k=i\epsilon^{ikj}$ is the angular momentum operator for the spin-$1$ gluonic excitation. The projectors $P^{i}_{\lambda}$ read 
\begin{align}
P^{i}_{0}&=\hat{r}_0^i= \hat{r}^i\,,\label{pr10}\\
P^{i}_{\pm 1}&=\hat{r}^i_{\pm}=\mp\left(\hat{\theta}^i\pm i\hat{\phi}^i\right)/\sqrt{2}\,,\label{pr11}
\end{align}
with
\begin{align}
\hat{\bm{r}}&=(\sin(\theta)\cos(\phi),\,\sin(\theta)\sin(\phi)\,,\cos(\theta)) \,, \nonumber \\
\hat{\bm\theta}&=(\cos(\theta)\cos(\phi),\,\cos(\theta)\sin(\phi)\,,-\sin(\theta)) \,, \nonumber \\
\hat{\bm\phi}&=(-\sin(\phi),\,\cos(\phi)\,,0)\,.
\end{align}
The $1/m$ operators in Eq.~(\ref{sdm}), with coefficients $V_{SK}(r)$ and $V_{SKb}(r)$\footnote{The operator with coefficient $V_{SKb}(r)$ contains the tensor $r^ir^j$ contracted to other vectors. In the case of standard quarkonia, for which the symmetry group is $SO(3)$, it is natural to decompose the tensor $r^ir^j$ into a sum of a trace part and a traceless symmetric part, each of which corresponds to an irreducible representation of $SO(3)$. This is done for the operators with coefficients $V_{S^2}(r)$ and $V_{S_{12}a}(r)$, since they also appear in the case of standard quarkonia. In the case of hybrid states here, since the symmetry group is $D_{\infty h}$ instead of $SO(3)$, this decomposition is not of particular relevance and we decided to write the $V_{SKb}(r)$-operator without substracting the trace part.}, couple the angular momentum of the gluonic excitation with the total spin of the heavy-quark-antiquark pair. These operators are characteristic of the hybrid states and are absent for standard quarkonia. 
Among the $1/m^2$ operators in Eq.~(\ref{sdm2}), the operators with coefficients $V_{SLa}(r)$, $V_{S^2}(r)$, and $V_{S_{12}a}$ are the standard spin-orbit, total spin squared, and tensor spin operators respectively, which appear for standard quarkonia also. In addition to them,
three novel operators appear at order $1/m^2$. The operators with coefficients $V_{SLb}(r)$ and $V_{SLc}(r)$ are generalizations of the spin-orbit operator to the hybrid states. Similarly, the operator with coefficient $V_{S_{12}b}(r)$ is generalization of the tensor spin operator to the hybrid states. It should be noted that there are contributions from the operators with coefficients $V_{SK}(r)$ and $V_{SKb}(r)$ at order $1/m^2$. For conciseness of presentation, we choose to treat these contributions as $1/m$-terms in $V_{SK}(r)$ and $V_{SKb}(r)$, instead of showing the $V_{SK}$- and $V_{SKb}$-operators again in Eq.~(\ref{sdm2}).

The coefficients $V_i(r)$ on the right-hand side of Eqs.~(\ref{sdm}) and (\ref{sdm2}) have the form $V_i(r)=V_{oi}(r)+V_i^{np}(r)$, where $V_{oi}(r)$ is the perturbative octet potential and $V_i^{np}(r)$ is the nonperturbative contribution. From the multipole expansion, $V_i^{np}(r)$ is a power series in $r^2$, $V_i^{np}(r)=V_i^{np(0)}+V_i^{np(1)}r^2+\dots$.  We will work at next-to-leading order in the multipole expansion for the $1/m$-potentials and leading order in the multipole expansion for the $1/m^2$-potentials. Therefore, up to the precision we work at, we have
\begin{align}
V_{SK}(r)&=V^{np\,(0)}_{SK}+\frac{V^{np\,(0)}_{SK2}}{m}+V^{np\,(1)}_{SK}r^2\,,\label{eq:V_SK_first}\\
V_{SKb}(r)&=V^{np\,(0)}_{SKb}\,,\label{eq:V_SKb_first}\\
V_{SLa}(r)&=V_{o\,SL}(r)+V^{np\,(0)}_{SLa}\,,\label{eq:V_SLa_first}\\
V_{SLb}(r)&=V^{np\,(0)}_{SLb}\,,\label{eq:V_SLb_first}\\
V_{SLc}(r)&=V^{np\,(0)}_{SLc}\,,\label{eq:V_pxs_first}\\
V_{S^2}(r)&=V_{o\,S^2}(r)+V^{np\,(0)}_{S^2}\,, \label{eq:V_S2_first}\\
V_{S_{12a}}(r)&=V_{o\,S_{12}}(r)\,, \label{eq:V_S12_first}\\
V_{S_{12b}}(r)&=V^{np\,(0)}_{S_{12}b}\,.\label{eq:V_S12b_first}
\end{align}
In Eqs.~(\ref{eq:V_SLa_first}),~(\ref{eq:V_S2_first}), and~(\ref{eq:V_S12_first}), $V_{o\,SL}(r)$, $V_{o\,S^2}(r)$, and $V_{o\,S_{12}}(r)$ are the perturbative tree-level spin-dependent octet potentials given by Eqs.~(\ref{vols})-(\ref{vos12}). The constants $V^{np\,(0)}_{SK}$, 
$V^{np\,(0)}_{SK2}$, $V^{np\,(1)}_{SK}$, $V^{np\,(0)}_{SKb}$, $V^{np\,(0)}_{SLa}$,
$V^{np\,(0)}_{SLb}$, $V^{np\,(0)}_{SLc}$, $V^{np\,(0)}_{S^2}$, and $V^{np\,(0)}_{S_{12}b}$ are obtained from diagrams (b), (c), (d), (e), (f), and (g) in Fig.~\ref{match} with insertions of spin-dependent operators with a chromomagnetic field or a chromoelectric field in the pNRQCD Lagrangian Eq.~(\ref{pnrqcd1})
, and are expressed as nonperturbative purely gluonic correlators. It should be emphasized that the expressions of 
$V_{SK}(r)$, $V_{SKb}(r)$, $V_{SLa}(r)$, $V_{SLb}(r)$, $V_{SLc}(r)$, $V_{S^2}(r)$, $V_{S_{12}a}(r)$, and $V_{S_{12}b}(r)$ in Eqs.~(\ref{eq:V_SK_first})-(\ref{eq:V_S12b_first}) are valid only for $1/r\gg\Lambda_{\textrm{QCD}}$. For arbitrary values of $r$, they are given by generalized Wilson loops.

To demonstrate the essential steps for obtaining the $V^{np\,(j)}_i$'s in terms of purely gluonic correlators, here we will go through the derivation for the simpliest one, $V^{np\,(0)}_{SK}$, and leave the details of the derivations for the remaining ones in Appendix~\ref{app_matching}. In these derivations, relations among gluonic correlators derived from transformation properties of the gluonic operators under $C$, $P$, $T$ are used, which are summarized in Appendix~\ref{app_CPT}.

Consider diagram (b) in Fig.~\ref{match}, with an insertion of the $c_F$-term in Eq.~(\ref{pnrqcd1}). Its contribution to $\delta V_{\la\la'}$ is given by
\begin{align}
\delta V^{c_F}_{\la\la'}&=
i\frac{c_F}{2m}\hat{r}_{\lambda}^{i\,\dag}\hat{r}_{\la^{\prime}}^{k}\left(U_B\right)^{ijk}_{bcd}\left(h^{bcd}{S}_1^j-h^{bdc}{S}_2^j\right)\,,
\label{eq:V_cF}
\end{align}
where 
\begin{align}
\left(U_B\right)^{ijk}_{bcd}&\equiv\lim_{T\to\infty}\frac{ie^{i\Lambda T}}{T}
\int^{T/2}_{-T/2}dt\,\langle 0|{G}^{ia}(T/2) \phi^{ab}(T/2,t)g {B}^{jc}(t)\phi^{de}(T/2,t) {G}^{ke}(-T/2)|0\rangle\,\label{eq:UB}
\end{align}
and $h^{abc}=\frac{2}{T_F}{\rm Tr}[T^aT^bT^c]$.
In Eq.~(\ref{eq:UB}), repeated color indices are summed and all fields are understood as evaluated at $\bm{R}$. The structure of the gluonic correlator in Eq.~(\ref{eq:UB}) can be read off from diagram (b) in Fig.~\ref{match}. The gluelump operators create and destroy the gluonic excitation at times $-T/2$ and $T/2$ respectively. The adjoint Wilson lines correspond to the propagation of the octet fields due to the covariant derivative in the first line of Eq.~(\ref{pnrqcd1}). The insertion of $\bm{B}$ correspond to the emission vertex in diagram (b) of Fig.~\ref{match} denoted by a solid black dot. All possible times of insertion must be taken into account and therefore the time of insertion is integrated over from $-T/2$ to $T/2$. The exponential factor $e^{i\Lambda T}$ in front of the correlator is the result of factoring out the gluelump mass $\Lambda$ in the potential as indicated in Eq.~(\ref{eq:matching_potential}). The expression on the right-hand side of Eq.~(\ref{eq:UB}) is finite in the limit $T\to\infty$, since the large $T$ behavior of the time integral is compensated by the factor $1/T$ and the exponential factor $e^{i\Lambda T}$ compensates for the time evolution of the gluelump operator from $-T/2$ to $T/2$. 

Using Eqs.~(\ref{eq:sym_UB_C2}), which is derived from the charge conjugation properties of the gluon fields, Eq.~(\ref{eq:V_cF}) becomes
\begin{align}
\delta V^{c_F}_{\la\la'}&=
i\frac{c_F}{2m}\hat{r}_{\lambda}^{i\,\dag}\hat{r}_{\la^{\prime}}^{k}\left(U_B\right)^{ijk}_{bcd}h^{bcd}{S}^j\,.\label{eq:V_cF_2}
\end{align}
The color combination
\begin{align}
(\hat{U}_B)^{ijk}&\equiv(U_B)^{ijk}_{bcd}h^{bcd}\,,\label{eq:hat_UB}
\end{align}
being a rotationally invariant tensor, can be written as
\begin{align}
(\hat{U}_B)^{ijk}&=\tilde{U}_B\epsilon^{ijk} \,.\label{eq:tensor_1}
\end{align}
Therefore, Eq.~(\ref{eq:V_cF_2}) becomes
\begin{align}
\delta V^{c_F}_{\la\la'}
&=i\frac{c_F}{2m}\hat{r}_{\lambda}^{i\,\dag}\hat{r}_{\la^{\prime}}^{k}\tilde{U}_B
\epsilon^{ijk}{S}^j\,,
\end{align} 
from which it follows that
\begin{align}
V^{np(0)}_{SK}&=\frac{c_F}{2}\tilde{U}_B\,.\label{eq:V_SK_0}
\end{align}

The detailed derivations of $V^{np\,(0)}_{SK2}$, $V^{np\,(1)}_{SK}$, $V^{np\,(0)}_{SKb}$, $V^{np\,(0)}_{SLa}$, $V^{np\,(0)}_{SLb}$, $V^{np\,(0)}_{SLc}$,  $V^{np\,(0)}_{S^2}$, and $V^{np\,(0)}_{S_{12}b}$ in terms of gluonic correlators are shown in Appendix~\ref{app_matching}.
Here we list the final results. Similar to Eq.~(\ref{eq:UB}), we have to define the relevant gluonic correlators that appear in the matching calculation of the two-point function. All of these gluonic correlators involve a gluelump operator at $t=-T/2$ and another gluelump operator at $t=-T/2$. The relevant gluonic correlators that correspond to diagram (c) in Fig.~\ref{match} are
\begin{align}
\left(U^{ss}_{EE}\right)^{ijkl}&\equiv\lim_{T\to\infty}\frac{ie^{i\Lambda T}}{T}
\int^{T/2}_{-T/2}dt\int^{t}_{-T/2}dt'\,\langle 0|{G}^{ia}(T/2)\phi^{ab}(T/2,t)g {E}^{jb}(t) \nonumber\\
&\quad \quad \quad\quad\times gE^{kc}(t')\phi^{cd}(t',-T/2){G}^{ld}(-T/2)|0\rangle\,,\label{eq:UssEE}\\
\left(U^{ss}_{BB}\right)^{ijkl}&\equiv\lim_{T\to\infty}\frac{ie^{i\Lambda T}}{T}
\int^{T/2}_{-T/2}dt\int^{t}_{-T/2}dt'\,\langle 0|{G}^{ia}(T/2)\phi^{ab}(T/2,t)g {B}^{jb}(t) \nonumber\\
&\quad \quad \quad\quad\times gB^{kc}(t')\phi^{cd}(t',-T/2){G}^{ld}(-T/2)|0\rangle\,.\label{eq:UssBB}
\end{align}
The correlator $\left(U^{ss}_{EE}\right)^{ijkl}$ in Eq.~(\ref{eq:UssEE}) arises from insertions of two singlet-octet vertices with a chromoelectric field from the pNRQCD Lagrangian Eq.~(\ref{pnrqcd1}). The adjoint Wilson lines connecting the gluelump operators
to the chromoelectric fields arise from the two propagators of the octet field in diagram (c) of Fig.~\ref{match}. Similarly, $\left(U^{ss}_{BB}\right)^{ijkl}$ in Eq.~(\ref{eq:UssBB}) is defined like $\left(U^{ss}_{EE}\right)^{ijkl}$ with the chromoelectric field replaced by the chromomagnetic field.
The relevant gluonic correlators that correspond to diagram (d) in Fig.~\ref{match} are
\begin{align}
\left(U^{oo}_{EE}\right)^{ijkl}_{bcdefg}&\equiv\lim_{T\to\infty}\frac{ie^{i\Lambda T}}{T}
\int^{T/2}_{-T/2}dt\int^{t}_{-T/2}dt'\,\langle 0|{G}^{ia}(T/2)\phi^{ab}(T/2,t)g {E}^{jc}(t) \phi^{de}(t,t')\nonumber\\
&\quad \quad \quad\quad\times gE^{kf}(t')\phi^{gh}(t',-T/2){G}^{lh}(-T/2)|0\rangle\,,\label{eq:UooEE}\\
\left(U^{oo}_{BB}\right)^{ijkl}_{bcdefg}&\equiv\lim_{T\to\infty}\frac{ie^{i\Lambda T}}{T}
\int^{T/2}_{-T/2}dt\int^{t}_{-T/2}dt'\,\langle 0|{G}^{ia}(T/2)\phi^{ab}(T/2,t)g {B}^{jc}(t) \phi^{de}(t,t')\nonumber\\
&\quad \quad \quad\quad\times gB^{kf}(t')\phi^{gh}(t',-T/2){G}^{lh}(-T/2)|0\rangle\,,\label{eq:UooBB}\\
\left(U^{oo}_{BDE}\right)^{ijklm}_{bcdefg}&\equiv\lim_{T\to\infty}\frac{ie^{i\Lambda T}}{T}
\int^{T/2}_{-T/2}dt\int^{t}_{-T/2}dt'\,\langle 0|{G}^{ia}(T/2)\phi^{ab}(T/2,t)g {B}^{jc}(t) \phi^{de}(t,t')\nonumber\\
&\quad \quad \quad\quad\times [\bm{D}^kgE^{l}(t')]^f\phi^{gh}(t',-T/2){G}^{mh}(-T/2)|0\rangle\,,\label{eq:UooBDE}\\
\left(U^{oo}_{DEB}\right)^{ijklm}_{bcdefg}&\equiv\lim_{T\to\infty}\frac{ie^{i\Lambda T}}{T}
\int^{T/2}_{-T/2}dt\int^{t}_{-T/2}dt'\,\langle 0|{G}^{ia}(T/2)\phi^{ab}(T/2,t)g [\bm{D}^j{E}^{k}(t)]^c \phi^{de}(t,t')\nonumber\\
&\quad \quad \quad\quad\times gB^{lf}(t')\phi^{gh}(t',-T/2){G}^{mh}(-T/2)|0\rangle\,.\label{eq:UooDEB}
\end{align}
The correlator $\left(U^{oo}_{EE}\right)^{ijkl}_{bcdefg}$ in Eq.~(\ref{eq:UooEE}) arises from insertions of two octet-octet vertices with a chromoelectric field from the pNRQCD Lagrangian. The three adjoint Wilson lines arise from the three propagators of the octet field
in diagram (d) of Fig.~\ref{match}. The correlator
$\left(U^{oo}_{BDE}\right)^{ijklm}_{bcdefg}$ in Eq.~(\ref{eq:UooBDE}) arises from 
insertions of two octet-octet vertices, one with a chromomagnetic field at time $t$ and another with a covariant derivative of the chromoelectric field at time $t'<t$. $\left(U^{oo}_{BB}\right)^{ijkl}_{bcdefg}$ in Eq.~(\ref{eq:UooBB}) 
and $\left(U^{oo}_{DEB}\right)^{ijklm}_{bcdefg}$ in Eq.~(\ref{eq:UooDEB}) are similarly defined.
The relevant gluonic correlators that correspond to diagram (e) in Fig.~\ref{match} are
\begin{align}
(U^{sso}_{BEE})^{ijklm}_{def}&\equiv\lim_{T\to\infty}\frac{ie^{i\Lambda T}}{T}
\int^{T/2}_{-T/2}dt\int^{t}_{-T/2}dt'\int^{t'}_{-T/2}dt''\langle 0|{G}^{ia}(T/2)\phi^{ab}(T/2,t)g {B}^{jb}(t)\nonumber\\
&\quad\quad\quad \times gE^{kc}(t')\phi^{cd}(t',t'')gE^{le}(t'')\phi^{fg}(t'',-T/2){G}^{mg}(-T/2)|0\rangle\,,\label{eq:UssoBEE}\\
(U^{sso}_{EBE})^{ijklm}_{def}&\equiv\lim_{T\to\infty}\frac{ie^{i\Lambda T}}{T}
\int^{T/2}_{-T/2}dt\int^{t}_{-T/2}dt'\int^{t'}_{-T/2}dt''\langle 0|{G}^{ia}(T/2)\phi^{ab}(T/2,t)g {E}^{jb}(t)\nonumber\\
&\quad\quad\quad \times gB^{kc}(t')\phi^{cd}(t',t'')gE^{le}(t'')\phi^{fg}(t'',-T/2){G}^{mg}(-T/2)|0\rangle\,,\label{eq:UssoEBE}\\
(U^{sso}_{EEB})^{ijklm}_{def}&\equiv\lim_{T\to\infty}\frac{ie^{i\Lambda T}}{T}
\int^{T/2}_{-T/2}dt\int^{t}_{-T/2}dt'\int^{t'}_{-T/2}dt''\langle 0|{G}^{ia}(T/2)\phi^{ab}(T/2,t)g {E}^{jb}(t)\nonumber\\
&\quad\quad\quad \times gE^{kc}(t')\phi^{cd}(t',t'')gB^{le}(t'')\phi^{fg}(t'',-T/2){G}^{mg}(-T/2)|0\rangle\,.\label{eq:UssoEEB}
\end{align}
The relevant gluonic correlators that correspond to diagram (f) in Fig.~\ref{match} are
\begin{align}
(U^{oss}_{BEE})^{ijklm}_{bcd}&\equiv\lim_{T\to\infty}\frac{ie^{i\Lambda T}}{T}
\int^{T/2}_{-T/2}dt\int^{t}_{-T/2}dt'\int^{t'}_{-T/2}dt''\langle 0|{G}^{ia}(T/2)\phi^{ab}(T/2,t)g {B}^{jc}(t)\nonumber\\
&\quad\quad\quad \times \phi^{de}(t,t')gE^{ke}(t')gE^{lf}(t'')\phi^{fg}(t'',-T/2){G}^{mg}(-T/2)|0\rangle\,,\label{eq:UossBEE}\\
(U^{oss}_{EBE})^{ijklm}_{bcd}&\equiv\lim_{T\to\infty}\frac{ie^{i\Lambda T}}{T}
\int^{T/2}_{-T/2}dt\int^{t}_{-T/2}dt'\int^{t'}_{-T/2}dt''\langle 0|{G}^{ia}(T/2)\phi^{ab}(T/2,t)g {B}^{jc}(t)\nonumber\\
&\quad\quad\quad \times \phi^{de}(t,t')gE^{ke}(t')gE^{lf}(t'')\phi^{fg}(t'',-T/2){G}^{mg}(-T/2)|0\rangle\,,\label{eq:UossEBE}\\
(U^{oss}_{EEB})^{ijklm}_{bcd}&\equiv\lim_{T\to\infty}\frac{ie^{i\Lambda T}}{T}
\int^{T/2}_{-T/2}dt\int^{t}_{-T/2}dt'\int^{t'}_{-T/2}dt''\langle 0|{G}^{ia}(T/2)\phi^{ab}(T/2,t)g {B}^{jc}(t)\nonumber\\
&\quad \times \phi^{de}(t,t')gE^{ke}(t')gE^{lf}(t'')\phi^{fg}(t'',-T/2){G}^{mg}(-T/2)|0\rangle\,,\label{eq:UossEEB}
\end{align}
The relevant gluonic correlators that correspond to diagram (g) in Fig.~\ref{match} are
\begin{align}
(U^{ooo}_{BEE})^{ijklm}_{bcdefghpq}&\equiv\lim_{T\to\infty}\frac{ie^{i\Lambda T}}{T}
\int^{T/2}_{-T/2}dt\int^{t}_{-T/2}dt'\int^{t'}_{-T/2}dt''\langle 0|{G}^{ia}(T/2)\phi^{ab}(T/2,t)g {B}^{jc}(t)\nonumber\\
&\quad \times \phi^{de}(t,t')gE^{kf}(t')\phi^{gh}(t',t'')gE^{lp}(t'')\phi^{qr}(t'',-T/2){G}^{mr}(-T/2)|0\rangle\,,\label{eq:UoooBEE}\\
(U^{ooo}_{EBE})^{ijklm}_{bcdefghpq}&\equiv\lim_{T\to\infty}\frac{ie^{i\Lambda T}}{T}
\int^{T/2}_{-T/2}dt\int^{t}_{-T/2}dt'\int^{t'}_{-T/2}dt''\langle 0|{G}^{ia}(T/2)\phi^{ab}(T/2,t)g {E}^{jc}(t)\nonumber\\
&\quad \times \phi^{de}(t,t')gB^{kf}(t')\phi^{gh}(t',t'')gE^{lp}(t'')\phi^{qr}(t'',-T/2){G}^{mr}(-T/2)|0\rangle\,,\label{eq:UoooEBE}\\
(U^{ooo}_{EEB})^{ijklm}_{bcdefghpq}&\equiv\lim_{T\to\infty}\frac{ie^{i\Lambda T}}{T}
\int^{T/2}_{-T/2}dt\int^{t}_{-T/2}dt'\int^{t'}_{-T/2}dt''\langle 0|{G}^{ia}(T/2)\phi^{ab}(T/2,t)g {E}^{jc}(t)\nonumber\\
&\quad\times \phi^{de}(t,t')gE^{kf}(t')\phi^{gh}(t',t'')gB^{lp}(t'')\phi^{qr}(t'',-T/2){G}^{mr}(-T/2)|0\rangle\,.\label{eq:UoooEEB}
\end{align}
In Eqs.~(\ref{eq:UssoBEE}) to (\ref{eq:UoooEEB}), the correlators arise from insertions of three vertices, each being a singlet-octet or an octet-octet vertex, with a chromoelectric field or a chromomagnetic field.
Analogous to Eq.~(\ref{eq:hat_UB}), we define the color combinations
\begin{align}
(\hat{U}_{EE})^{ijkl}&\equiv (U^{oo}_{EE})^{ijkl}_{bcdefg}d^{bcd}d^{efg}+\frac{4T_F}{N_c}(U^{ss}_{EE})^{ijkl}\,,\label{eq:hat_UEE}\\
(\hat{U}_{BB\,a})^{ijkl}&\equiv (U^{oo}_{BB})^{ijkl}_{bcdefg}f^{bcd}f^{efg}\,,\label{eq:hat_UBBa}\\
(\hat{U}_{BB\,b})^{ijkl}&\equiv (U^{oo}_{BB})^{ijkl}_{bcdefg}h^{bcd}h^{egf}+\frac{4T_F}{N_c}(U^{ss}_{BB})^{ijkl}\,,\label{eq:hat_UBBb}\\
(\hat{U}_{BDE})^{ijklm}&\equiv (U^{oo}_{BDE})^{ijklm}_{bcdefg}h^{bcd}f^{efg}\,,\label{eq:hat_UBDE}\\
(\hat{U}_{DEB})^{ijklm}&\equiv (U^{oo}_{DEB})^{ijklm}_{bcdefg}f^{bcd}h^{efg}\,,\label{eq:hat_UDEB}\\
(\hat{U}_{BEE})^{ijklm}&\equiv (U^{ooo}_{BEE})^{ijklm}_{bcdefghpq}h^{bcd}d^{efg}d^{hpq}+\frac{4T_F}{N_c}({U}^{oss}_{BEE})^{ijklm}_{bcd}h^{bcd}\,,\label{eq:hat_UBEE}\\
(\hat{U}_{EBE})^{ijklm}&\equiv({U}^{ooo}_{EBE})^{ijklm}_{bcdefghpq}d^{bcd}h^{efg}d^{hpq}\,,\label{eq:hat_UEBE}\\
(\hat{U}_{EEB})^{ijklm}&\equiv({U}^{ooo}_{EEB})^{ijklm}_{bcdefghpq}d^{bcd}d^{efg}h^{hpq}+\frac{4T_F}{N_c}({U}^{sso}_{EEB})^{ijklm}_{def}h^{def}
\,,\label{eq:hat_UEEB}
\end{align}
where $h^{abc}$ is as defined below Eq.~(\ref{eq:UB}), $d^{abc}\equiv\frac{1}{2}(h^{abc}+h^{acb})$ and $f^{abc}=-\frac{i}{2}(h^{abc}-h^{acb})$.
The tensors defined in Eqs.~(\ref{eq:hat_UEE})-(\ref{eq:hat_UEEB}) have the form $\hat{U}^{ijkl}$ or $\hat{U}^{ijklm}$, which being rotationally invariant, have the tensor decompositions given by
\begin{align}
\hat{U}^{ijkl}&=\tilde{U}^{\rm I}\delta^{ij}\delta^{kl}+\tilde{U}^{\rm II}\delta^{ik}\delta^{jl}+\tilde{U}^{\rm III}\delta^{il}\delta^{jk}\,,\label{eq:tensor_2}\\
\hat{U}^{ijklm}&=\tilde{U}^{\rm i}\epsilon^{ikm}\delta^{jl}+\tilde{U}^{\rm ii}\epsilon^{jlm}\delta^{ik}
+\tilde{U}^{\rm iii}\epsilon^{jkl}\delta^{im}+\tilde{U}^{\rm iv}\epsilon^{ijl}\delta^{km}+\tilde{U}^{\rm v}\epsilon^{klm}\delta^{ij}\nonumber\\
&\quad +\tilde{U}^{\rm vi}\epsilon^{jkm}\delta^{il}
+\tilde{U}^{\rm vii}\epsilon^{ikl}\delta^{jm}+\tilde{U}^{\rm viii}\epsilon^{ijk}\delta^{lm}
+\tilde{U}^{\rm ix}\epsilon^{ijm}\delta^{kl}+\tilde{U}^{\rm x}\epsilon^{ilm}\delta^{jk}\,.\label{eq:tensor_3}
\end{align}
The nonperturbative coefficients $V^{np\,(0)}_{SK2}$, $V^{np\,(1)}_{SK}$, $V^{np\,(0)}_{SKb}$, $V^{np\,(0)}_{SLa}$, $V^{np\,(0)}_{SLb}$,
 $V^{np\,(0)}_{SLc}$, $V^{np\,(0)}_{S^2}$, and $V^{np\,(0)}_{S_{12}b}$ are then given by
\begin{align}
V^{np\,(0)}_{SK2}&=\frac{c_s}{4}\tilde{U}^{\rm I}_{EE}\,,\label{eq:V_SK_2}\\
V^{np\,(1)}_{SK}&=\frac{c_F}{8} \left[-\left(\tilde{U}^{\rm i}_{EBE}+2\tilde{U}^{\rm ix}_{EBE}\right)
-2\left(\tilde{U}^{\rm i}_{BEE}+\tilde{U}^{\rm ix}_{BEE}+\tilde{U}^{\rm x}_{BEE}\right)\right.\nonumber\\
&\quad\quad\,\,\left.
+\left(\tilde{U}^{\rm i}_{BDE}+\tilde{U}^{\rm ix}_{BDE}+\tilde{U}^{\rm x}_{BDE}\right)
\right]\,,\label{eq:V_SK_1}\\
V^{np\,(0)}_{SKb}&=\frac{c_F}{16} \left[2\left(\tilde{U}_{EBE}^{\rm v}-\tilde{U}^{\rm vi}_{EBE}+2\tilde{U}^{\rm ix}_{EBE}\right)\right.\nonumber\\
&\quad\quad\,\,+2\left(2\tilde{U}^{\rm i}_{BEE}
+\tilde{U}^{\rm ii}_{BEE}-\tilde{U}^{\rm iv}_{BEE}+\tilde{U}^{\rm vi}_{BEE}
-\tilde{U}^{\rm viii}_{BEE}+2\tilde{U}^{\rm x}_{BEE}\right)\nonumber\\
&\quad\quad\,\,\left.
-\left(2\tilde{U}^{\rm i}_{BDE}
+\tilde{U}^{\rm ii}_{BDE}-\tilde{U}^{\rm iv}_{BDE}+\tilde{U}^{\rm vi}_{BDE}
-\tilde{U}^{\rm viii}_{BDE}+2\tilde{U}^{\rm x}_{BDE}\right)
\right]\,,\label{eq:V_SKb}\\
V_{SLa}^{np\,(0)}&=\frac{1}{4}\left[c_F \tilde{U}^{\rm III}_{BB\,a}-c_s\tilde{U}^{\rm III}_{EE}\right]
\,,\label{eq:V_SLa}\\
V_{SLb}^{np\,(0)}&=\frac{c_F}{4} \tilde{U}^{\rm I}_{BB\,a}
\,,\label{eq:V_SLb}\\
V_{SLc}^{np\,(0)}&=-\frac{c_s}{4}\tilde{U}^{\rm I}_{EE}
\,,\label{eq:V_pxs}\\
V_{S^2}^{np\,(0)}&=\frac{c_F^2}{4}\tilde{U}^{\rm III}_{BB\,b}\,,\label{eq:V_S2}\\
V_{S_{12}b}^{np\,(0)}&=\frac{c_F^2}{2}\tilde{U}^{\rm I}_{BB\,b}\,.\label{eq:V_S_12_b}
\end{align}
In the derivation of Eqs.~(\ref{eq:V_SK_2})-(\ref{eq:V_S_12_b}), identities for the gluonic correlators in Eqs.~(\ref{eq:UssEE})-(\ref{eq:UoooEEB}) derived from discrete symmetries are used, which are summarized in Appendix~\ref{app_CPT}. Note that some of the components in Eqs.~\eqref{eq:tensor_2}-\eqref{eq:tensor_3} do not appear in Eqs.~(\ref{eq:V_SK_2})-(\ref{eq:V_S_12_b}) since they can be related to the other components through the discrete symmetry relations of Appendix~\ref{app_CPT}. From Eqs.~(\ref{eq:V_SK_0}) and (\ref{eq:V_SK_2})-(\ref{eq:V_S_12_b}), we see that the $V^{np\,(j)}_i$'s are products of a perturbative NRQCD matching coefficient $c_F$ or $c_S$, for which we know the dependence on the heavy-quark flavor, and a nonperturbative purely gluonic correlator, which is independent of the heavy-quark flavor.

\section{Spin splittings in the hybrid spectra}\label{sec4}
We obtain the spin-dependent contributions to the quarkonium hybrid spectrum by applying time-independent perturbation theory to the spin-dependent potentials in Eqs.~\eqref{sdm}-\eqref{sdm2}. We carry out perturbation theory to second order for the terms 
$V^{np\,(0)}_{SK}+\frac{V^{np\,(0)}_{SK2}}{m}$ in Eqs.~(\ref{sdm}) and~(\ref{eq:V_SK_first}), and to first order for the $V^{np\,(1)}_{SK}$ term and the $V^{np\,(0)}_{SKb}$ term in Eqs.~(\ref{sdm}),~(\ref{eq:V_SK_first}) and~(\ref{eq:V_SKb_first}) and the $1/m^2$-suppressed operators in Eqs.~(\ref{sdm2}),~(\ref{eq:V_SLa_first})-(\ref{eq:V_S12b_first}).

The zeroth-order wave functions are obtained following the procedure described in Ref.~\cite{Berwein:2015vca}, by solving the coupled Schr\"odinger equations involving the potentials $V^{(0)}_{\Sigma^-_u}(r)$ and $V^{(0)}_{\Pi_u}(r)$ generated by the $1^{+-}$ gluelump at short distances. The
Schr\"odinger wave functions $\left(\Psi^{N j m_j l s}\right)_\lambda(\bm{r},t)$ is related to the field operator $\hat{\Psi}_\lambda(\bm{r},\bm{R},t)$ by 
\begin{align}
\left(\Psi^{N j m_j l s}\right)_\lambda(\bm{r},t)&=\langle 0|\hat{\Psi}_\lambda(\bm{r},\bm{R}=0,t)|N\, j\, m_j\, l\, s\rangle\,.
\end{align} 
There are two types of solution corresponding to states with opposite parity $\left(\Psi^{N j m_j l s }_{+}\right)_{\la}$ and $\left(\Psi^{N j m_j l s }_{-}\right)_{\la}$\footnote{Note that the sign in the sub-index refers to relative sign of the $\la=+1$ and $\la=-1$ components and not to the parity of the state which depends also on $l$~\cite{Berwein:2015vca}.}:
\begin{align}
&\Psi^{N j m_j l s }_{+}(\bm{r})=\sum_{m_l m_s} \mathcal{C}^{j m_j}_{l\,m_l\,s\,m_s}\left(
\begin{array}{c}
 \psi_0^{(N)}(r)v_{l\,m_l}^0(\theta,\phi) \\
 \frac{1}{\sqrt{2}}\psi_{+}^{(N)}(r)v_{l\,m_l}^{+1}(\theta,\phi) \\
 \frac{1}{\sqrt{2}}\psi_{+}^{(N)}(r)v^{-1}_{l\,m_l}(\theta,\phi) \\
\end{array}
\right)\chi_{s\,m_s}\,,\label{psip}  \\
&\Psi^{N j m_j l s }_{-}(\bm{r})=\sum_{m_l m_s} \mathcal{C}^{j m_j}_{l\,m_l\,s\,m_s}\left(
\begin{array}{c}
0 \\
 \frac{1}{\sqrt{2}}\psi_{-}^{(N)}(r)v_{l\,m_l}^{+1}(\theta,\phi) \\
-\frac{1}{\sqrt{2}}\psi_{-}^{(N)}(r)v^{-1}_{l\,m_l}(\theta,\phi) \\
\end{array}
\right)\chi_{s\,m_s}\,,\label{psim}
\end{align}
where the components from top to bottom correspond to $\la=0,+1,-1$. We define $\bm{L}=\bm{L}_{Q\bar{Q}}+\bm{K}$, the sum of the orbital angular momentum of the heavy-quark-antiquark pair and the angular momentum of the gluelump, and $\bm{J}=\bm{L}+\bm{S}$, the spin of the quarkonium hybrid.
The quantum numbers are as follows: $l(l+1)$ is the eigenvalue of $\bm{L}^2$, $j(j+1)$ and $m_j$ the eigenvalues of $\bm{J}^2$ and $J_3$ respectively, and $s(s+1)$ the eigenvalue of $\bm{S}^2$. $\mathcal{C}^{j m_j}_{l\,m_l\,s\,m_s}$ are the Clebsch-Gordan coefficients. The angular eigenfunctions $v_{l\,m_l}^{\lambda}$ are generalizations of the spherical harmonics for systems with cylindrical symmetry. Their derivation can be found in textbooks such as Ref.~\cite{LandauLifshitz}. The $\chi_{s\,m_s}$ are the spin wave functions. The radial wave functions $\psi_{0}^{(N)},\,\psi_{+}^{(N)},\,\psi_{-}^{(N)}$ are obtained numerically by solving the coupled Schr\"odinger equations, with $N$ labeling the radially excited states. 

The angular wave functions $v_{l\,m_l}^{\lambda}$ are eigenfunctions of $\bm{L}^2$ and not of $\bm{L}_{Q\bar{Q}}^2$. As a result, the evaluation of matrix elements of operators involving $\bm{L}_{Q\bar{Q}}$ is not totally straightforward. The details of the calculation of these matrix elements can be found in Appendix \ref{app}. We will present the results for the four lowest-lying spin-multiplets shown in Table~\ref{tb:spin_multiplet}. Matrix elements of the spin-dependent operators in Eqs.~(\ref{sdm}) and~(\ref{sdm2}) for the angular part of the wave functions of the states in Table~\ref{tb:spin_multiplet} are listed in Appendix~\ref{app_elements}. 
\begin{table}[!t]
\caption{Lowest-lying quarkonium hybrid multiplets}
\begin{center}
\begin{tabular}{c|c|c|c}
\hline
\hline
Multiplet & $\,\,\,l\,\,\,$ & $J^{PC}(s=0)$ & $J^{PC}(s=1)$\\
\hline
$H_1$& $1$ & $1^{--}$ & $(0,1,2)^{-+}$ \\
$H_2$& $1$ & $1^{++}$ & $(0,1,2)^{+-}$ \\
$H_3$& $0$ & $0^{++}$ & $1^{+-}$ \\
$H_4$& $2$ & $2^{++}$ & $(1,2,3)^{+-}$ \\
\hline
\hline
\end{tabular}
\label{tb:spin_multiplet}
\end{center}
\end{table}
The eight nonperturbative parameters $\tilde{V}^{np\,(0)}_{SK}\equiv {V}^{np\,(0)}_{SK}+\frac{{V}^{np\,(0)}_{SK2}}{m}$, $V^{np\,(1)}_{SK}$, $V^{np\,(0)}_{SKb}$, $V^{np\,(0)}_{SLa}$, $V^{np\,(0)}_{SLb}$, $V^{np\,(0)}_{SLc}$, $V^{np\,(0)}_{S^2}$, and $V^{np\,(0)}_{S_{12}b}$ that appear in the spin-dependent potentials Eqs.~\eqref{eq:V_SK_first}-\eqref{eq:V_S12b_first} are obtained by fitting the spin-splittings to corresponding splittings from the lattice determinations of the charmonium hybrid spectrum. Two sets of lattice data from the Hadron Spectrum Collaboration have been used, one set from Ref.~\cite{Liu:2012ze} with a pion mass of $m_{\pi}\approx 400$ MeV and a more recent set from Ref.~\cite{Cheung:2016bym} with a pion mass of $m_{\pi}\approx 240$ MeV. We take the values $m^{RS}_c(1{\rm GeV})=1.477$~GeV \cite{Pineda:2001zq} and $\alpha_s$ at $4$-loops with three light flavors, $\alpha_s(2.6\textrm{~GeV})=0.26$. In the fit the lattice data is weighed by $(\Delta^2_{\textrm{lattice}}+\Delta^2_{\textrm{high-order}})^{-1/2}$, where $\Delta_{\textrm{lattice}}$ is the uncertainty of the lattice data and $\Delta_{\textrm{high-order}}=(m_{\textrm{lattice}}
-m_{\textrm{lattice spin-average}})\times\Lambda_{\rm QCD}/m$ is the estimated error due to higher-order terms in the potential. The $V^{np\,(j)}_i$'s in units of their natural size as powers of $\Lambda_{\rm QCD}$ are introduced to the fit through a prior. We take $\Lambda_{\rm QCD}=0.5$~GeV. 
\begin{figure*}[!t]
\begin{center}
 \includegraphics[height=0.25\textheight,width=0.40\textwidth]{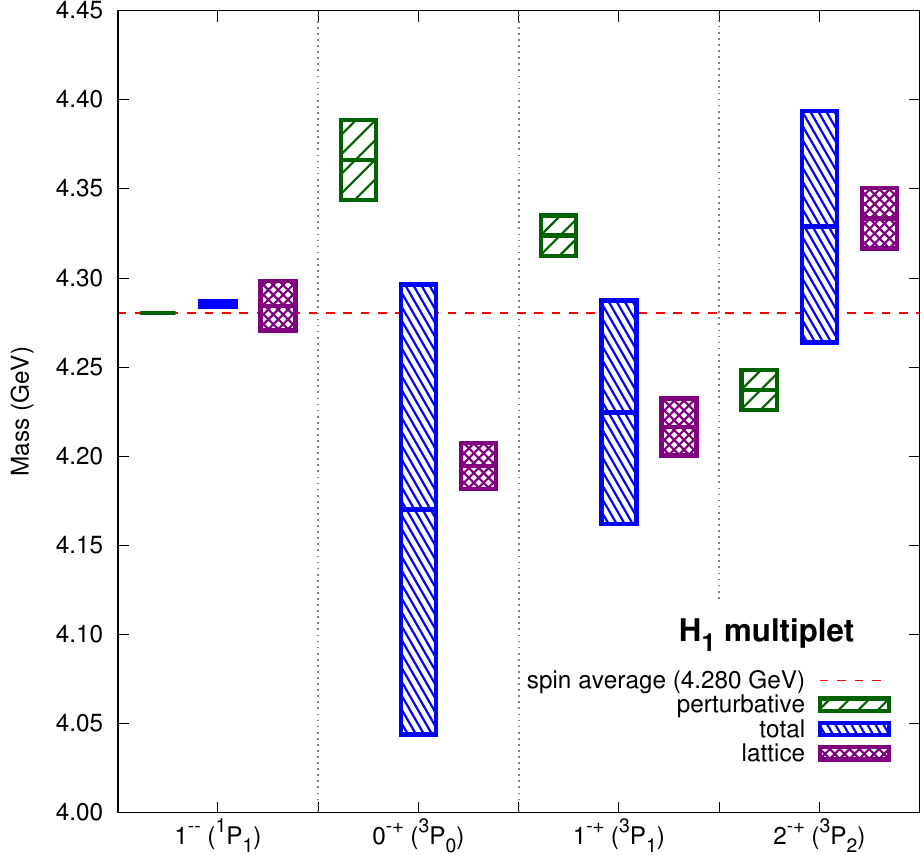}
\hspace*{0.50cm}
\includegraphics[height=0.25\textheight,width=0.40\textwidth]{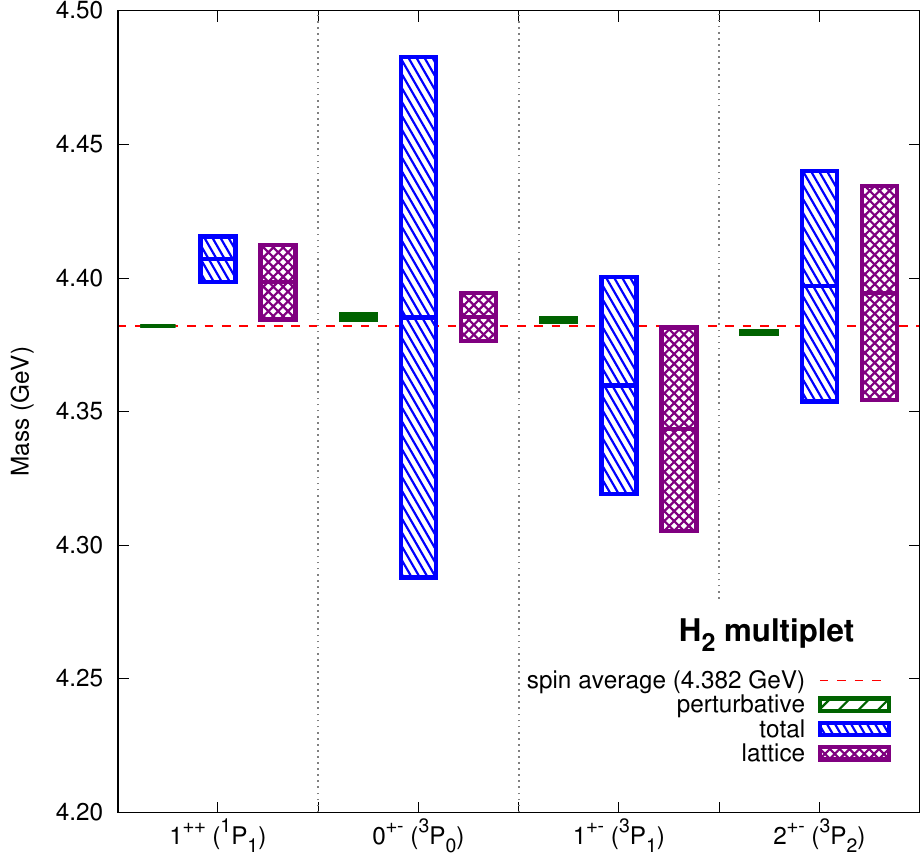}
\\[2ex]
\includegraphics[height=0.25\textheight,width=0.40\textwidth]{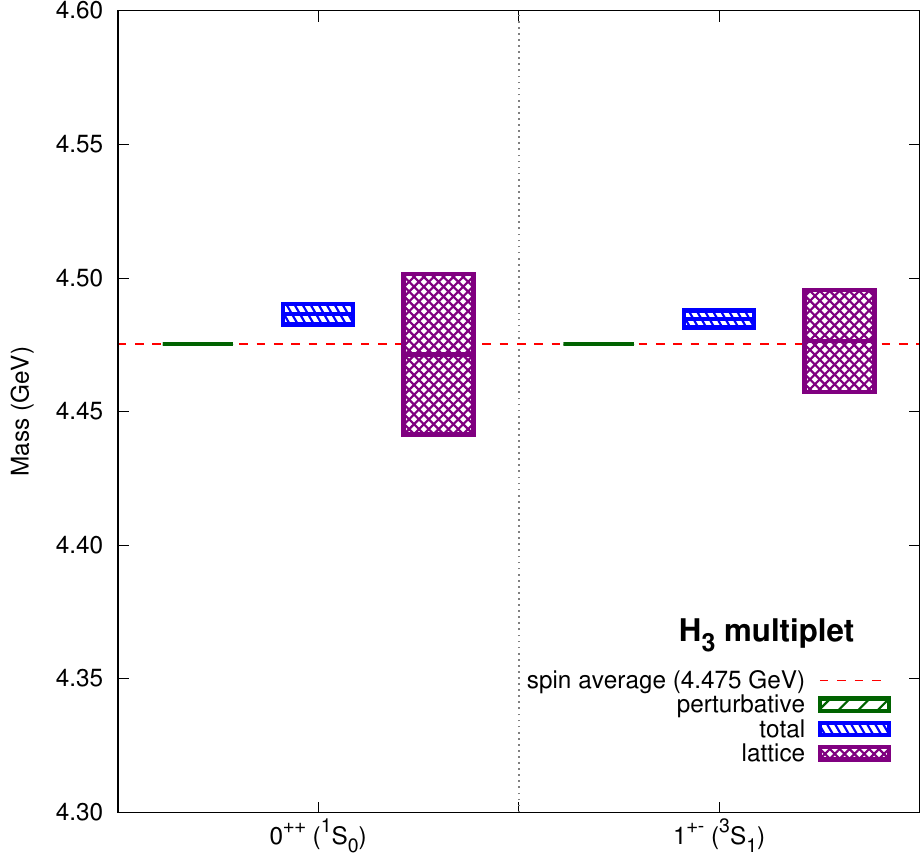}
\hspace*{0.50cm}
\includegraphics[height=0.25\textheight,width=0.40\textwidth]{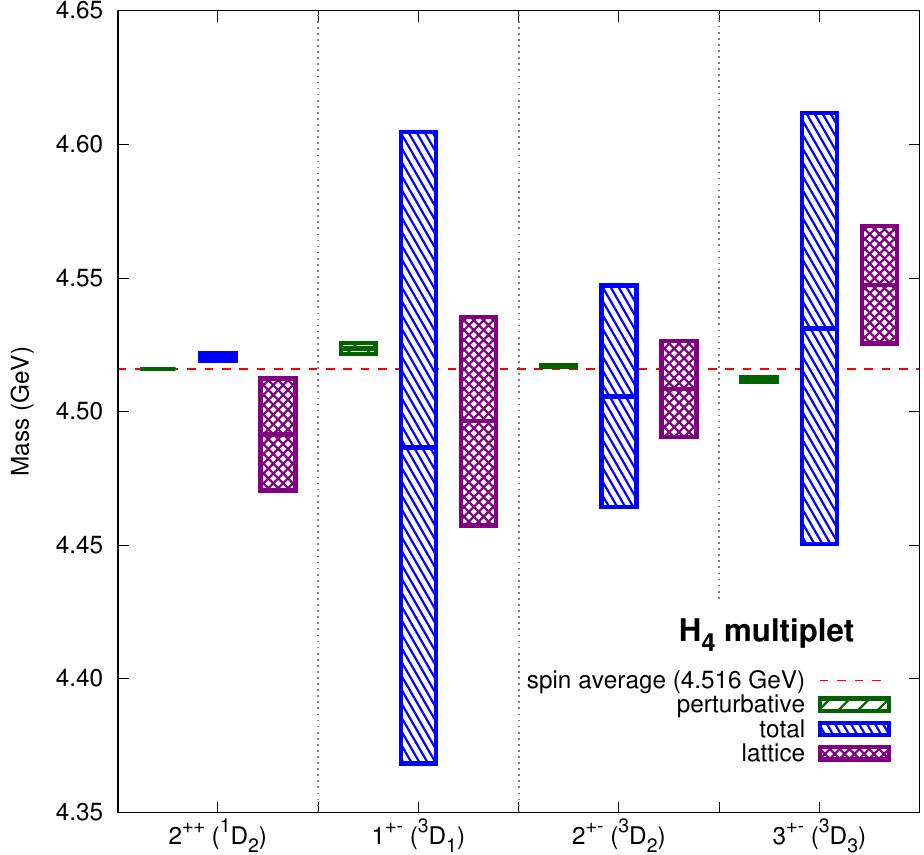}
\caption{Spectrum of the four lowest-lying charmonium hybrid multiplets. The lattice results from Ref.~\cite{Liu:2012ze} with $m_{\pi}\approx 400$~MeV are plotted in purple. In green we plotted the perturbative contributions to the spin-dependent operators in Eq.~\eqref{sdm2} added to the spin average of the lattice results (red dashed line). In blue we show the full result of the spin-dependent operators of Eqs.~\eqref{sdm}-\eqref{sdm2} including perturbative and nonperturbative contributions. The unknown nonperturbative matching coefficients are fitted to reproduce the lattice data. The height of the boxes indicate the uncertainty as detailed in the text.}
\label{fg:ccg_Liu}
\end{center}
\end{figure*}
\begin{figure*}[!t]
\begin{center}
\includegraphics[height=0.25\textheight,width=0.40\textwidth]{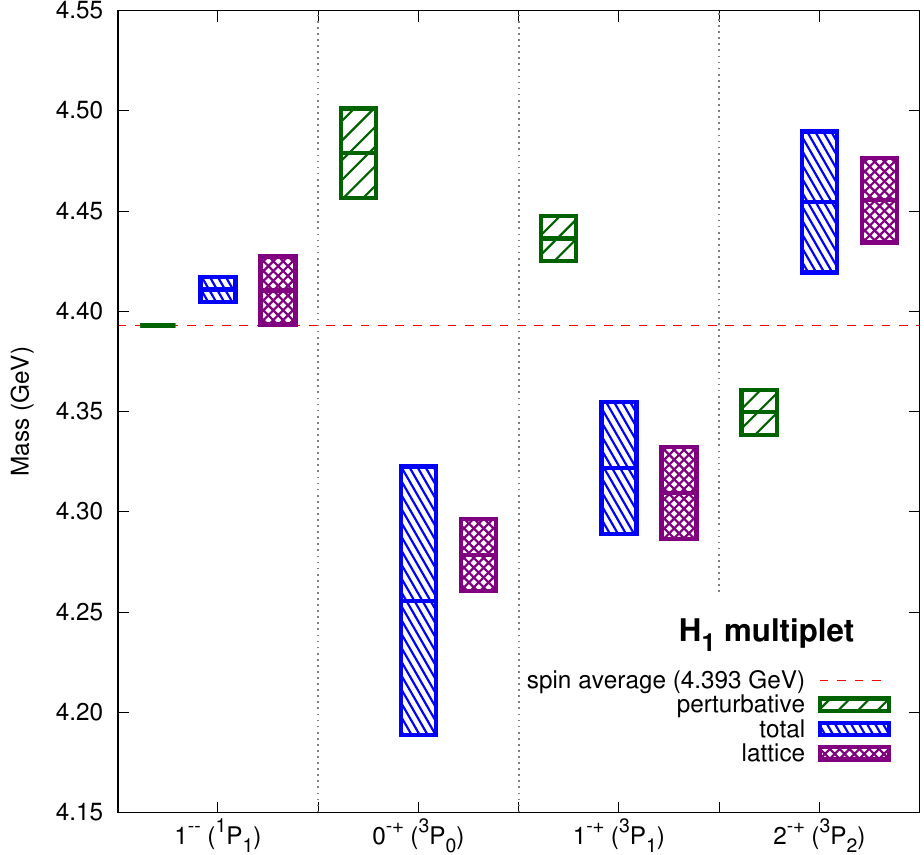}
\hspace*{0.50cm}
\includegraphics[height=0.25\textheight,width=0.40\textwidth]{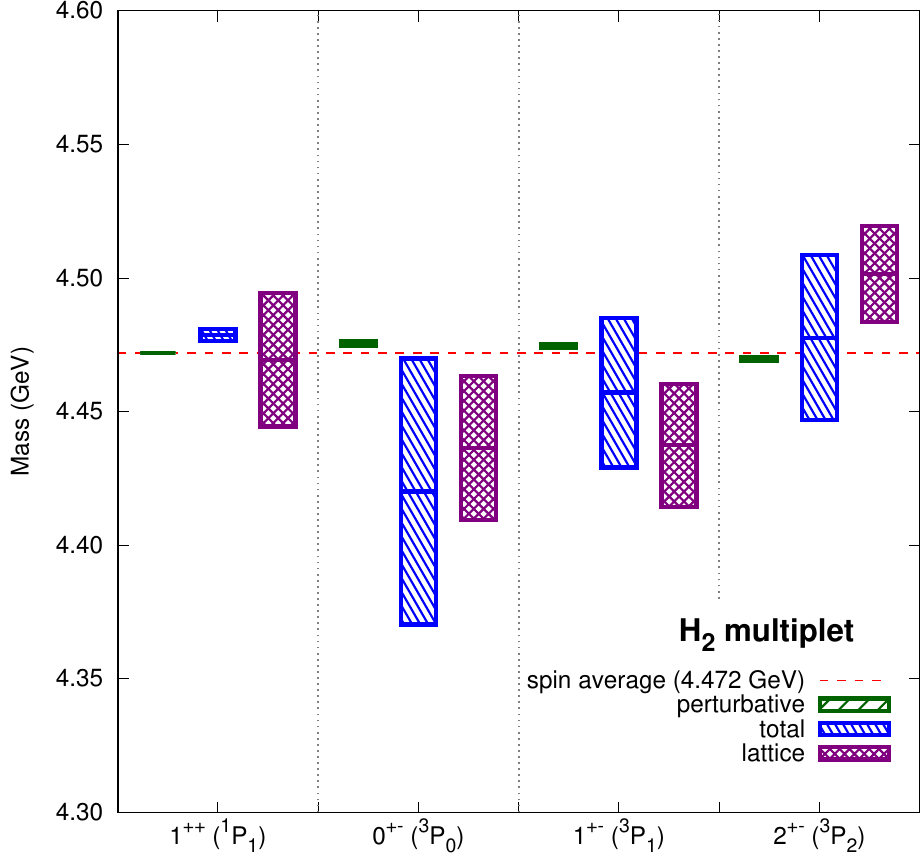} 
\\[2ex]
\includegraphics[height=0.25\textheight,width=0.40\textwidth]{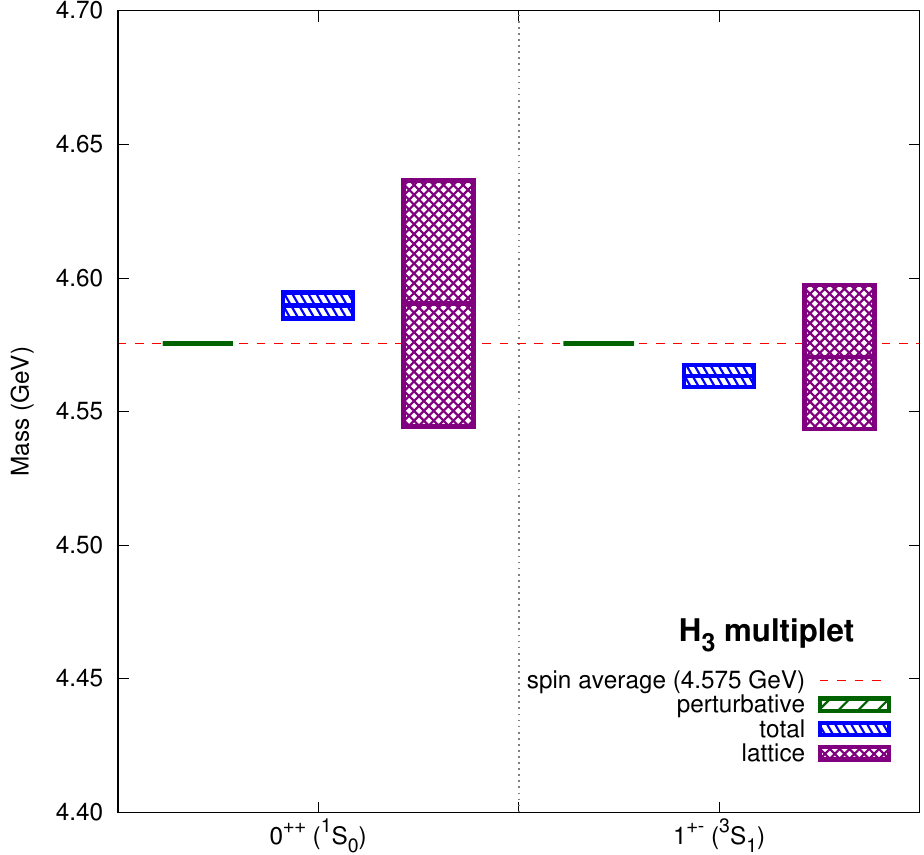}
\hspace*{0.50cm}
\includegraphics[height=0.25\textheight,width=0.40\textwidth]{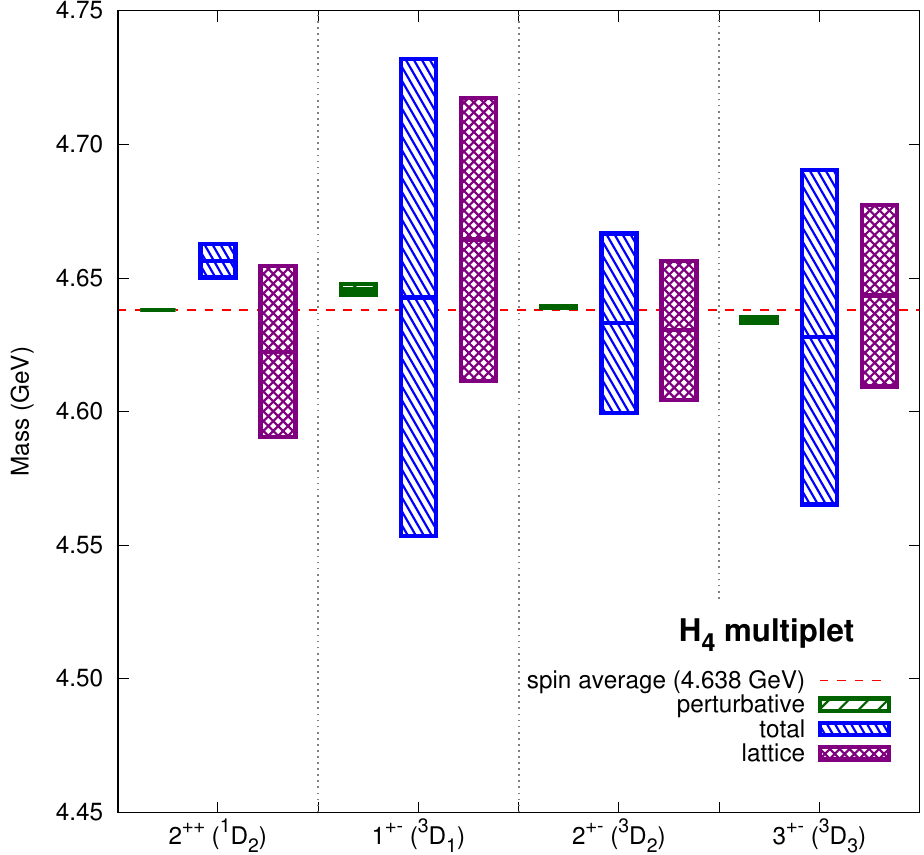}
\caption{Spectrum of the four lowest-lying charmonium hybrid multiplets. 
Same as in Figure~\ref{fg:ccg_Liu}, except using the lattice results from Ref.~\cite{Cheung:2016bym} with $m_\pi\approx 240$ MeV.}
\label{fg:ccg_Cheung}
\end{center}
\end{figure*}
\begin{table}[!t]
\caption{Nonperturbative matching coefficients determined by fitting charmonium hybrid spectrum obtained from the hybrid BOEFT to the lattice spectrum from the Hadron Spectrum Collaboration data of Refs.~\cite{Liu:2012ze} and~\cite{Cheung:2016bym} with pion masses of $m_{\pi}\approx 400\textrm{ MeV}$ and $m_{\pi}\approx 240\textrm{ MeV}$ respectively. The matching coefficients are normalized to their parametric natural size. We take the value $\Lambda_{\rm QCD}=0.5$ GeV.}
\begin{center}
\begin{tabular}{c|c|c}\hline\hline
                               & Ref.~\cite{Liu:2012ze}    & Ref.~\cite{Cheung:2016bym}\\ \hline
$\tilde{V}^{np\,(0)}_{SK}/\Lambda^2_{\rm QCD}$      &  $+1.50$  & $+1.03$\\
$V^{ np\,(1)}_{SK}/\Lambda^4_{\rm QCD}$      &  $-0.65$  & $-0.51$\\
$V^{ np\,(0)}_{SKb}/\Lambda^4_{\rm QCD}$      &  $+0.22$  & $+0.28$\\
$V^{ np\,(0)}_{SLa}/\Lambda^3_{\rm QCD}$     &  $+0.81$  & $-1.32$\\
$V^{np\,(0)}_{SLb}/\Lambda^3_{\rm QCD}$     &  $+1.18$  & $+2.44$\\
$V^{ np\,(0)}_{p\times S}/\Lambda^3_{\rm QCD}$      &  $+0.75$  & $+0.87$\\
$V^{np\, (0)}_{S^2}/\Lambda^3_{\rm QCD}$     &  $-0.26$  & $-0.33$\\
$V^{np\,(0)}_{S_{12}b}/\Lambda^3_{\rm QCD}$ &  $+0.69$  & $-0.39$\\
\hline\hline
\end{tabular}
\label{tb:npfit}
\end{center}
\end{table}
The results of the fit are shown in Figs.~\ref{fg:ccg_Liu} and~\ref{fg:ccg_Cheung} for the lattice data of Refs.~\cite{Liu:2012ze} and~\cite{Cheung:2016bym} respectively, and the obtained values of the nonperturbative matching coefficients are shown in Table~\ref{tb:npfit}. Each panel in Figs.~\ref{fg:ccg_Liu} and~\ref{fg:ccg_Cheung} corresponds to one of the multiplets of Table~\ref{tb:spin_multiplet}. The purple boxes indicate the lattice results: the middle line corresponds to the mass of the state obtained from the lattice and the height of the box corresponds to the uncertainty. The red dashed line indicates the spin average mass of the lattice results. The green boxes correspond to the contribution to the spin-splittings from the perturbative contributions to Eqs.~\eqref{eq:V_SK_first}-\eqref{eq:V_S12b_first}, i.e, the contributions from the spin-dependent terms of the octet potential in Eqs.~\eqref{vols}-\eqref{vos12}. The height of the green box ($\Delta_{\rm p}$) is an estimate on the uncertainty given by the parametric size of higher order corrections, $\mathcal{O}(m\alpha_s^5)$, to the potentials in Eqs.~\eqref{vols}-\eqref{vos12}. The blue boxes are the full results including the nonperturbative contributions after fitting the eight nonperturbative parameters to the lattice data. The height of the blue box corresponds to the uncertainty of the full result. This uncertainty is given by $\Delta_{\rm full}=(\Delta_{\rm p}^2 +\Delta^2_{\rm np}+\Delta^2_{\rm fit})^{1/2}$, where the uncertainty of the nonperturbative contribution $\Delta_{\rm np}$ is estimated to be of parametric size of higher order corrections, $\mathcal{O}(\Lambda_{\rm QCD}(\Lambda_{\rm QCD}/m)^3)$, to the matching coefficients. $\Delta_{\rm fit}$ is the statistical error of the fit. For the fits to both sets of lattice data, the resulting $\chi^2$/d.o.f. for the eight $V^{np\,(j)}_i$'s is $0.999$. It should be noted that the leading contribution $V^{ np\,(0)}_{ SK}$ has the most dominant effect. 

An interesting feature is that for the spin-triplets, the value of the perturbative contributions decreases with $J$. This trend is opposite to that of the lattice results. This discrepancy can be reconciled thanks to the nonperturbative contributions, in particular due to the contribution from $V^{np\,(0)}_{SK}$, which is only suppressed by $1/m$, and has no perturbative counterpart. A consequence of the countervail of the perturbative contribution is a relatively large uncertainty on the full result with respect its absolute value caused by a large nonperturbative contribution. Due to this uncertainty the mass hierarchies among the spin-triplet states of the multiplets $H_2$ and $H_4$ are not firmly determined. This is reflected on the change of the mass hierarchies for the central values of the lattice data from Ref.~\cite{Liu:2012ze} to Ref.~\cite{Cheung:2016bym}.

All the dependence on the heavy-quark mass of the $V^{np\,(j)}_i$'s in Eqs.~\eqref{eq:V_SK_0} and \eqref{eq:V_SK_2}-\eqref{eq:V_S_12_b} is encoded in the NRQCD matching coefficients $c_F$ and $c_s$. At leading order in $\alpha_s$ these coefficients are known to be equal to $1$ and the dependence on the heavy-quark mass only appears when the next-to-leading order is considered \cite{Manohar:1997qy}. Hence, at the order we are working, only the 
heavy-quark mass dependence of $c_F$ in Eq.~\eqref{eq:V_SK_0} is relevant. We use the one-loop expression of $c_F$ in Eq.~(\ref{eq:V_SK_0}), with the renormalization scale set as the heavy-quark mass. Taking this mass dependence into account, we can use the set of nonperturbative parameters to predict the spin contributions in the bottomonium hybrid sector, for which lattice determinations are yet not available due to their larger difficulty compared to the charm sector.

\begin{figure*}[!t]
\begin{center}
\includegraphics[height=0.25\textheight,width=0.40\textwidth]{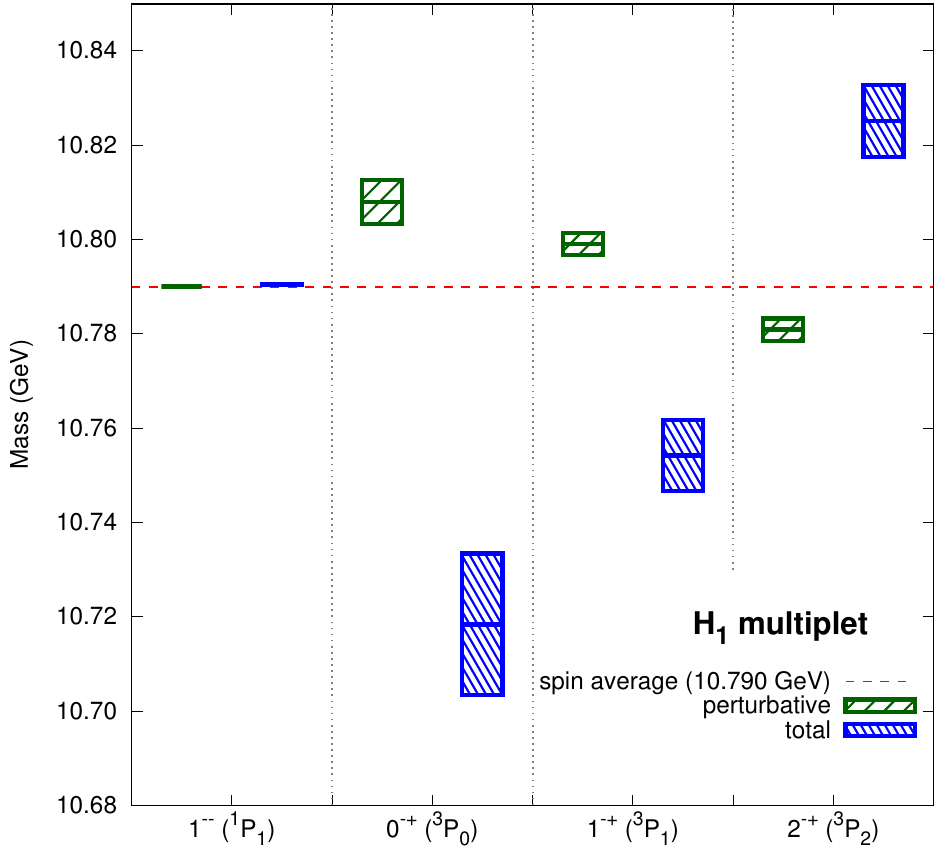}
\hspace*{0.50cm}
\includegraphics[height=0.25\textheight,width=0.40\textwidth]{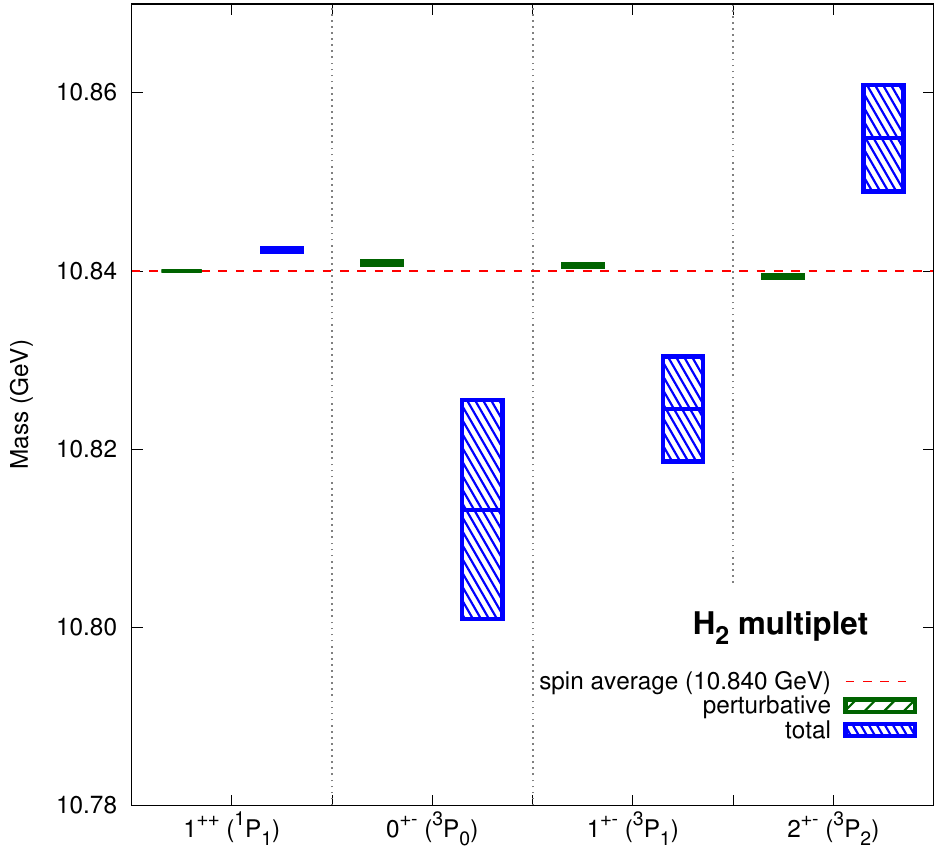} 
\\[2ex]
\includegraphics[height=0.25\textheight,width=0.40\textwidth]{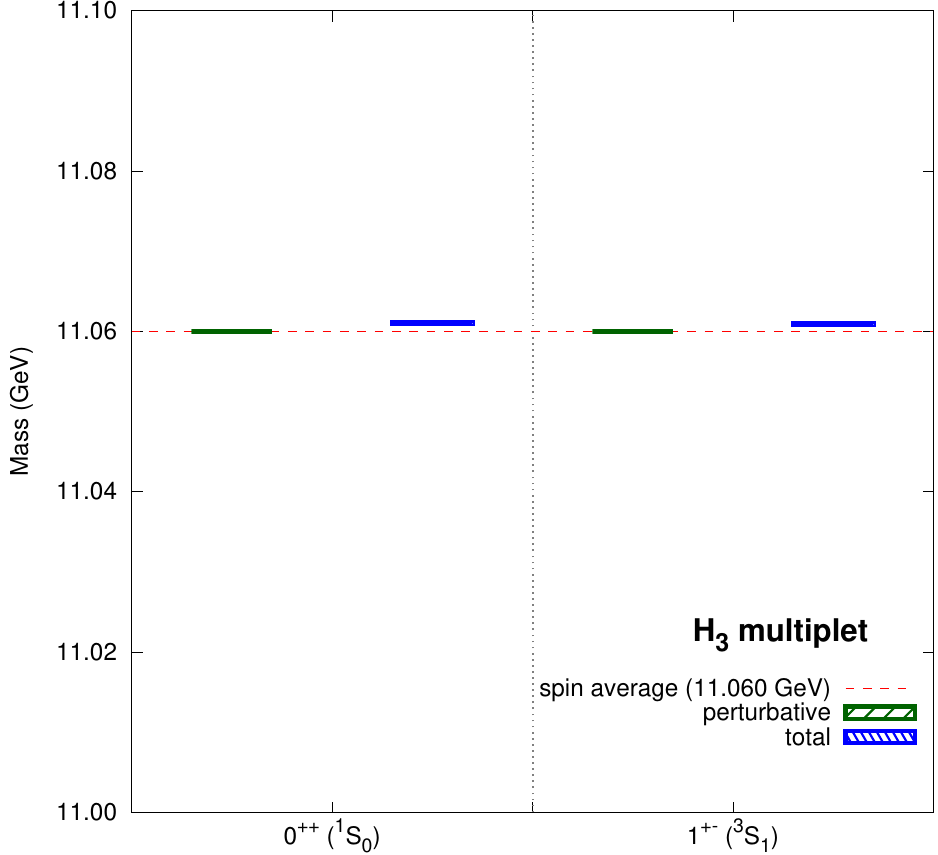}
\hspace*{0.50cm}
\includegraphics[height=0.25\textheight,width=0.40\textwidth]{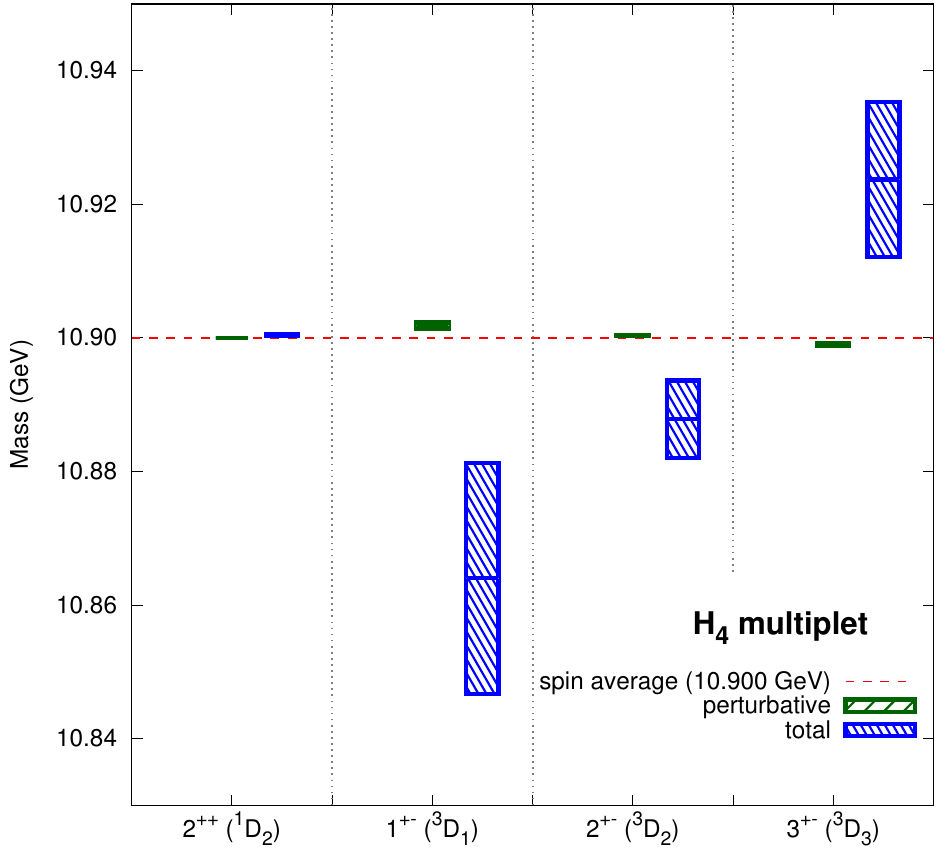}
\caption{Spectrum of the four lowest-lying bottomonium hybrids computed by adding the spin-dependent contributions from Eqs.~\eqref{eq:V_SK_first}-\eqref{eq:V_S12b_first} to the spectrum obtained in Ref.~\cite{Berwein:2015vca}. The values of nonperturbative contribution to the matching coefficients are determined from the fit of the charmonium hybrids spectrum obtained from the BOEFT to the lattice data of Ref.~\cite{Liu:2012ze} shown in Fig.~\ref{fg:ccg_Liu}. The average mass for each multiplet is shown as a red dashed line. The results with only the perturbative contributions and the full results for the matching coefficients are shown as green and blue boxes respectively. The height of the boxes indicates the uncertainty as detailed in the text.}
\label{fig:bbg_Liu}
\end{center}
\end{figure*}

\begin{figure*}[!t]
\begin{center}
\includegraphics[height=0.25\textheight,width=0.40\textwidth]{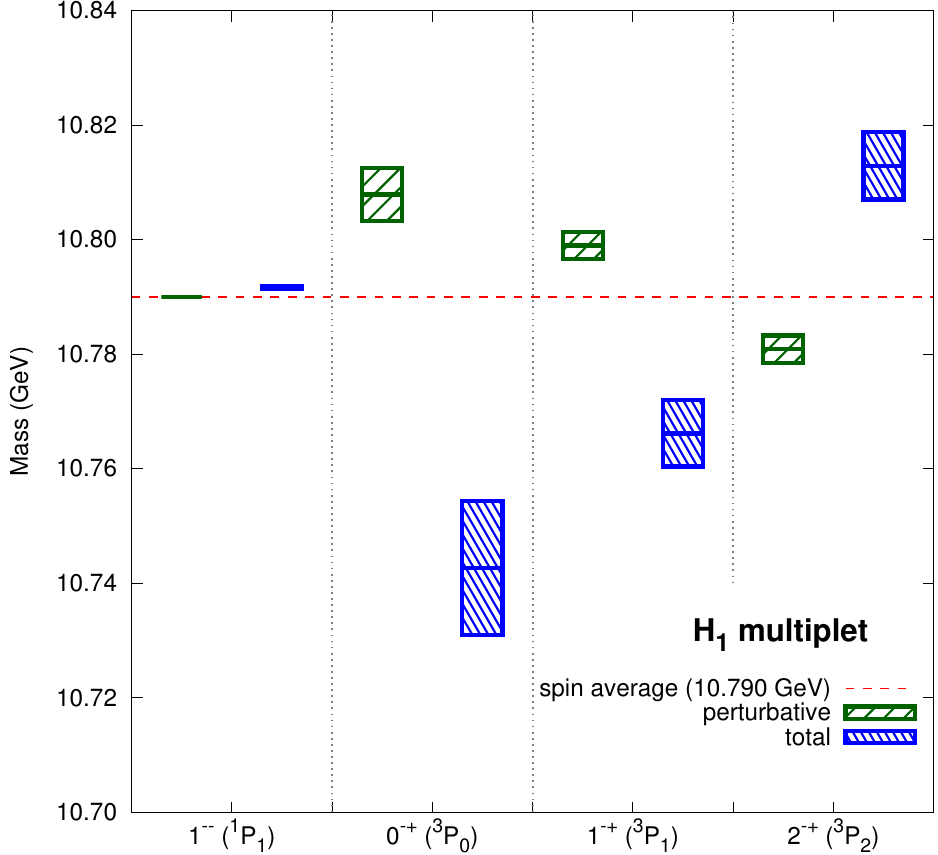}
\hspace*{0.50cm}
\includegraphics[height=0.25\textheight,width=0.40\textwidth]{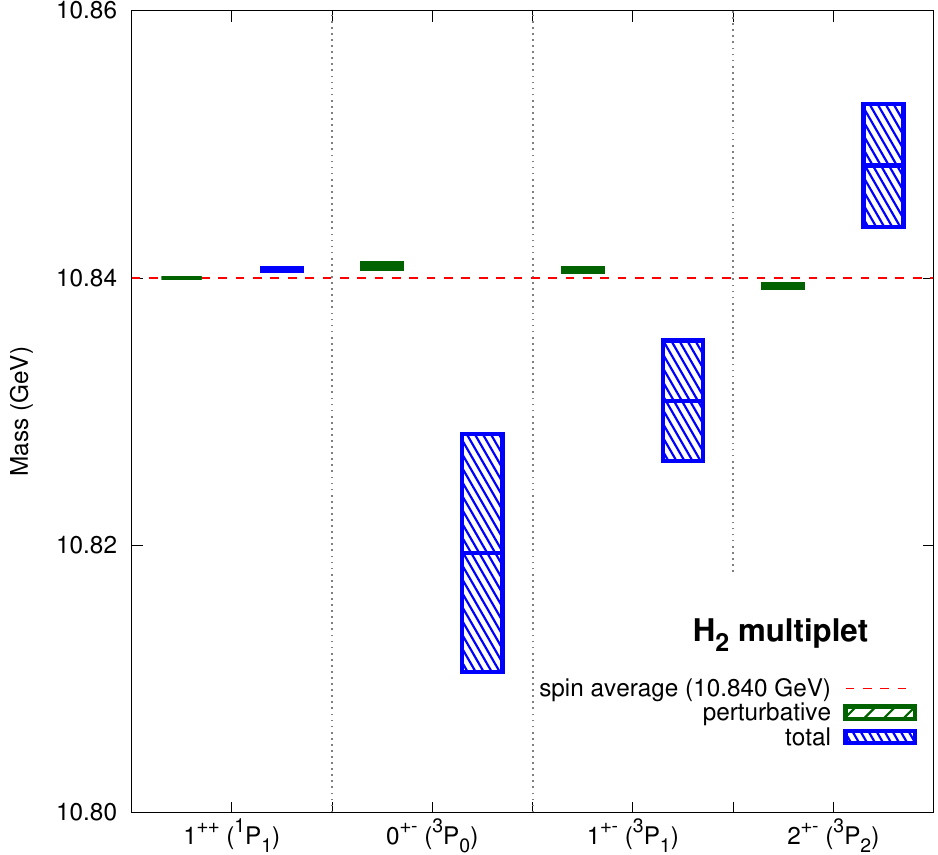} 
\\[2ex]
\includegraphics[height=0.25\textheight,width=0.40\textwidth]{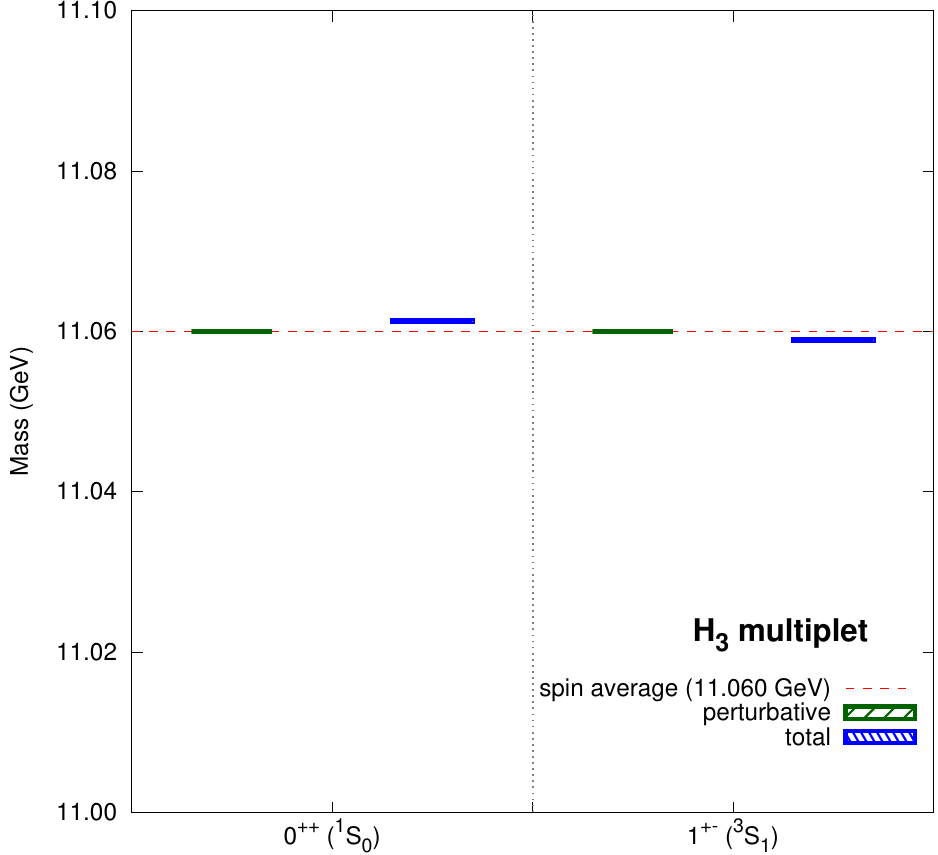}
\hspace*{0.50cm}
\includegraphics[height=0.25\textheight,width=0.40\textwidth]{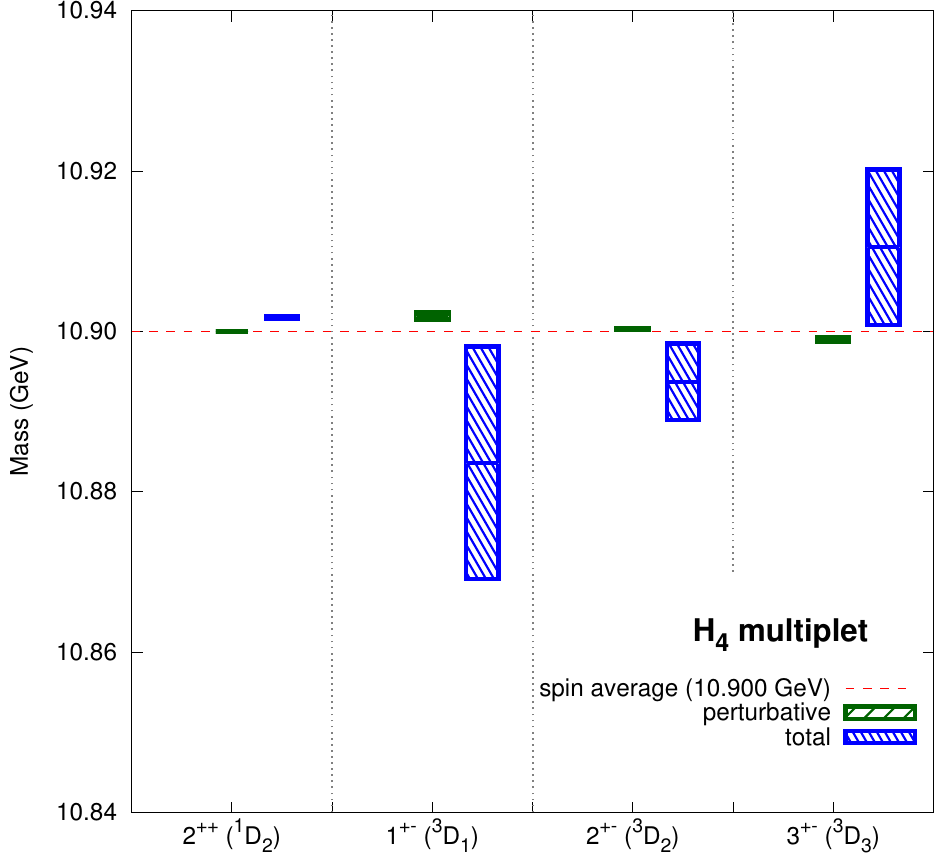}
\caption{Spectrum of the four lowest-lying bottomonium hybrids computed by adding the spin-dependent contributions from Eqs.~\eqref{eq:V_SK_first}-\eqref{eq:V_S12b_first} to the spectrum obtained in Ref.~\cite{Berwein:2015vca}. 
Same as in Figure~\ref{fig:bbg_Liu}, except
using the result of the fit in Fig.~\ref{fg:ccg_Cheung} with the lattice data from Ref.~\cite{Cheung:2016bym}.}
\label{fig:bbg_Cheung}
\end{center}
\end{figure*}

We compute the bottomonium hybrids spectrum by adding the spin-dependent contributions from Eqs.~\eqref{eq:V_SK_first}-\eqref{eq:V_S12b_first} to the spectrum obtained in Ref.~\cite{Berwein:2015vca}. We show the results thus obtained in Figs.~\ref{fig:bbg_Liu} and~\ref{fig:bbg_Cheung} for the values in the second and third columns of Table~\ref{tb:npfit} respectively. We use the value of the bottom mass $m_b^{RS}(\textrm{1 GeV})=4.863$GeV. 
\section{Conclusions}\label{conc}

The spin-dependent operators for heavy quarkonium hybrids up to order $1/m^2$ were presented in Ref.~\cite{Brambilla:2018pyn}. The most prominent feature is the appearance of two spin-dependent operators already at order $1/m$, unlike standard quarkonia, in which case the spin-dependent operators appear at order $1/m^2$. These operators, in Eq.~\eqref{sdm}, couple the total spin of the heavy-quark-antiquark pair with the spin of the gluonic degrees of freedom that generates the hybrid state. At order $1/m^2$, we have the spin-orbit, total spin squared, and tensor spin operators familiar in the studies of standard quarkonia. In addition, three new operators appear at order $1/m^2$, which involve the projection operators that project the gluonic degrees of freedom onto representations of $D_{\infty h}$, and can be viewed as generalizations of the spin-orbit and tensor spin operators to hybrid states. All the $1/m^2$ spin-dependent operators are shown in Eq.~\eqref{sdm2}. The structure of the spin-dependent operators is valid for both $r\ll 1/\Lambda_{\rm QCD}$ and $r\sim 1/\Lambda_{\rm QCD}$, however the power counting and the form of the potentials are different in these two regimes. Here we have explicitly worked out the case 
$r\ll 1/\Lambda_{\rm QCD}$.

In the short heavy-quark-antiquark distance regime, $r\ll 1/\Lambda_{\rm QCD}$,  the matching coefficients, i.e. the potentials, of the spin-dependent operators of the BOEFT, the EFT for hybrids, are obtained by matching the two-point functions for the hybrid states in weakly-coupled pNRQCD and the BOEFT. Two types of contributions arise. The perturbative one correspond to the spin-dependent terms of the octet potential, Eqs.~\eqref{vols}-\eqref{vos12}, and are generated when the soft scale is integrated out and NRQCD matched to pNRQCD, which we review in Sec.~\ref{s1}. The nonperturbative contribution can be organized as a polynomial series in $r^2$ with coefficients encoding the gluon dynamics. In this paper, we present the expressions for these coefficients, Eqs.~(\ref{eq:V_SK_0}) and (\ref{eq:V_SK_2})-(\ref{eq:V_S_12_b}), in terms of integrals over the insertion time of chromoelectric or chromomagnetic fields in gauge invariant correlators of the gluonic degrees of freedom, Eqs.~(\ref{eq:UB}) and~(\ref{eq:UssEE})-(\ref{eq:UoooEEB}). In these nonperturbative contributions, the dependence on heavy quark mass appears only in NRQCD matching coefficients, such as $c_F$ and $c_s$, and is factored out from the gluonic correlators. The details of the calculation can be found in Sec.~\ref{pot} and Appendix~\ref{app_matching}. To reduce the two-point function in pNRQCD to a form matching the one in
the BOEFT, it is necessary to use relations between different gluonic correlators derived from the transformation properties of the gluon fields under $C$, $P$ and $T$ . These relations can be found in
Appendix~\ref{app_CPT}. 

The values of the nonperturbative contributions can be obtained by evaluating on the lattice the gluonic correlators we provide. These computations are at the moment not available. Nevertheless, these values can be estimated by comparing with direct lattice computations of the hybrid charmonium spectrum. To do so, we compute the contributions of the spin-dependent operators to the hybrid spectrum to $\mathcal{O}(\Lambda^3_{\rm QCD}/m^2)$ using standard time-independent perturbation theory in Sec.~\ref{sec4}. We have used the charmonium hybrid spectrum computed on the lattice by the Hadron Spectrum Collaboration in Refs.~\cite{Liu:2012ze} and \cite{Cheung:2016bym} and fit the values of the nonperturbative contributions to the matching coefficients to reproduce the lattice spectrum. The results are shown in Figs.~\ref{fg:ccg_Liu} and \ref{fg:ccg_Cheung} and Table~\ref{tb:npfit}. We found that it is possible to reproduce the lattice data of the charmonium hybrid spectrum with nonperturbative matching coefficients of natural sizes. The values of the pion mass utilized in Refs.~\cite{Liu:2012ze} and \cite{Cheung:2016bym} are $m_{\pi}\approx 400$ MeV and $m_{\pi}\approx 240$ MeV respectively. The variation of the values of the nonperturbative matching coefficients obtained by the fits for the two lattice data sets can be tentatively attributed to the light-quark mass dependence of the gluon correlators, in particular for the matching coefficient of the leading spin-dependent operator $V^{np\,(0)}_{SK}$.

Finally, we have taken advantage of the fact that the gluonic correlators are independent of the heavy-quark flavor to compute the mass spin-splittings for the bottomonium hybrid spectrum. The results are shown in Figs.~\ref{fig:bbg_Cheung} and \ref{fig:bbg_Liu}. The bottomonium hybrid spectrum including spin-dependent contributions has not yet been computed on the lattice\footnote{In Ref.~\cite{Liao:2001yh} three states were identified as hybrids.}. Calculations of the bottomonium hybrid spectrum on the lattice are difficult due to the widely-separated scales of the system, i.e. the bottom-quark mass being much larger than $\Lambda_{\rm QCD}$. Therefore, precision calculations of the bottomonium hybrid spectrum on the lattice would require both large volume and small lattice spacing, which is computationally challenging. On the other hand, an EFT approach can take advantage of the same wide separation of scales and is, like lattice QCD, a model-independent approach rooted in QCD. Combining both approaches opens a promising path towards the understanding of exotic quarkonia.

The impact of this calculation is manifold. First, as observed above, the spin dependence of the operators for quarkonium hybrids is significantly different from that for standard quarkonia. This has an important impact on the phenomenological calculation. Second, we have obtained for the first time the expressions of the nonperturbative contributions to the spin-dependent potential for quarkonium hybrids in terms of gauge-invariant correlators depending only on the gluonic degrees of freedom, which are suitable for computation on the lattice or evaluation in QCD vacuum models. The technology to calculate these correlators in lattice QCD already exists \cite{Capitani:2018rox,Mueller:2019mkh,Bicudo:2018jbb,Bicudo:2018yhk,Yanagihara:2019foh} and could be readily applied. Third, we emphasize that the obtained nonperturbative correlators depend only on the gluonic degrees of freedom and not on the heavy-quark flavor. This allows us to extract the unknown nonperturbative parameters in the spin-dependent potential from the charmonium hybrid spectrum and use them for bottomonium hybrids. Finally, since the BOEFT can be generalized to considering also light quarks as the light degrees of freedom~\cite{TarrusCastella:2019rit}, the spin dependent operators will likely have similar characteristics also in that case. Therefore, we supply the full list of matrix elements of the spin-dependent operators in Appendix~\ref{app_elements} to facilitate applications to the spectrum of XYZ states for phenomenologists and model builders. Since most of the phenomenological applications for XYZ states up to date either do not contain such spin-dependent terms or construct them in a way inspired by the traditional quarkonium case, we believe that this result can prove to be very useful. 

The next step forward in the BOEFT framework will be to release the assumption $mv\gg\Lambda_{\rm QCD}$ we used in this paper, and work out the spin-dependent corrections using only the hierarchy $\Lambda_{\rm QCD} \gg mv^2$ underlying the Born-Oppenheimer approximation~\cite{Soto:2017one}. In this case the spin decomposition of the potential will be the same
as obtained here but the actual form of the $r$-dependent potentials will be given in terms of generalized Wilson loops. This is analogous to the computation of the spin-dependent potential for traditional quarkonia as generalized Wilson loops in strongly-coupled pNRQCD~\cite{Eichten:1980mw,Brambilla:2000gk,Pineda:2000sz} which was later used by lattice groups to obtain the form of the nonperturbative spin-dependent potentials~\cite{Bali:1996bw,Koma:2007jq,Koma:2012bc} and can be addressed with the technology developed in~\cite{Juge:1997nc,Juge:2002br,Capitani:2018rox}.

\section*{Acknowledgements}

We are grateful to Antonio Vairo for suggesting us the computation of the spin-dependent potentials in the short-distance regime and for the collaboration in the early stage of this work. We also thank 
Antonio Pineda for useful comments on the spin-dependent terms of the octet potential. This work has been supported by the DFG and the NSFC through funds provided to the Sino-German CRC 110 ``Symmetries and the Emergence of Structure in QCD'', and by the DFG cluster of excellence ``Origin and structure of the universe'' (www.universe-cluster.de). 
N.B. and W.K.L. acknowledge support from the DFG cluster of excellence
``Origins" (www.origins-cluster.de).
J.S. acknowledges support by Spanish MINECO's grant No. FPA2017-86380-P and the Junta de Andaluc\'ia grant No. FQM-370.  J.T.C. has been partially supported in part by the Spanish grants FPA2017-86989-P and SEV-2016-0588 from the Ministerio de Ciencia, Innovaci\'on y Universidades, and the grant 2017SGR1069 from the Generalitat de Catalunya. J.T.C acknowledges the financial support from the European Union's Horizon 2020 research and innovation programme under the Marie Sk\l{}odowska--Curie Grant Agreement No. 665919.

\appendix
\section{Identities for gluonic correlators from \texorpdfstring{$C$, $P$, $T$}{CPT}}\label{app_CPT}
In this appendix, we list the identities derived from the transformation properties of the fields in the correlators in Eqs.~(\ref{eq:UB}) and
(\ref{eq:UssEE})-(\ref{eq:UoooEEB}) under $C$, $P$, $T$ that have been used in the matching calculation of the spin-dependent BOEFT potentials, shown in detail in Sec.~\ref{pot} and Appendix~\ref{app_matching}. 

Under $C$, $P$, $T$, the operators ${E}^{ia}(\bm{R},t)$, ${B}^{ia}(\bm{R},t)$, and $\phi^{ab}(\bm{R};t,t')$ 
transform as
\begin{align}
{C}E^{ia}(\bm{R},t){C}^{-1}&=-(-)^aE^{ia}(\bm{R},t)\,,\\ 
{C}B^{ia}(\bm{R},t){C}^{-1}&=-(-)^aB^{ia}(\bm{R},t)\,,\\
C\phi^{ab}(\bm{R};t,t')C^{-1}&=(-)^a \phi^{ab}(\bm{R};t,t') (-)^b\,,\\
{P}E^{ia}(\bm{R},t){P}^{-1}&=-E^{ia}(-\bm{R},t)\,,\\ 
{P}B^{ia}(\bm{R},t){P}^{-1}&=B^{ia}(-\bm{R},t)\,,\\
P\phi^{ab}(\bm{R};t,t')P^{-1}&= \phi^{ab}(-\bm{R};t,t')\,,\\
{T}E^{ia}(\bm{R},t){T}^{-1}&=(-)^aE^{ia}(\bm{R},-t)\,,\\ 
{T}B^{ia}(\bm{R},t){T}^{-1}&=-(-)^aB^{ia}(\bm{R},-t)\,,\\
T\phi^{ab}(\bm{R};t,t')T^{-1}&=(-)^b \phi^{ba\,\dagger}(\bm{R};-t',-t) (-)^a\,,
\end{align}
where $(-)^a\equiv 1$ for $a=1,3,4,6,8$ and $(-)^a\equiv -1$ for $a=2,5,7$. 
For $\kappa=1^{+-}$, the gluelump operator ${G}^{ia}(\bm{R},t)$ transforms under $C$, $P$, $T$ as
\begin{align}
{C}G^{ia}(\bm{R},t){C}^{-1}&=-(-)^aG^{ia}(\bm{R},t)\,,\\
{P}G^{ia}(\bm{R},t){P}^{-1}&=G^{ia}(-\bm{R},t)\,,\\
{T}G^{ia}(\bm{R},t){T}^{-1}&=-(-)^aG^{ia}(\bm{R},-t)\,.
\end{align} 
Inserting the identity operator $C^{-1}C$ between the fields in the correlators in Eqs.~(\ref{eq:UB}) and (\ref{eq:UooEE})-(\ref{eq:UoooEEB}) and assuming $C$-invariance of the vaccuum give
\begin{align}
\left(U_B\right)^{ijk}_{bcd}T^b_{\alpha\beta}T^c_{\gamma\delta}T^d_{\rho\sigma}&=
-\left(U_B\right)^{ijk}_{bcd}(T^b)^T_{\alpha\beta}(T^c)^T_{\gamma\delta}(T^d)^T_{\rho\sigma}\,,\label{eq:sym_UB_C}\\
\left(U^{oo}_{EE}\right)^{ijkl}_{bcdefg}T^b_{\alpha\beta}T^c_{\gamma\delta}T^d_{\rho\sigma}T^e_{\eta\tau}T^{f}_{\omega\kappa}T^{g}_{\mu\nu}
&=\left(U^{oo}_{EE}\right)^{ijkl}_{bcdefg}(T^b)^T_{\alpha\beta}
(T^c)^T_{\gamma\delta}(T^d)^T_{\rho\sigma}(T^e)^T_{\eta\tau}(T^f)^T_{\omega\kappa}(T^g)^T_{\mu\nu}\,,\label{eq:sym_UooEE_C}\\
\left(U^{oo}_{BB}\right)^{ijkl}_{bcdefg}T^b_{\alpha\beta}T^c_{\gamma\delta}T^d_{\rho\sigma}T^e_{\eta\tau}T^{f}_{\omega\kappa}T^{g}_{\mu\nu}
&=\left(U^{oo}_{BB}\right)^{ijkl}_{bcdefg}(T^b)^T_{\alpha\beta}
(T^c)^T_{\gamma\delta}(T^d)^T_{\rho\sigma}(T^e)^T_{\eta\tau}(T^f)^T_{\omega\kappa}(T^g)^T_{\mu\nu}\,,\label{eq:sym_UooBB_C}\\
\left(U^{oo}_{BDE}\right)^{ijklm}_{bcdefg}T^b_{\alpha\beta}T^c_{\gamma\delta}T^d_{\rho\sigma}T^e_{\eta\tau}T^{f}_{\omega\kappa}T^{g}_{\mu\nu}
&=\left(U^{oo}_{BDE}\right)^{ijklm}_{bcdefg}(T^b)^T_{\alpha\beta}
(T^c)^T_{\gamma\delta}(T^d)^T_{\rho\sigma}(T^e)^T_{\eta\tau}(T^f)^T_{\omega\kappa}(T^g)^T_{\mu\nu}\,,\label{eq:sym_UooBDE_C}\\
\left(U^{oo}_{DEB}\right)^{ijklm}_{bcdefg}T^b_{\alpha\beta}T^c_{\gamma\delta}T^d_{\rho\sigma}T^e_{\eta\tau}T^{f}_{\omega\kappa}T^{g}_{\mu\nu}
&=\left(U^{oo}_{DEB}\right)^{ijklm}_{bcdefg}(T^b)^T_{\alpha\beta}
(T^c)^T_{\gamma\delta}(T^d)^T_{\rho\sigma}(T^e)^T_{\eta\tau}(T^f)^T_{\omega\kappa}(T^g)^T_{\mu\nu}\,,\label{eq:sym_UooDEB_C}\\
(U^{sso}_{BEE})^{iljmk}_{def}T^d_{\alpha\beta}T^e_{\gamma\delta}T^f_{\rho\sigma}&=-(U^{sso}_{BEE})^{iljmk}_{def}(T^d)^T_{\alpha\beta}(T^e)^T_{\gamma\delta}(T^f)^T_{\rho\sigma}
\,,\label{eq:sym_UssoBEE_C}\\
(U^{sso}_{EBE})^{iljmk}_{def}T^d_{\alpha\beta}T^e_{\gamma\delta}T^f_{\rho\sigma}&=-(U^{sso}_{EBE})^{iljmk}_{def}(T^d)^T_{\alpha\beta}(T^e)^T_{\gamma\delta}(T^f)^T_{\rho\sigma}
\,,\label{eq:sym_UssoEBE_C}\\
(U^{sso}_{EEB})^{iljmk}_{def}T^d_{\alpha\beta}T^e_{\gamma\delta}T^f_{\rho\sigma}&=-(U^{sso}_{EEB})^{iljmk}_{def}(T^d)^T_{\alpha\beta}(T^e)^T_{\gamma\delta}(T^f)^T_{\rho\sigma}
\,,\label{eq:sym_UssoEEB_C}\\
(U^{oss}_{BEE})^{iljmk}_{bcd}T^b_{\alpha\beta}T^c_{\gamma\delta}T^d_{\rho\sigma}&=-(U^{oss}_{BEE})^{iljmk}_{bcd}(T^b)^T_{\alpha\beta}(T^c)^T_{\gamma\delta}(T^d)^T_{\rho\sigma}
\,,\label{eq:sym_UossBEE_C}\\
(U^{oss}_{EBE})^{iljmk}_{bcd}T^b_{\alpha\beta}T^c_{\gamma\delta}T^d_{\rho\sigma}&=-(U^{oss}_{EBE})^{iljmk}_{bcd}(T^b)^T_{\alpha\beta}(T^c)^T_{\gamma\delta}(T^d)^T_{\rho\sigma}
\,,\label{eq:sym_UossEBE_C}\\
(U^{oss}_{EEB})^{iljmk}_{bcd}T^b_{\alpha\beta}T^c_{\gamma\delta}T^d_{\rho\sigma}&=-(U^{oss}_{EEB})^{iljmk}_{bcd}(T^b)^T_{\alpha\beta}(T^c)^T_{\gamma\delta}(T^d)^T_{\rho\sigma}
\,,\label{eq:sym_UossEEB_C}
\end{align}
\begin{align}
&(U^{ooo}_{BEE})^{iljmk}_{bcdefghpq}T^b_{\alpha\beta}T^c_{\gamma\delta}T^d_{\rho\sigma}T^e_{\eta\tau}T^{f}_{\omega\kappa}T^{g}_{\mu\nu}
T^h_{\epsilon\delta}T^p_{\lambda\xi}T^q_{\psi\chi}
\nonumber\\
=&\,-(U^{ooo}_{BEE})^{iljmk}_{bcdefghpq}(T^b)^T_{\alpha\beta}(T^c)^T_{\gamma\delta}(T^d)^T_{\rho\sigma}(T^e)^T_{\eta\tau}(T^f)^T_{\omega\kappa}(T^g)^T_{\mu\nu}
(T^h)^T_{\epsilon\delta}(T^p)^T_{\lambda\xi}(T^q)^T_{\psi\chi}\,,
\label{eq:sym_UoooBEE_C}\\
&(U^{ooo}_{EBE})^{iljmk}_{bcdefghpq}T^b_{\alpha\beta}T^c_{\gamma\delta}T^d_{\rho\sigma}T^e_{\eta\tau}T^{f}_{\omega\kappa}T^{g}_{\mu\nu}
T^h_{\epsilon\delta}T^p_{\lambda\xi}T^q_{\psi\chi}
\nonumber\\
=&\,-(U^{ooo}_{EBE})^{iljmk}_{bcdefghpq}(T^b)^T_{\alpha\beta}(T^c)^T_{\gamma\delta}(T^d)^T_{\rho\sigma}(T^e)^T_{\eta\tau}(T^f)^T_{\omega\kappa}(T^g)^T_{\mu\nu}
(T^h)^T_{\epsilon\delta}(T^p)^T_{\lambda\xi}(T^q)^T_{\psi\chi}\,,
\label{eq:sym_UoooEBE_C}\\
&(U^{ooo}_{EEB})^{iljmk}_{bcdefghpq}T^b_{\alpha\beta}T^c_{\gamma\delta}T^d_{\rho\sigma}T^e_{\eta\tau}T^{f}_{\omega\kappa}T^{g}_{\mu\nu}
T^h_{\epsilon\delta}T^p_{\lambda\xi}T^q_{\psi\chi}
\nonumber\\
=&\,-(U^{ooo}_{EEB})^{iljmk}_{bcdefghpq}(T^b)^T_{\alpha\beta}(T^c)^T_{\gamma\delta}(T^d)^T_{\rho\sigma}(T^e)^T_{\eta\tau}(T^f)^T_{\omega\kappa}(T^g)^T_{\mu\nu}
(T^h)^T_{\epsilon\delta}(T^p)^T_{\lambda\xi}(T^q)^T_{\psi\chi}\,.
\label{eq:sym_UoooEEB_C}
\end{align}
From Eqs.~(\ref{eq:sym_UB_C})-(\ref{eq:sym_UoooEEB_C}), taking appropriate summations on the fundamental color indices, and utilizing the cyclic symmetry and the total symmetry of indices of $h^{abc}$ and $d^{abc}$ respectively, we have
\begin{align}
\left(U_B\right)^{ijk}_{bcd}h^{bcd}&=-\left(U_B\right)^{ijk}_{bcd}h^{bdc}\,,\label{eq:sym_UB_C2}\\
\left(U^{oo}_{EE}\right)^{ijkl}_{bcdefg}h^{bcd}h^{efg}&=\left(U^{oo}_{EE}\right)^{ijkl}_{bcdefg} h^{bdc}h^{egf}\,,\label{eq:sym_UooEE_C2}\\
\left(U^{oo}_{EE}\right)^{ijkl}_{bcdefg}h^{bcd}h^{egf}&=\left(U^{oo}_{EE}\right)^{ijkl}_{bcdefg} h^{bdc}h^{efg}\,,\label{eq:sym_UooEE_C3}\\
\left(U^{oo}_{BB}\right)^{ijkl}_{bcdefg}h^{bcd}h^{efg}&=\left(U^{oo}_{BB}\right)^{ijkl}_{bcdefg} h^{bdc}h^{egf}\,,\label{eq:sym_UooBB_C2}\\
\left(U^{oo}_{BB}\right)^{ijkl}_{bcdefg}h^{bcd}h^{egf}&=\left(U^{oo}_{BB}\right)^{ijkl}_{bcdefg} h^{bdc}h^{efg}\,,\label{eq:sym_UooBB_C3}\\
\left(U^{oo}_{BDE}\right)^{ijklm}_{bcdefg}h^{bcd}f^{efg}&=\left(U^{oo}_{BDE}\right)^{ijklm}_{bcdefg} h^{bdc}f^{egf}\,,\label{eq:sym_UooBDE_C2}\\
\left(U^{oo}_{DEB}\right)^{ijklm}_{bcdefg}f^{bcd}h^{efg}&=\left(U^{oo}_{DEB}\right)^{ijklm}_{bcdefg} f^{bdc}h^{egf}\,,\label{eq:sym_UooDEB_C2}\\
(U^{sso}_{BEE})^{ijklm}_{def}d^{def}&=0\,,\label{eq:sym_UssoBEE_C2}\\
(U^{sso}_{EBE})^{ijklm}_{def}d^{def}&=0\,, \label{eq:sym_UssoEBE_C2}\\
(U^{oss}_{EBE})^{ijklm}_{bcd}d^{bcd}&=0\,, \label{eq:sym_UossEBE_C2}\\
(U^{oss}_{EEB})^{ijklm}_{bcd}d^{bcd}&=0\,. \label{eq:sym_UossEEB_C2}\\
(U^{sso}_{EEB})^{ijklm}_{def}h^{def}&=-(U^{sso}_{EEB})^{ijklm}_{def}h^{dfe}\,, \label{eq:sym_UssoEEB_C2}\\
(U^{oss}_{BEE})^{ijklm}_{bcd}h^{bcd}&=-(U^{oss}_{BEE})^{ijklm}_{bcd}h^{bdc}\,, \label{eq:sym_UossBEE_C2}\\
(U^{ooo}_{BEE})^{ijklm}_{bcdefghpq}h^{bcd}d^{efg}d^{hpq}
&=-(U^{ooo}_{BEE})^{ijklm}_{bcdefghpq}h^{bdc}d^{efg}d^{hpq}\,,\label{eq:sym_UoooBEE_C2}\\
(U^{ooo}_{EBE})^{ijklm}_{bcdefghpq}d^{bcd}h^{efg}d^{hpq}
&=-(U^{ooo}_{EBE})^{ijklm}_{bcdefghpq}d^{bcd}h^{egf}d^{hpq}\,,\label{eq:sym_UoooEBE_C2}\\
(U^{ooo}_{EEB})^{ijklm}_{bcdefghpq}d^{bcd}d^{efg}h^{hpq}
&=-(U^{ooo}_{EEB})^{ijklm}_{bcdefghpq}d^{bcd}d^{efg}h^{hqp}\,.\label{eq:sym_UoooEEB_C2}
\end{align}
Inserting the identity operator $T^{-1}T$ between the fields in the correlators in Eqs.~(\ref{eq:UooBDE}),(\ref{eq:UossBEE}), (\ref{eq:UoooBEE}), and (\ref{eq:UoooEBE}) and assuming
$T$-invariance of the vaccuum give
\begin{align}
\left(U^{oo}_{BDE}\right)^{ijklm}_{bcdefg}T^b_{\alpha\beta}T^c_{\gamma\delta}T^d_{\rho\sigma}T^e_{\eta\tau}T^{f}_{\omega\kappa}T^{g}_{\mu\nu}
&=-\left(U^{oo}_{DEB}\right)^{mklji}_{gfedcb}(T^b)^T_{\alpha\beta}
(T^c)^T_{\gamma\delta}(T^d)^T_{\rho\sigma}(T^e)^T_{\eta\tau}(T^f)^T_{\omega\kappa}(T^g)^T_{\mu\nu}\,,\label{eq:sym_UooBDE_T}\\
(U^{oss}_{BEE})^{iljmk}_{def}T^d_{\alpha\beta}T^e_{\gamma\delta}T^f_{\rho\sigma}
&=-(U^{sso}_{EEB})^{kmjli}_{fed}(T^d)^T_{\alpha\beta}(T^e)^T_{\gamma\delta}(T^f)^T_{\rho\sigma}\,,\label{eq:sym_UossBEE_T}
\end{align}
\begin{align}
&(U^{ooo}_{BEE})^{iljmk}_{bcdefghpq}T^b_{\alpha\beta}T^c_{\gamma\delta}T^d_{\rho\sigma}T^e_{\eta\tau}T^{f}_{\omega\kappa}T^{g}_{\mu\nu}
T^h_{\epsilon\delta}T^p_{\lambda\xi}T^q_{\psi\chi}
\nonumber\\
=&\,-(U^{ooo}_{EEB})^{kmjli}_{qphgfedcb}(T^b)^T_{\alpha\beta}(T^c)^T_{\gamma\delta}(T^d)^T_{\rho\sigma}(T^e)^T_{\eta\tau}(T^f)^T_{\omega\kappa}(T^g)^T_{\mu\nu}
(T^h)^T_{\epsilon\delta}(T^p)^T_{\lambda\xi}(T^q)^T_{\psi\chi}\,,
\label{eq:sym_UoooBEE_T}\\
&(U^{ooo}_{EBE})^{iljmk}_{bcdefghpq}T^b_{\alpha\beta}T^c_{\gamma\delta}T^d_{\rho\sigma}T^e_{\eta\tau}T^{f}_{\omega\kappa}T^{g}_{\mu\nu}
T^h_{\epsilon\delta}T^p_{\lambda\xi}T^q_{\psi\chi}
\nonumber\\
=&\,-(U^{ooo}_{EBE})^{kmjli}_{qphgfedcb}(T^b)^T_{\alpha\beta}(T^c)^T_{\gamma\delta}(T^d)^T_{\rho\sigma}(T^e)^T_{\eta\tau}(T^f)^T_{\omega\kappa}(T^g)^T_{\mu\nu}
(T^h)^T_{\epsilon\delta}(T^p)^T_{\lambda\xi}(T^q)^T_{\psi\chi}\,.
\label{eq:sym_UoooEBE_T}
\end{align}
From Eqs.~(\ref{eq:sym_UooBDE_T})-(\ref{eq:sym_UoooEBE_T}), taking appropriate summations on the fundamental color indices, and utilizing the cyclic symmetry and the total symmetry of indices
of $h^{abc}$ and $d^{abc}$ respectively, we have
\begin{align}
(U^{oo}_{BDE})^{ijklm}_{bcdefg} h^{bcd}f^{efg}
&=-(U^{oo}_{DEB})^{mklji}_{bcdefg}f^{bcd}h^{efg}\,.\label{eq:sym_UooBDE_T2}\\
(U^{oss}_{BEE})^{ijklm}_{bcd} h^{bcd}
&=-(U^{sso}_{EEB})^{mlkji}_{def}h^{def}\,.\label{eq:sym_UossBEE_T2}\\
(U^{ooo}_{BEE})^{ijklm}_{bcdefghpq} h^{bcd}d^{efg}d^{hpq}
&=-(U^{ooo}_{EEB})^{mlkji}_{bcdefghpq} d^{bcd}d^{efg}h^{hpq}\,,\label{eq:sym_UoooEBE_T2}\\
(U^{ooo}_{EBE})^{ijklm}_{bcdefghpq} d^{bcd}h^{efg}d^{hpq}
&=-(U^{ooo}_{EBE})^{mlkji}_{bcdefghpq}d^{bcd}h^{efg}d^{hpq}\,,\label{eq:sym_UoooBEE_T2}
\end{align}
which imply that the tensor components (Eq.~(\ref{eq:tensor_3})) of $(\hat{U}_{BDE})^{ijklm}$, $(\hat{U}_{DEB})^{ijklm}$, $(\hat{U}_{BEE})^{ijklm}$,
 $(\hat{U}_{EBE})^{ijklm}$, and $(\hat{U}_{EEB})^{ijklm}$ in 
Eqs.~(\ref{eq:hat_UBDE})-(\ref{eq:hat_UEEB}) satisfy
\begin{align}
\tilde{U}^{\rm i}_{BDE}&=\tilde{U}^{\rm ix}_{DEB}\,,\quad
\tilde{U}^{\rm ii}_{BDE}=\tilde{U}^{\rm vii}_{DEB}\,,\nonumber\\
\tilde{U}^{\rm iii}_{BDE}&=\tilde{U}^{\rm iii}_{DEB}\,,\quad
\tilde{U}^{\rm iv}_{BDE}=\tilde{U}^{\rm v}_{DEB}\,,\nonumber\\
\tilde{U}^{\rm v}_{BDE}&=-\tilde{U}^{\rm viii}_{DEB}\,,\quad
\tilde{U}^{\rm vi}_{BDE}=\tilde{U}^{\rm iv}_{DEB}\,,\nonumber\\
\tilde{U}^{\rm vii}_{BDE}&=-\tilde{U}^{\rm vi}_{DEB}\,,\quad
\tilde{U}^{\rm viii}_{BDE}=\tilde{U}^{\rm ii}_{DEB}\,,\nonumber\\
\tilde{U}^{\rm ix}_{BDE}&=\tilde{U}^{\rm x}_{DEB}\,,\quad
\tilde{U}^{\rm x}_{BDE}=\tilde{U}^{\rm i}_{DEB}\,,\label{eq:sym_tilde_UBDE_T}
\end{align}
\begin{align}
\tilde{U}^{\rm ii}_{EBE}&=\tilde{U}^{\rm iv}_{EBE}\,,\quad
\tilde{U}^{\rm v}_{EBE}=\tilde{U}^{\rm viii}_{EBE}\,,\nonumber\\
\tilde{U}^{\rm vi}_{EBE}&=\tilde{U}^{\rm vii}_{EBE}\,,\quad
\tilde{U}^{\rm ix}_{EBE}=\tilde{U}^{\rm x}_{EBE}\,.\label{eq:sym_tilde_UEBE_T}
\end{align}
and
\begin{align}
\tilde{U}^{\rm i}_{BEE}&=\tilde{U}^{\rm i}_{EEB}\,,\quad
\tilde{U}^{\rm ii}_{BEE}=\tilde{U}^{\rm iv}_{EEB}\,,\nonumber\\
\tilde{U}^{\rm iii}_{BEE}&=\tilde{U}^{\rm iii}_{EEB}\,,\quad
\tilde{U}^{\rm iv}_{BEE}=\tilde{U}^{\rm ii}_{EEB}\,,\nonumber\\
\tilde{U}^{\rm v}_{BEE}&=\tilde{U}^{\rm viii}_{EEB}\,,\quad
\tilde{U}^{\rm vi}_{BEE}=\tilde{U}^{\rm vii}_{EEB}\,,\nonumber\\
\tilde{U}^{\rm vii}_{BEE}&=\tilde{U}^{\rm vi}_{EEB}\,,\quad
\tilde{U}^{\rm viii}_{BEE}=\tilde{U}^{\rm v}_{EEB}\,,\nonumber\\
\tilde{U}^{\rm ix}_{BEE}&=\tilde{U}^{\rm x}_{EEB}\,,\quad
\tilde{U}^{\rm x}_{BEE}=\tilde{U}^{\rm ix}_{EEB}\,.\label{eq:sym_tilde_UBEE_T}
\end{align}
Inserting the identity operator $P^{-1}P$ between the fields in the correlators $(U^{ss}_{EE})^{ijkl}$, $(U^{oo}_{EE})^{ijkl}_{bcdefg}$, $(U^{ss}_{BB})^{ijkl}$, and $(U^{oo}_{BB})^{ijkl}_{bcdefg}$, and assuming $P$-invariance of the vaccuum give
\begin{align}
\hat{r}^{i\dagger}_+\hat{r}^{l}_+\left(U^{ss}_{EE}\right)^{ijkl}
=\hat{r}^{i\dagger}_-\hat{r}^{l}_-\left(U^{ss}_{EE}\right)^{ijkl}\,,\label{eq:sym_UssEE_P}\\
\hat{r}^{i\dagger}_+\hat{r}^{l}_+\left(U^{oo}_{EE}\right)^{ijkl}_{bcdefg}
=\hat{r}^{i\dagger}_-\hat{r}^{l}_-\left(U^{oo}_{EE}\right)^{ijkl}_{bcdefg}\,,\label{eq:sym_UooEE_P}\\
\hat{r}^{i\dagger}_+\hat{r}^{l}_+\left(U^{ss}_{BB}\right)^{ijkl}
=\hat{r}^{i\dagger}_-\hat{r}^{l}_-\left(U^{ss}_{BB}\right)^{ijkl}\,,\label{eq:sym_UssBB_P}\\
\hat{r}^{i\dagger}_+\hat{r}^{l}_+\left(U^{oo}_{BB}\right)^{ijkl}_{bcdefg}
=\hat{r}^{i\dagger}_-\hat{r}^{l}_-\left(U^{oo}_{BB}\right)^{ijkl}_{bcdefg}\,.\label{eq:sym_UooBB_P}
\end{align}
Then, using that $\hat{r}^{\dagger}_{\pm}=-\hat{r}_{\mp}$ and $\delta^{ij}=\sum_\la \hat{r}^{\dagger\,i}_{\la}\hat{r}^{j}_{\la}$, together with Eqs.~\eqref{eq:sym_UssEE_P}-\eqref{eq:sym_UooBB_P}, we can derive that the tensor components in Eqs.~(\ref{eq:tensor_2}) satisfy
\begin{align}
\tilde{U}^{\rm I}_{EE}&=\tilde{U}^{\rm II}_{EE}\,,\quad \tilde{U}^{\rm I}_{BB\,a}=\tilde{U}^{\rm II}_{BB\,a}\,,\quad
\tilde{U}^{\rm I}_{BB\,b}=\tilde{U}^{\rm II}_{BB\,b}\,.\label{eq:sym_tilde_U_P}
\end{align}

\section{Matching weakly-coupled pNRQCD to the BOEFT}\label{app_matching}
In this appendix, we show the derivations of Eqs.~(\ref{eq:V_SK_2})-(\ref{eq:V_S_12_b}) in detail.
Consider diagram (d) in Fig.~\ref{match}, with insertion of one $c_F$-vertex and one $\bm{L}_{Q\bar{Q}}\cdot\bm{B}$-vertex. Its contribution to $\delta V_{\la\la'}$ is given by
\footnote{Note that in deriving Eq.~(\ref{eq:V_cF_LB_1}), exponential factors of the form $e^{-ih_o(T/2-t)}$, $e^{-ih_o(t-t')}$, and $e^{-ih_o(t'+T/2)}$ originating from the octet-field propagators are approximated by $1$, as justified by the fact that $h_o\sim mv^2$, $T\sim 1/\Lambda_{\rm{QCD}}$, and $\Lambda_{\rm{QCD}}\gg mv^2$. Similar approximations are used in deriving 
Eqs.~(\ref{eq:V_cs_rE_1}), (\ref{eq:V_cF_cF_1}), (\ref{eq:V_cF_rrDE_0}) and (\ref{eq:V_e_BEE})-(\ref{eq:V_g_EEB}).}
\begin{align}
\delta V^{c_F,\bm{L}\cdot \bm{B}}_{\la\la'}
&=\,-\frac{c_F}{16m^2}\hat{r}_{\la}^{i\,\dag}\left(U^{oo}_{BB}\right)^{ijkl}_{bcdefg}
\left[{S}^j_1{L}_{Q\bar{Q}}^k\left(h^{bcd}h^{efg}-h^{bcd}h^{egf}\right)
+L^j_{Q\bar{Q}}S_1^k\left(h^{bcd}h^{efg}-h^{bdc}h^{efg}\right) \right.\nonumber \\
&\quad\quad
\left.+{S}^j_2{L}_{Q\bar{Q}}^k\left(h^{bdc}h^{egf}-h^{bdc}h^{efg}\right)
+L^j_{Q\bar{Q}}S_2^k\left(h^{bdc}h^{egf}-h^{bcd}h^{egf}\right)\right]\hat{r}^l_{\la'}\,.\label{eq:V_cF_LB_0}
\end{align}
Similar to Eqs.~(\ref{eq:tensor_2}) and (\ref{eq:sym_tilde_U_P}), rotational invariance and parity imply that the coefficients 
$\left(U^{oo}_{BB}\right)^{ijkl}_{bcdefg}\left(h^{bcd}h^{efg}-h^{bdc}h^{efg}\right)$ 
and $\left(U^{oo}_{BB}\right)^{ijkl}_{bcdefg}\left(h^{bdc}h^{egf}-h^{bcd}h^{egf}\right)$ in Eq.~(\ref{eq:V_cF_LB_0}) have tensor decomposition of the form
$\hat{U}^{ijkl}=\tilde{U}^{\rm I}(\delta^{ij}\delta^{kl}+\delta^{ik}\delta^{jl})+\tilde{U}^{\rm III}\delta^{il}\delta^{jk}$, and thus is symmetric in the indices $jk$.
Therefore, Eq.~(\ref{eq:V_cF_LB_0}) becomes
\begin{align}
\delta V^{c_F,\bm{L}\cdot \bm{B}}_{\la\la'}&=\,-\frac{c_F}{16m^2}\hat{r}_{\la}^{i\,\dag}\left(U^{oo}_{BB}\right)^{ijkl}_{bcdefg}
\left[{S}^j_1{L}_{Q\bar{Q}}^k\left(2h^{bcd}h^{efg}-h^{bcd}h^{egf}-h^{bdc}h^{efg}\right) \right.\nonumber \\
&\quad\quad
\left.+{S}^j_2{L}_{Q\bar{Q}}^k\left(2h^{bdc}h^{egf}-h^{bcd}h^{egf}-h^{bdc}h^{efg}\right)\right]\hat{r}^l_{\la'}\,,\label{eq:V_cF_LB_1}
\end{align}
which with Eqs.~(\ref{eq:sym_UooBB_C2}) and (\ref{eq:hat_UBBa}) is simplified to
\begin{align}
\delta V^{c_F,\bm{L}\cdot \bm{B}}_{\la\la'}
&=\frac{c_F}{4m^2}\hat{r}_{\la}^{i\,\dag}(\hat{U}_{BB\,a})^{ijkl}{S}^j{L}_{Q\bar{Q}}^k\hat{r}^l_{\la'}\,.\label{eq:V_cF_LB}
\end{align}
With the tensor decomposition Eq.~(\ref{eq:tensor_2}) and Eq.~(\ref{eq:sym_tilde_U_P}), Eq.~(\ref{eq:V_cF_LB}) becomes
\begin{align}
\delta V^{c_F,\bm{L}\cdot \bm{B}}_{\la\la'}&=\frac{c_F}{4m^2}\tilde{U}^{\rm III}_{BB\,a}\left(\hat{r}^{i\,\dag}_{\la}\bm{L}_{Q\bar{Q}}\hat{r}^i_{\lap}\right)\cdot\bm{S}
+\frac{c_F}{4m^2}\tilde{U}^{\rm I}_{BB\,a}\hat{r}^{i\,\dag}_{\la}\left(L_{Q\bar{Q}}^iS^j+S^iL_{Q\bar{Q}}^j\right)\hat{r}^{j}_{\lap}\,.\label{eq:delta_V_cF_LB}
\end{align}

Consider diagrams (c) and (d) in Fig.~\ref{match}, with insertion of one $c_s$-vertex and one $\bm{r}\cdot\bm{E}$-vertex. Its contribution to $\delta V_{\la\la'}$ is given by
\begin{align}
\delta V^{c_s,\bm{r}\cdot\bm{E}}_{\la\la'}
=&\,-\frac{c_s}{2m^2}\hat{r}_{\la}^{i\,\dag}\Biggl\{\frac{T_F}{N_c}\left(U^{ss}_{EE}\right)^{ijkl}
\left[(\bm{p}\times\bm{S})^jr^k+r^j(\bm{p}\times\bm{S})^k\right]\nonumber\\
&\,+\frac{\left(U^{oo}_{EE}\right)^{ijkl}_{bcdefg}}{4} 
\left[(\bm{p}\times\bm{S}_1)^jr^k h^{bcd}d^{efg}+(\bm{p}\times\bm{S}_2)^jr^k h^{bdc}d^{efg}\right.\nonumber\\
&\,\left.
+r^j(\bm{p}\times\bm{S}_1)^k d^{bcd}h^{efg}+r^j(\bm{p}\times\bm{S}_2)^k d^{bcd}h^{egf}\right]\Biggl\}\hat{r}^{l}_{\lap}
\,,\label{eq:V_cs_rE_1}
\end{align}
which with Eqs.~(\ref{eq:sym_UooEE_C2}) and (\ref{eq:sym_UooEE_C3}) is simplified to
\begin{align}
\delta V^{c_s,\bm{r}\cdot\bm{E}}_{\la\la'}
=&\,-\frac{c_s}{8m^2}\hat{r}_{\la}^{i\,\dag}(\hat{U}_{EE})^{ijkl}
\left[r^j(\bm{p}\times\bm{S})^k+(\bm{p}\times\bm{S})^j r^k\right]
\hat{r}^l_{\la'}\,,\label{eq:V_cs_rE_2}
\end{align}
where $(\hat{U}_{EE})^{ijkl}$ is defined in Eq.~(\ref{eq:hat_UEE}).

With the tensor decomposition Eq.~(\ref{eq:tensor_2}) and Eq.~(\ref{eq:sym_tilde_U_P}) and using the commutation relation  $[r^i,p^j]=i\delta^{ij}$, Eq.~(\ref{eq:V_cs_rE_2}) becomes
\begin{align}
\delta V^{c_s,\bm{r}\cdot\bm{E}}_{\la\la'}=&-\frac{c_s}{4m^2}\left\{
\tilde{U}_{EE}^{\rm I}\left[\hat{\bm{r}}^\dagger_\lambda\cdot(\bm{p}\times\bm{S})(\bm{r}\cdot \hat{\bm{r}}_{\lambda'})
+(\hat{\bm{r}}^{\dagger}_\lambda\cdot \bm{r})(\bm{p}\times\bm{S})\cdot\hat{\bm{r}}_{\lambda'}+i(\hat{\bm{r}}^{\dagger}_\lambda\times\hat{\bm{r}}_{\lambda'})\cdot\bm{S}\right]\right.\nonumber\\
&\left.+\tilde{U}_{EE}^{\rm{III}}\hat{r}_\lambda^{i\dagger}\bm{L}_{Q\bar{Q}}\cdot\bm{S}\hat{r}^i_{\lambda'}
\right\}
\label{eq:delta_V_cs_rE}
\end{align}
Adding up Eqs.~(\ref{eq:delta_V_cF_LB}) and (\ref{eq:delta_V_cs_rE}), we obtain Eqs.~\eqref{eq:V_SK_2} and~\eqref{eq:V_SLa}-\eqref{eq:V_pxs}. 

Consider diagrams (c) and (d) in Fig.~\ref{match}, with insertion of two $c_F$-vertices. Its contribution to $\delta V_{\la\la'}$ is given by
\begin{align}
\delta V^{c_F,c_F}_{\la\la'}=&\,
-\frac{c_F^2}{m^2}\hat{r}_{\la}^{i\,\dag}\hat{r}^l_{\la'}\left[\frac{T_F}{N_c}(U^{ss}_{BB})^{ijkl}\left({S}_1-{S}_2\right)^j\left({S}_1-{S}_2\right)^k \right.\nonumber\\
&\,\left.+\frac{\left(U^{oo}_{BB}\right)^{ijkl}_{bcdefg}}{4}\left({S}^j_1{S}^k_1 h^{bcd}h^{efg}+{S}^j_2{S}^k_2h^{bdc}h^{egf}
-{S}^j_1{S}^k_2 h^{bcd}h^{egf}-{S}^j_2{S}^k_1h^{bdc}h^{efg}\right)\right]\,,\label{eq:V_cF_cF_1}
\end{align}
which with Eqs.~(\ref{eq:sym_UooBB_C2}), (\ref{eq:sym_UooBB_C3}), and (\ref{eq:hat_UBBb}) is simplified to
\begin{align}
\delta V^{c_F,c_F}_{\la\la'}=&\,
-\frac{c_F^2}{4m^2}\hat{r}_{\la}^{i\,\dag}\hat{r}^l_{\la'}
\left[(\hat{U}_{BB\,c})^{ijkl}({S}^j_1{S}^k_1+{S}^j_2{S}^k_2)
-(\hat{U}_{BB\,b})^{ijkl}({S}^j_1{S}^k_2+{S}^j_2{S}^k_1)\right]\,,\label{eq:V_cF_cF}
\end{align}
where 
\begin{align}
(\hat{U}_{BB\,c})^{ijkl}&\equiv (U^{oo}_{BB})^{ijkl}_{bcdefg}h^{bcd}h^{efg}+\frac{4T_F}{N_c}(U^{ss}_{BB})^{ijkl}\,.
\end{align}
Similar to Eqs.~(\ref{eq:tensor_2}) and (\ref{eq:sym_tilde_U_P}), rotational invariance and parity imply that $(\hat{U}_{BB\,c})^{ijkl}$,
also has tensor decomposition of the form
$\hat{U}^{ijkl}=\tilde{U}^{\rm I}(\delta^{ij}\delta^{kl}+\delta^{ik}\delta^{jl})+\tilde{U}^{\rm III}\delta^{il}\delta^{jk}$. 
In Eq.~(\ref{eq:V_cF_cF}), the terms ${S}^j_1{S}^k_1$ and ${S}^j_2{S}^k_2$ can be rewritten using $\sigma^i\sigma^j=i\epsilon^{ijk}\sigma^k+\delta^{ij}$, the first term of which gives
zero when contracted with $(\hat{U}_{BB\,c})^{ijkl}$, since $(\hat{U}_{BB\,c})^{ijkl}=(\hat{U}_{BB\,c})^{ikjl}$. Therefore, after applying the tensor decomposition of $(\hat{U}_{BB\,c})^{ijkl}$ and $(\hat{U}_{BB\,b})^{ijkl}$, the spin-dependent terms in Eq.~(\ref{eq:V_cF_cF}) are given by
\begin{align}
\delta V^{c_F,c_F}_{\la\la'\,SD}&=
\frac{c_F^2}{4m^2}\tilde{U}^{\rm III}_{BB\,b}\bm{S}^2\de_{\la\lap}
+\frac{c_F^2}{2m^2}\tilde{U}^{\rm I}_{BB\,b} \hat{r}^{i\dag}_{\la}\hat{r}^j_{\lap}\left(S^i_1S^j_2+S^i_2S^j_1\right)\,,\label{eq:delta_V_cF_cF}
\end{align}
which gives Eqs.~(\ref{eq:V_S2}) and (\ref{eq:V_S_12_b}).

Consider diagram (d) in Fig.~\ref{match} with insertion of one $c_F$-vertex and one $r^ir^j\bm{D}^iE^j$-vertex. Its contribution to $\delta V_{\la\la'}$ is given by
\begin{align}
\delta V^{c_F, r^ir^j \bm{D}^i E^j}_{\lambda\lambda'}&=\frac{ic_F}{16m}\hat{r}^{i\dagger}_\lambda \hat{r}^m_{\lambda'}
\left[r^kr^l(S_1^jh^{bcd}-S_2^jh^{bdc})(U^{oo}_{BDE})^{ijklm}_{bcdefg}f^{efg}\right.\nonumber\\
&\quad\quad\left.+r^jr^k(S_1^lh^{efg}-S_2^lh^{egf})(U^{oo}_{DEB})^{ijklm}_{bcdefg}f^{bcd}
\right]\,,\label{eq:V_cF_rrDE_0}
\end{align}
which with the help of Eqs.~(\ref{eq:sym_UooBDE_C2}), (\ref{eq:sym_UooDEB_C2}), (\ref{eq:hat_UBDE}), and (\ref{eq:hat_UDEB}) becomes
\begin{align}
\delta V^{c_F, r^ir^j \bm{D}^i E^j}_{\lambda\lambda'}&=\frac{ic_F}{16m}\hat{r}^{i\dagger}_\lambda \hat{r}^m_{\lambda'}
\left[r^kr^lS^j(\hat{U}_{BDE})^{ijklm}+r^jr^kS^l(\hat{U}_{DEB})^{ijklm}\right]\,.\label{eq:V_cF_rrDE}
\end{align}
Consider diagrams (e), (f), and (g) in Fig.~\ref{match}, with insertion of one $c_F$-vertex and two $\bm{r}\cdot\bm{E}$-vertices. Its contribution to $\delta V_{\la\la'}$ is given by
\begin{align}
\delta V^{c_F,\bm{r}\cdot\bm{E},\bm{r}\cdot\bm{E}}_{\la\la'}&=\delta V_e+\delta V_f+\delta V_g\,,
\end{align}
where each term is the sum of the three possible diagrams with the insertion of $c_F$-vertex in a different location
\begin{align}
\delta V_e&=(\delta V_e)_{BEE}+(\delta V_e)_{EBE}+(\delta V_e)_{EEB}\,,\\
\delta V_f&=(\delta V_f)_{BEE}+(\delta V_f)_{EBE}+(\delta V_f)_{EEB}\,,\\
\delta V_g&=(\delta V_g)_{BEE}+(\delta V_g)_{EBE}+(\delta V_g)_{EEB}\,,
\end{align}
with
\begin{align}
(\delta V_e)_{BEE}&=-\frac{ic_F}{2m}\left(\frac{T_F}{N_c}\right)\hat{r}^{i\dagger}_\lambda \hat{r}^m_{\lambda'}r^k r^l(S_1^j-S_2^j)(U^{sso}_{BEE})^{ijklm}_{def}d^{def}
\,,\label{eq:V_e_BEE}\\
(\delta V_e)_{EBE}&=-\frac{ic_F}{2m}\left(\frac{T_F}{N_c}\right)\hat{r}^{i\dagger}_\lambda \hat{r}^m_{\lambda'}r^j r^l(S_1^k-S_2^k)(U^{sso}_{EBE})^{ijklm}_{def}d^{def}\,,\\
(\delta V_e)_{EEB}&=-\frac{ic_F}{2m}\left(\frac{T_F}{N_c}\right)\hat{r}^{i\dagger}_\lambda \hat{r}^m_{\lambda'}r^j r^k
(h^{def}S_1^l-h^{dfe}S_2^l)(U^{sso}_{EEB})^{ijklm}_{def}\,,\\
(\delta V_f)_{BEE}&=-\frac{ic_F}{2m}\left(\frac{T_F}{N_c}\right)\hat{r}^{i\dagger}_\lambda \hat{r}^m_{\lambda'}r^k r^l
(h^{bcd}S_1^j-h^{bdc}S_2^j)(U^{oss}_{BEE})^{ijklm}_{bcd}\,,\\
(\delta V_f)_{EBE}&=-\frac{ic_F}{2m}\left(\frac{T_F}{N_c}\right)\hat{r}^{i\dagger}_\lambda \hat{r}^m_{\lambda'}r^j r^l(S_1^k-S_2^k)
(U^{oss}_{EBE})^{ijklm}_{bcd}d^{bcd}\,,\\
(\delta V_f)_{EEB}&=-\frac{ic_F}{2m}\left(\frac{T_F}{N_c}\right)\hat{r}^{i\dagger}_\lambda \hat{r}^m_{\lambda'}r^j r^k
(S_1^l-S_2^l)(U^{oss}_{EEB})^{ijklm}_{bcd}d^{bcd}\,,\\
(\delta V_g)_{BEE}&=-\frac{ic_F}{8m}\hat{r}^{i\dagger}_\lambda \hat{r}^m_{\lambda'}r^k r^l
(h^{bcd}S_1^j-h^{bdc}S_2^j)(U^{ooo}_{BEE})^{ijklm}_{bcdefghpq}d^{efg}d^{hpq}\,,\\
(\delta V_g)_{EBE}&=-\frac{ic_F}{8m}\hat{r}^{i\dagger}_\lambda \hat{r}^m_{\lambda'}r^j r^l
(h^{efg}S_1^k-h^{egf}S_2^k)(U^{ooo}_{EBE})^{ijklm}_{bcdefghpq}d^{bcd}d^{hpq}\,,\\
(\delta V_g)_{EEB}&=-\frac{ic_F}{8m}\hat{r}^{i\dagger}_\lambda \hat{r}^m_{\lambda'}r^j r^k
(h^{hpq}S_1^l-h^{hqp}S_2^l)(U^{ooo}_{EEB})^{ijklm}_{bcdefghpq}d^{bcd}d^{efg}\,.\label{eq:V_g_EEB}
\end{align}
From Eqs.~(\ref{eq:sym_UssoBEE_C2})-(\ref{eq:sym_UossEEB_C2}), 
\begin{align}
(\delta V_e)_{BEE}&=(\delta V_e)_{EBE}=0\,,\\
(\delta V_f)_{EBE}&=(\delta V_f)_{EEB}=0\,.
\end{align}
Using Eqs.~(\ref{eq:sym_UssoEEB_C2})-(\ref{eq:sym_UoooEEB_C2}) and~(\ref{eq:hat_UBEE})-(\ref{eq:hat_UEEB}) , we have
\begin{align}
(\delta V_f)_{BEE}+(\delta V_g)_{BEE}
&=-\frac{ic_F}{8m}\hat{r}^{i\dagger}_\lambda \hat{r}^m_{\lambda'}r^k r^l
S^j(\hat{U}_{BEE})^{ijklm}
\,,\label{eq:Vf_Vg}\\
(\delta V_g)_{EBE}&=
-\frac{ic_F}{8m}\hat{r}^{i\dagger}_\lambda \hat{r}^m_{\lambda'}r^j r^l
S^k(\hat{U}_{EBE})^{ijklm}
\,,\label{eq:Vg}\\
(\delta V_e)_{EEB}+(\delta V_g)_{EEB}
&=-\frac{ic_F}{8m}\hat{r}^{i\dagger}_\lambda \hat{r}^m_{\lambda'}r^j r^k
S^l(\hat{U}_{EEB})^{ijklm}
\,.\label{eq:Ve_Vg}
\end{align}
Adding up Eqs.~(\ref{eq:V_cF_rrDE}) and (\ref{eq:Vf_Vg})-(\ref{eq:Ve_Vg}), applying the tensor decomposition Eq.~(\ref{eq:tensor_3}) and
using Eqs.~(\ref{eq:sym_tilde_UBDE_T}), (\ref{eq:sym_tilde_UEBE_T}) and (\ref{eq:sym_tilde_UBEE_T}), we have
\begin{align}
&\delta V^{c_F, r^ir^j \bm{D}^i E^j}_{\lambda\lambda'}+\delta V^{c_F,\bm{r}\cdot\bm{E},\bm{r}\cdot\bm{E}}_{\la\la'}
=\,\frac{c_F}{16m}\left\{ \left[-2\tilde{U}^{\rm i}_{EBE}-4\tilde{U}^{\rm ix}_{BEE}+2\tilde{U}_{BDE}^{\rm ix}\right]
r^2\left(\hat{r}^{i\dagger}_{\lambda}\bm{K}^{ij}\hat{r}^j_{\lambda'}\right)\cdot\bm{S} \right.\nonumber\\
&\,+
\left[2\left(\tilde{U}^{\rm v}_{EBE}-\tilde{U}^{\rm vi}_{EBE}+\tilde{U}^{\rm ii}_{BEE}-\tilde{U}^{\rm iv}_{BEE}+\tilde{U}^{\rm vi}_{BEE}
-\tilde{U}^{\rm viii}_{BEE}\right)
-\left(\tilde{U}^{\rm ii}_{BDE}-\tilde{U}^{\rm iv}_{BDE}
+\tilde{U}^{\rm vi}_{BDE}-\tilde{U}^{\rm viii}_{BDE}\right)\right]\nonumber\\
&\,\times
\left[\left(\bm{r}\cdot \hat{\bm{r}}^{\dagger}_{\lambda}\right)\left(r^i\bm{K}^{ij}\hat{r}^j_{\lambda'}\right)\cdot\bm{S}
-\left(r^i\bm{K}^{ij}\hat{r}^{j\dagger }_{\lambda}\right)\cdot\bm{S} \left(\bm{r}\cdot \hat{\bm{r}}_{\lambda'}\right)\right]\nonumber\\
&\,\left. 
+\left[-4\left(\tilde{U}^{\rm ix}_{EBE}+\tilde{U}^{\rm i}_{BEE}+\tilde{U}^{\rm x}_{BEE}\right)+
2\left(\tilde{U}^{\rm i}_{BDE}+\tilde{U}^{\rm x}_{BDE}\right)\right]
\left(\bm{S}\cdot\bm{r}\right)\left(\hat{r}^{i\dagger}_{\lambda}\bm{K}^{ij}\hat{r}^j_{\lambda'}\right)\cdot\bm{r}\right\}\,.\label{eq:V_cF_rE_rE}
\end{align}
We can eliminate the last term on the right-hand side of Eq.~(\ref{eq:V_cF_rE_rE}) using the relation
\begin{align}
\left(\bm{S}\cdot\bm{r}\right)\left(\hat{r}^{i\dagger}_{\lambda}\bm{K}^{ij}\hat{r}^j_{\lambda'}\right)\cdot\bm{r}
&=r^2\left(\hat{r}^{i\dagger}_{\lambda}\bm{K}^{ij}\hat{r}^j_{\lambda'}\right)\cdot\bm{S}
-\left(\bm{r}\cdot \hat{\bm{r}}^{\dagger}_{\lambda}\right)\left(r^i\bm{K}^{ij}\hat{r}^j_{\lambda'}\right)\cdot\bm{S}\nonumber\\
&\,\,\,\,\,\,+\left(r^i\bm{K}^{ij}\hat{r}^{j\dagger }_{\lambda}\right)\cdot\bm{S} \left(\bm{r}\cdot \hat{\bm{r}}_{\lambda'}\right)\,.
\end{align}
Therefore, we have
\begin{align}
\delta V^{c_F, r^ir^j \bm{D}^i E^j}_{\lambda\lambda'}+\delta V^{c_F,\bm{r}\cdot\bm{E},\bm{r}\cdot\bm{E}}_{\la\la'}
=&\,\frac{V^{np(1)}_{SK}}{m}r^2\left(\hat{r}^{i\dagger}_{\lambda}\bm{K}^{ij}\hat{r}^j_{\lambda'}\right)\cdot\bm{S}\nonumber\\
&\,+\frac{V^{np(0)}_{SKb}}{m}\left[\left(\bm{r}\cdot \hat{\bm{r}}^{\dagger}_{\lambda}\right)\left(r^i\bm{K}^{ij}\hat{r}^j_{\lambda'}\right)\cdot\bm{S}
-\left(r^i\bm{K}^{ij}\hat{r}^{j\dagger}_{\lambda}\right)\cdot\bm{S} \left(\bm{r}\cdot \hat{\bm{r}}_{\lambda'}\right)\right]\,,
\end{align}
where $V^{np(1)}_{SK}$ and $V^{np(0)}_{SKb}$ are given by Eqs.~(\ref{eq:V_SK_1}) and (\ref{eq:V_SKb}).
%
%
\section{Matrix elements of operators involving \texorpdfstring{$\bm{L}_{Q\bar{Q}}$}{LQQb}} \label{app}
In this appendix, we rewrite the operators for $V_{SLa}$, $V_{SLb}$, and $V_{SLc}$ in Eq.~(\ref{sdm2}) in a way such that the matrix elements of them sandwiched by the wave functions in Eqs.~(\ref{psip}) and (\ref{psim}) can be readily computed.

The angular momentum operator in spherical coordinates is
\begin{align}
\bm{L}_{Q\bar{Q}}=-i\hat{\bm{\phi}}\pa_{\theta}+\frac{i}{\sin\theta}\hat{\bm{\theta}}\pa_{\phi}\,.
\end{align}
Let us compute the commutators of the angular momentum operator and the unit vectors in spherical coordinates
\begin{align}
\left[{L}_{Q\bar{Q}}^i,\hat{r}_{0}^j\right]&=\hat{r}^i_+\hat{r}^j_--\hat{r}^i_-\hat{r}^j_+\,, \\
\left[{L}_{Q\bar{Q}}^i,\hat{\theta}^j\right]&=i \hat{\phi}^i\hat{r}_0^j+i\cot(\theta)\hat{\theta}^i\hat{\phi}^j\,, \\
\left[{L}_{Q\bar{Q}}^i,\hat{\phi}^j\right]&=-i\hat{\theta}^i\left(\hat{r}_0^j+\cot(\theta)\hat{\theta}^j\right)\,,
\end{align}
from which one can obtain 
\begin{align}
\left[{L}_{Q\bar{Q}}^i,\hat{r}_{\pm}^j\right]=\pm\left(\hat{r}^j_0\hat{r}^i_{\pm}+\cot(\theta)\hat{\theta}^i\hat{r}^j_{\pm}\right)\,.
\end{align}
Therefore, for any $\la$
\begin{align}
\left[{L}_{Q\bar{Q}}^i,\hat{r}_{\la}^j\right]&=\la\cot(\theta)\hat{r}_{\la}^j\hat{\theta}^i+\sqrt{1-\frac{\la(\la-1)}{2}}\hat{r}^j_{\la-1}\hat{r}^i_{+}-\sqrt{1-\frac{\la(\la+1)}{2}}\hat{r}^j_{\la+1}\hat{r}^i_{-}\,.
\end{align}
To compute the matrix elements of the $V_{SLa}$-operator, we rewrite the operator in the following way (for a matrix in $\lambda\lambda'$, indices run as $0, +1, -1$ from left to right and top to bottom),
\begin{align}
&\left(\hat{r}_{\la}^{i\dag}\bm{L}_{Q\bar{Q}}\hat{r}^{i}_{\lap}\right)\cdot \bm{S}=\left(\bm{L}_{Q\bar{Q}}\de_{\la\lap}+\hat{r}_{\la}^{i\dag}\left[\bm{L}_{Q\bar{Q}},\,\hat{r}^{i}_{\lap}\right]\right)\cdot \bm{S}=\left(
\begin{array}{ccc}
 \bm{L}_{Q\bar{Q}} & \hat{\bm{r}}_{+} & -\hat{\bm{r}}_{-} \\
 \hat{\bm{r}}_{+}^{\dag}  & \bm{L}_{Q\bar{Q}}+\cot\theta \hat{\bm{\theta}} &  0 \\
 -\hat{\bm{r}}_{-}^{\dag} & 0 & \bm{L}_{Q\bar{Q}}-\cot\theta \hat{\bm{\theta}} \\
\end{array}\right)\cdot \bm{S}\nonumber\\
&=\de_{\la\lap}\left[\bm{L}_{Q\bar{Q}}+\la\left(\cot\theta \hat{\bm{\theta}}+\hat{\bm{r}}_0\right)\right]\cdot\bm{S}+i\left(\hat{\bm{r}}^{\dag}_{\la}\times \hat{\bm{r}}_{\lap}\right)\cdot\bm{S}=\de_{\la\lap}\bm{L}\cdot\bm{S}-\left(\hat{r}_{\la}^{i\dag}\bm{K}^{ij}\hat{r}^{j}_{\lap}\right)\cdot \bm{S}\nonumber\\
&=\frac{\de_{\la\lap}}{2}\left(\bm{J}^2-\bm{L}^2-\bm{S}^2\right)-\left(\hat{r}_{\la}^{i\dag}\bm{K}^{ij}\hat{r}^{j}_{\lap}\right)\cdot \bm{S}\,,
\label{eq:reduce_V_SLa}
\end{align}
where we have used that
\begin{align}
&\left[\bm{L}_{Q\bar{Q}}+\la\left(\cot\theta \hat{\bm{\theta}}+\hat{\bm{r}}_0\right)\right]^2=\bm{L}^2_{Q\bar{Q}}+\frac{\la^2}{\sin^2\theta}+2i\la\frac{\cos\theta}{\sin^2\theta}\partial_{\phi}\equiv\bm{L}^2\,,
\end{align}
which is the operator whose eigenfunctions are our angular wave functions,
\begin{align}
\left(\bm{L}^2_{Q\bar{Q}}+\frac{\la^2}{\sin^2\theta}+2i\la\frac{\cos\theta}{\sin^2\theta}\partial_{\phi}\right)v^{\la}_{lm}(\theta,\phi)=l(l+1)v^{\la}_{lm}(\theta,\phi)\,,
\end{align}
with
\begin{align}
v^{\lambda}_{lm}(\theta,\phi)&=\frac{(-1)^{m+\lambda}}{2^l}\sqrt{\frac{2l+1}{4\pi}\frac{(l-m)!}{(l+m)!(l-\lambda)!(l+\lambda)!}}\,P^\lambda_{lm}(\cos\theta)e^{im\phi}\,,\\
P^\lambda_{lm}(x)&=(1-x)^{\frac{m-\lambda}{2}}(1+x)^{\frac{m+\lambda}{2}}
\partial_x^{l+m}(x-1)^{l+\lambda}(x+1)^{l-\lambda}\,,
\end{align}
with $|m|\le l$ and $|\lambda|\le l$.

Next, we consider the $V_{SLb}$-operator,
\begin{align}
\hat{r}_{\la}^{\dag i}\left({L}_{Q\bar{Q}}^i{S}^l+{S}^i{L}_{Q\bar{Q}}^l\right)\hat{r}_{\lap}^l\,.\label{eq:VSLb_op}
\end{align}
The first term in Eq.~\eqref{eq:VSLb_op} can be manipulated as follows:
\begin{align}
\hat{r}_{\la}^{\dag i}{S}^i{L}_{Q\bar{Q}}^l\hat{r}_{\lap}^l&=\left(\hat{\bm r}_{\la}^{\dag}\cdot\bm{S}\right)\left(\hat{\bm r}_{\lap}\cdot\bm{L}_{Q\bar{Q}}+[{L}_{Q\bar{Q}}^l,\,\hat{r}_{\lap}^l]\right)=\left(\hat{\bm r}_{\la}^{\dag}\cdot\bm{S}\right)\left(\hat{\bm r}_{\lap}\cdot\bm{L}_{Q\bar{Q}}-(\lap)^2\frac{\cot\theta}{\sqrt{2}}\right)\,.
\end{align}
This expression vanishes for $\lap=0$. In the case $\lap=\pm1$,
\begin{align}
\hat{r}_{\la}^{\dag i}{S}^i{L}_{Q\bar{Q}}^l\hat{r}_{\pm}^l&=\left(\hat{\bm r}_{\la}^{\dag}\cdot\bm{S}\right)\left(\hat{\bm r}_{\pm}\cdot\bm{L}_{Q\bar{Q}}-\frac{\cot\theta}{\sqrt{2}}\right)=\mp\frac{(\hat{\bm r}_{\la}^{\dag}\cdot\bm{S})}{\sqrt{2}}\left(\pm\pa_{\theta}+\frac{i}{\sin\theta}\pa_{\phi}\pm\cot\theta\right)\nonumber \\
&=\mp\frac{(\hat{\bm r}_{\la}^{\dag}\cdot\bm{S})}{\sqrt{2}}\mathcal{K}^{\prime}_{\mp}\,.\label{eq:SLb_prime}
\end{align}
The operators $\mathcal{K}_{\pm}$, defined by
\begin{align}
\mathcal{K}_{\pm}&\equiv\mp\pa_{\theta}+\frac{i}{\sin\theta}\pa_{\phi}\mp\cot\theta\,,
\end{align}
act as the $\la$-raising and -lowering operators for the angular wave functions $v^{\la}_{l m_l}$,
\begin{align}
\mathcal{K}_{\pm}v^{\la}_{l m_l}(\theta,\phi)=\sqrt{l(l+1)-\la(\la\pm 1)}v^{\la \pm 1}_{l m_l}(\theta,\phi)\,,
\end{align}
and the prime in $\mathcal{K}^{\prime}_{\mp}$ in Eq.~(\ref{eq:SLb_prime}) indicates that the operator lowers the index $\lap$ according to the value of $\lap=\pm 1$. The second piece of the operator in Eq.~(\ref{eq:VSLb_op}) can be written in a similar way,
\begin{align}
\hat{r}_{\la}^{\dag i}{L}_{Q\bar{Q}}^i{S}^l\hat{r}_{\lap}^l&=\left([\hat{r}_{\la}^{\dag i},{L}_{Q\bar{Q}}^i]+\bm{L}_{Q\bar{Q}}\cdot\hat{\bm r}_{\la}^{\dag}\right)\left(\bm{S}\cdot\hat{\bm r}_{\lap}\right)=\left(\bm{L}_{Q\bar{Q}}\cdot\hat{\bm r}_{\la}^{\dag}-\la^2\frac{\cot(\theta)}{\sqrt{2}}\right)\left(\hat{\bm r}_{\lap}\cdot\bm{S}\right)\,.
\end{align}
In this case the operator vanishes for $\la=0$. For $\la=\pm1$ we have
\begin{align}
\hat{r}_{\pm}^{\dag i}{L}_{Q\bar{Q}}^i{S}^l\hat{r}_{\lap}^l&=\left(\left(\hat{\bm r}_{\pm}\cdot\bm{L}_{Q\bar{Q}}\right)^{\dag}-\frac{\cot\theta}{\sqrt{2}}\right)\left(\hat{\bm r}_{\lap}\cdot\bm{S}\right)=\mp\left(\pm\pa_{\theta}+\frac{i}{\sin(\theta)}\pa_{\phi}\pm\cot\theta\right)^{\dag}\frac{\hat{\bm r}_{\lap}\cdot\bm{S}}{\sqrt{2}}\nonumber \\
&=\mp \mathcal{K}^{\dag}_{\mp}\frac{(\hat{\bm r}_{\lap}\cdot\bm{S})}{\sqrt{2}}\,.
\end{align}
Adding up both contributions give
\begin{align}
&\hat{r}_{\la}^{\dag i}\left({L}_{Q\bar{Q}}^i{S}^l+{S}^i{L}_{Q\bar{Q}}^l\right)\hat{r}_{\lap}^l=\mp \mathcal{K}^{\dag}_{\mp}\frac{(\hat{\bm r}_{\lap}\cdot\bm{S})}{\sqrt{2}}\de_{\la\pm 1}
\mp\frac{(\hat{\bm r}_{\la}^{\dag}\cdot\bm{S})}{\sqrt{2}}\mathcal{K}^{\prime}_{\mp}\de_{\lap\pm 1}\,.\label{eq:matrix_SLb}
\end{align}

Finally, we consider the $V_{SLc}$-operator, which is written as a matrix in $\lambda\lambda'$ as
\begin{align}
(\bm{r}\cdot\hat{\bm{r}}_\lambda^\dagger)(\bm{p}\times\bm{S})\cdot \hat{\bm{r}}_{\lambda'}
+\hat{\bm{r}}_\lambda^\dagger\cdot(\bm{p}\times \bm{S})(\bm{r}\cdot\hat{\bm{r}}_{\lambda'})&=
\begin{pmatrix}
2\bm{L}_{Q\bar{Q}}\cdot\bm{S} & r(\bm{p}\times\bm{S})\cdot\hat{\bm{r}}_+ & r(\bm{p}\times\bm{S})\cdot\hat{\bm{r}}_-\\
\hat{\bm{r}}_+^\dagger\cdot(\bm{p}\times \bm{S})r & 0 & 0\\
\hat{\bm{r}}_-^\dagger\cdot(\bm{p}\times \bm{S})r  & 0 & 0 
\end{pmatrix}\,.
\end{align}
The entry $2\bm{L}_{Q\bar{Q}}\cdot\bm{S}$ can be rewritten using Eq.~\ref{eq:reduce_V_SLa} as
\begin{align}
2\bm{L}_{Q\bar{Q}}\cdot\bm{S}&=\bm{J}^2-\bm{L}^2-\bm{S}^2\,.
\end{align}
For the entry $r(\bm{p}\times\bm{S})\cdot \hat{\bm{r}}_+$, we have
\begin{align}
r(\bm{p}\times\bm{S})\cdot \hat{\bm{r}}_+=&\,-r(\bm{p}\times \hat{\bm{r}}_+)\cdot\mathbf{S}\nonumber\\
=&\,-r\left[\left(-i\hat{\bm{r}}\partial_r-\frac{1}{r}\hat{\bm{r}}\times\bm{L}_{Q\bar{Q}}\right)\times\hat{\bm{r}}_+\right]\cdot \bm{S}\nonumber\\
=&\,\left[ir\hat{\bm{r}}\times\partial_r\hat{\bm{r}}_++(\hat{\bm{r}}\times\bm{L}_{Q\bar{Q}})\times\hat{\bm{r}}_+\right]\cdot\bm{S}\nonumber\\
=&\,\left[(\hat{\bm{r}}\times\bm{L}_{Q\bar{Q}})\times\hat{\bm{r}}_+\right]\cdot\bm{S} \nonumber\\
=&\, \left[\hat{r}^jL^i_{Q\bar{Q}}\hat{r}_+^j-\hat{r}^i(\bm{L}_{Q\bar{Q}}\cdot\hat{\bm{r}}_+)\right]S^i\nonumber\\
=&\,\left\{\hat{r}^j[L^i_{Q\bar{Q}},\hat{r}_+^j]+\hat{r}^j\hat{r}^j_+L^i_{Q\bar{Q}}-\hat{r}^i[L^j_{Q\bar{Q}},\hat{r}_+^j]-\hat{r}^i\hat{\bm{r}}_+\cdot\bm{L}_{Q\bar{Q}}\right\}
S^i\nonumber\\
=&\,\left[\hat{r}_+^i+\frac{1}{\sqrt{2}}\hat{r}^i\left(\cot\theta+\frac{i}{\sin\theta}\partial_\phi+\partial_\theta\right)\right]S^i\nonumber\\
=&\left(\hat{\bm{r}}_++\frac{1}{\sqrt{2}}\hat{\bm{r}}{\mathcal{K}'_-}\right)\cdot\bm{S} \,.
\end{align}
The other entries can be manipulated similarly. The result is 
\begin{align}
&(\bm{r}\cdot\hat{\bm{r}}_\lambda^\dagger)(\bm{p}\times\bm{S})\cdot \hat{\bm{r}}_{\lambda'}
+\hat{\bm{r}}_\lambda^\dagger\cdot(\bm{p}\times \bm{S})(\bm{r}\cdot\hat{\bm{r}}_{\lambda'})\nonumber\\
=&\,
\begin{pmatrix}
2\bm{L}_{Q\bar{Q}}\cdot\bm{S} & \left(\hat{\bm{r}}_++\frac{1}{\sqrt{2}}\hat{\bm{r}}{\mathcal{K}'_-}\right)\cdot\bm{S} 
& -\left(\hat{\bm{r}}_-+\frac{1}{\sqrt{2}}\hat{\bm{r}}{\mathcal{K}'_+}\right)\cdot \bm{S} \\
 \left(\hat{\bm{r}}_+^\dagger+\frac{1}{\sqrt{2}}{\mathcal{K}_-^\dagger}\hat{\bm{r}}\right)\cdot\bm{S}& 0 & 0\\
-\left(\hat{\bm{r}}_-^\dagger+\frac{1}{\sqrt{2}}{\mathcal{K}_+^\dagger}\hat{\bm{r}}\right)\cdot \bm{S} & 0 & 0 
\end{pmatrix}\,.\label{eq:matrix_pxS}
\end{align}

\section{List of matrix elements of spin-dependent operators}\label{app_elements}
In this appendix, we list the angular part of the matrix elements of the spin-dependent operators in Eqs.~(\ref{sdm}) and (\ref{sdm2})
which are required when applying perturbation theory. Let us write the angular wave functions in Eqs.~(\ref{psip}) and (\ref{psim}) as
\begin{align}
&\Phi^{j m_j l s}_\la(\theta,\phi)=\sum_{m_l m_s} \mathcal{C}^{j m_j}_{l\,m_l\,s\,m_s}v^{\la}_{l\,m_l}(\theta,\phi)\chi_{s\,m_s}\,. 
\end{align}
Define the angular matrix elements of the $V_{SK}$-term by
\begin{align}
M_{\la\lap}^{SK}(j, l,s)&=\int d\Omega\, \Phi^{j m_j l s\,\dagger}_\la(\theta,\phi)
\left(\hat{r}^{i\dag}_{\la}\bm{K}^{ik}\hat{r}^{k\dag}_{\lap}\right)\cdot\bm{S}\,
\Phi^{j m_j l s}_{\lap}(\theta,\phi)\,,
\end{align}
where there is no sum on $j,m_j,l,s,\lambda,\lambda'$.
Note that $M_{\la\lap}^{SK}(j, l,s)$ is independent of $m_j$ owing to rotational invariance.
For $s=0$, we have $M_{\lambda\lambda'}^{SK}(j, j,0)=0$. For $s=1$ and $l=0,1,2$,
$M_{\la\lap}^{SK}(j, l,1)$ are given by\\
\begin{center}
\begin{tabular}{|c|c|c|c|}
\multicolumn{4}{c}{$M_{\la\lap}^{SK}(1, 0, 1)$:}\\
\hline
\backslashbox{$\la$}{$\lap$} & $0$ & $+$ & $-$ \\
\hline
$0$& $0$ & $0$ & $0$ \\
\hline
$+$& $0$ & $0$ & $0$ \\
\hline
$-$& $0$ & $0$ & $0$\\
\hline
\end{tabular}\quad\quad
\begin{tabular}{|c|c|c|c|}
\multicolumn{4}{c}{$M_{\la\lap}^{SK}(0, 1, 1)$:}\\
\hline
\backslashbox{$\la$}{$\lap$} & $0$ & $+$ & $-$ \\
\hline
$0$& $0$ & $-1$ & $-1$ \\
\hline
$+$& $-1$ & $-1$ & $0$ \\
\hline
$-$& $-1$ & $0$ & $-1$\\
\hline
\end{tabular}\quad\quad
\begin{tabular}{|c|c|c|c|}
\multicolumn{4}{c}{$M_{\la\lap}^{SK}(1, 1,1)$:}\\
\hline
\backslashbox{$\la$}{$\lap$} & $0$ & $+$ & $-$ \\
\hline
$0$& $0$ & $-\frac{1}{2}$ & $-\frac{1}{2}$ \\
\hline
$+$& $-\frac{1}{2}$ & $-\frac{1}{2}$ & $0$ \\
\hline
$-$& $-\frac{1}{2}$ & $0$ & $-\frac{1}{2}$\\
\hline
\end{tabular}\quad\quad
\begin{tabular}{|c|c|c|c|}
\multicolumn{4}{c}{$M_{\la\lap}^{SK}(2, 1, 1)$:}\\
\hline
\backslashbox{$\la$}{$\lap$} & $0$ & $+$ & $-$ \\
\hline
$0$& $0$ & $\frac{1}{2}$ & $\frac{1}{2}$ \\
\hline
$+$& $\frac{1}{2}$ & $\frac{1}{2}$ & $0$ \\
\hline
$-$& $\frac{1}{2}$ & $0$ & $\frac{1}{2}$\\
\hline
\end{tabular}\\
\end{center}
\begin{center}
\begin{tabular}{|c|c|c|c|}
\multicolumn{4}{c}{$M_{\la\lap}^{SK}(1, 2, 1)$:}\\
\hline
\backslashbox{$\la$}{$\lap$} & $0$ & $+$ & $-$ \\
\hline
$0$& $0$ & $-\frac{\sqrt{3}}{2}$ & $-\frac{\sqrt{3}}{2}$ \\
\hline
$+$& $-\frac{\sqrt{3}}{2}$ & $-\frac{1}{2}$ & $0$ \\
\hline
$-$& $-\frac{\sqrt{3}}{2}$ & $0$ & $-\frac{1}{2}$\\
\hline
\end{tabular}\quad\quad
\begin{tabular}{|c|c|c|c|}
\multicolumn{4}{c}{$M_{\la\lap}^{SK}(2, 2, 1)$:}\\
\hline
\backslashbox{$\la$}{$\lap$} & $0$ & $+$ & $-$ \\
\hline
$0$& $0$ & $-\frac{1}{2\sqrt{3}}$ & $-\frac{1}{2\sqrt{3}}$ \\
\hline
$+$& $-\frac{1}{2\sqrt{3}}$ & $-\frac{1}{6}$ & $0$ \\
\hline
$-$& $-\frac{1}{2\sqrt{3}}$ & $0$ & $-\frac{1}{6}$\\
\hline
\end{tabular}\quad\quad
\begin{tabular}{|c|c|c|c|}
\multicolumn{4}{c}{$M_{\la\lap}^{SK}(3, 2,1)$:}\\
\hline
\backslashbox{$\la$}{$\lap$} & $0$ & $+$ & $-$ \\
\hline
$0$& $0$ & $\frac{1}{\sqrt{3}}$ & $\frac{1}{\sqrt{3}}$ \\
\hline
$+$& $\frac{1}{\sqrt{3}}$ & $\frac{1}{3}$ & $0$ \\
\hline
$-$& $\frac{1}{\sqrt{3}}$ & $0$ & $\frac{1}{3}$\\
\hline
\end{tabular}\\
\end{center}
For the $V_{SKb}$-term, we define the corresponding angular matrix elements by
\begin{align}
M_{\la\lap}^{SKb}(j, l,s)&=\int d\Omega\, \Phi^{j m_j l s\,\dagger}_\la(\theta,\phi)
\left[\left(\bm{r}\cdot \bm{P}^{\dag}_{\la}\right)\left(r^i\bm{K}^{ij}P^j_{\lap}\right)\cdot\bm{S}\right.\nonumber\\
&\quad\left.
-\left(r^i\bm{K}^{ij}P^{j\dag }_{\la}\right)\cdot\bm{S} \left(\bm{r}\cdot \bm{P}_{\lap}\right)\right]
\Phi^{j m_j l s}_{\lap}(\theta,\phi)\,.
\end{align}
For $s=0$, we have $M_{\lambda\lambda'}^{SKb}(j, j,0)=0$. For $s=1$ and $l=0,1,2$,
$M_{\la\lap}^{SKb}(j, l,1)$ are given by\\
\begin{center}
\begin{tabular}{|c|c|c|c|}
\multicolumn{4}{c}{$M_{\la\lap}^{SKb}(1, 0, 1)$:}\\
\hline
\backslashbox{$\la$}{$\lap$} & $0$ & $+$ & $-$ \\
\hline
$0$& $0$ & $0$ & $0$ \\
\hline
$+$& $0$ & $0$ & $0$ \\
\hline
$-$& $0$ & $0$ & $0$\\
\hline
\end{tabular}\quad\quad
\begin{tabular}{|c|c|c|c|}
\multicolumn{4}{c}{$M_{\la\lap}^{SKb}(0, 1, 1)$:}\\
\hline
\backslashbox{$\la$}{$\lap$} & $0$ & $+$ & $-$ \\
\hline
$0$& $0$ & $-r^2$ & $-r^2$ \\
\hline
$+$& $-r^2$ & $0$ & $0$ \\
\hline
$-$& $-r^2$ & $0$ & $0$\\
\hline
\end{tabular}\quad\quad
\begin{tabular}{|c|c|c|c|}
\multicolumn{4}{c}{$M_{\la\lap}^{SKb}(1, 1,1)$:}\\
\hline
\backslashbox{$\la$}{$\lap$} & $0$ & $+$ & $-$ \\
\hline
$0$& $0$ & $-\frac{r^2}{2}$ & $-\frac{r^2}{2}$ \\
\hline
$+$& $-\frac{r^2}{2}$ & $0$ & $0$ \\
\hline
$-$& $-\frac{r^2}{2}$ & $0$ & $0$\\
\hline
\end{tabular}\quad\quad
\begin{tabular}{|c|c|c|c|}
\multicolumn{4}{c}{$M_{\la\lap}^{SKb}(2, 1, 1)$:}\\
\hline
\backslashbox{$\la$}{$\lap$} & $0$ & $+$ & $-$ \\
\hline
$0$& $0$ & $\frac{r^2}{2}$ & $\frac{r^2}{2}$ \\
\hline
$+$& $\frac{r^2}{2}$ & $0$ & $0$ \\
\hline
$-$& $\frac{r^2}{2}$ & $0$ & $0$\\
\hline
\end{tabular}\\
\end{center}
\begin{center}
\begin{tabular}{|c|c|c|c|}
\multicolumn{4}{c}{$M_{\la\lap}^{SKb}(1, 2, 1)$:}\\
\hline
\backslashbox{$\la$}{$\lap$} & $0$ & $+$ & $-$ \\
\hline
$0$& $0$ & $-\frac{\sqrt{3}}{2}r^2$ & $-\frac{\sqrt{3}}{2}r^2$ \\
\hline
$+$& $-\frac{\sqrt{3}}{2}r^2$ & $0$ & $0$ \\
\hline
$-$& $-\frac{\sqrt{3}}{2}r^2$ & $0$ & $0$\\
\hline
\end{tabular}\quad\quad
\begin{tabular}{|c|c|c|c|}
\multicolumn{4}{c}{$M_{\la\lap}^{SKb}(2, 2, 1)$:}\\
\hline
\backslashbox{$\la$}{$\lap$} & $0$ & $+$ & $-$ \\
\hline
$0$& $0$ & $-\frac{r^2}{2\sqrt{3}}$ & $-\frac{r^2}{2\sqrt{3}}$ \\
\hline
$+$& $-\frac{r^2}{2\sqrt{3}}$ & $0$ & $0$ \\
\hline
$-$& $-\frac{r^2}{2\sqrt{3}}$ & $0$ & $0$\\
\hline
\end{tabular}\quad\quad
\begin{tabular}{|c|c|c|c|}
\multicolumn{4}{c}{$M_{\la\lap}^{SKb}(3, 2,1)$:}\\
\hline
\backslashbox{$\la$}{$\lap$} & $0$ & $+$ & $-$ \\
\hline
$0$& $0$ & $\frac{r^2}{\sqrt{3}}$ & $\frac{r^2}{\sqrt{3}}$ \\
\hline
$+$& $\frac{r^2}{\sqrt{3}}$ & $0$ & $0$ \\
\hline
$-$& $\frac{r^2}{\sqrt{3}}$ & $0$ & $0$\\
\hline
\end{tabular}\\
\end{center}
The matrix elements of $\bm{J}^2$, $\bm{L}^2$ and $\bm{S}^2$ are trivial: 
\begin{align}
\int d\Omega\, \Phi^{j m_j l s\,\dagger}_\la(\theta,\phi)\bm{J}^2\delta_{\la\lap}\Phi^{j m_j l s}_{\lap}(\theta,\phi)&=j(j+1)\delta_{\la\lap}\,,\label{imsot1}\\
\int d\Omega\, \Phi^{j m_j l s\,\dagger}_\la(\theta,\phi)\bm{L}^2\delta_{\la\lap}\Phi^{j m_j l s}_{\lap}(\theta,\phi)&=l(l+1)\delta_{\la\lap}\,,\label{imsot2}\\
\int d\Omega\, \Phi^{j m_j l s\,\dagger}_\la(\theta,\phi)\bm{S}^2\delta_{\la\lap}\Phi^{j m_j l s}_{\lap}(\theta,\phi)&=s(s+1)\delta_{\la\lap}\,.\label{imsot3}
\end{align}
The $V_{SLa}$-term can be reduced it to the sum of Eqs.~\eqref{imsot1}-\eqref{imsot3} and the $V_{SK}$-term using
Eq.~(\ref{eq:reduce_V_SLa}). The $V_{S^2}$-term corresponds to Eqs.~\eqref{imsot3}.
For the $V_{SLb}$-term, we define the corresponding angular matrix elements by
\begin{align}
M_{\la\lap}^{SLb}(j, l,s)&=\int d\Omega\, \Phi^{j m_j l s\,\dagger}_\la(\theta,\phi)
\hat{r}^{i\dag}_{\la}\left(L_{Q\bar{Q}}^iS^k+S^iL_{Q\bar{Q}}^k\right)\hat{r}^{k}_{\lap}
\Phi^{j m_j l s}_{\lap}(\theta,\phi)\,.
\end{align}
$M_{\la\lap}^{SLb}(j, l,s)$ is calculated using Eq.~(\ref{eq:matrix_SLb}).
For $s=0$, we have $M_{\la\lap}^{SLb}(j, j, 0)=0$. For $s=1$ and $l=0,1,2$, $M^{SLb}_{\la\lap}(j, l, 1)$ are given by\\
\begin{center}
\begin{tabular}{|c|c|c|c|}
\multicolumn{4}{c}{$M_{\la\lap}^{SLb}(1, 0, 1)$:}\\
\hline
\backslashbox{$\la$}{$\lap$} & $0$ & $+$ & $-$ \\
\hline
$0$& $0$ & $0$ & $0$ \\
\hline
$+$& $0$ & $0$ & $0$ \\
\hline
$-$& $0$ & $0$ & $0$\\
\hline
\end{tabular}\quad\quad
\begin{tabular}{|c|c|c|c|}
\multicolumn{4}{c}{$M_{\la\lap}^{SLb}(0, 1, 1)$:}\\
\hline
\backslashbox{$\la$}{$\lap$} & $0$ & $+$ & $-$ \\
\hline
$0$& $0$ & $0$ & $0$ \\
\hline
$+$& $0$ & $-2$ & $2$ \\
\hline
$-$& $0$ & $2$ & $-2$\\
\hline
\end{tabular}\quad\quad
\begin{tabular}{|c|c|c|c|}
\multicolumn{4}{c}{$M_{\la\lap}^{SLb}(1, 1,1)$:}\\
\hline
\backslashbox{$\la$}{$\lap$} & $0$ & $+$ & $-$ \\
\hline
$0$& $0$ & $0$ & $0$ \\
\hline
$+$& $0$ & $-1$ & $1$ \\
\hline
$-$& $0$ & $1$ & $-1$\\
\hline
\end{tabular}\quad\quad
\begin{tabular}{|c|c|c|c|}
\multicolumn{4}{c}{$M_{\la\lap}^{SLb}(2, 1,1)$:}\\
\hline
\backslashbox{$\la$}{$\lap$} & $0$ & $+$ & $-$ \\
\hline
$0$& $0$ & $0$ & $0$ \\
\hline
$+$& $0$ & $1$ & $-1$ \\
\hline
$-$& $0$ & $-1$ & $1$\\
\hline
\end{tabular}\\
\end{center}
\begin{center}
\begin{tabular}{|c|c|c|c|}
\multicolumn{4}{c}{$M_{\la\lap}^{SLb}(1, 2,1)$:}\\
\hline
\backslashbox{$\la$}{$\lap$} & $0$ & $+$ & $-$ \\
\hline
$0$& $0$ & $0$ & $0$ \\
\hline
$+$& $0$ & $-3$ & $3$ \\
\hline
$-$& $0$ & $3$ & $-3$\\
\hline
\end{tabular}\quad\quad
\begin{tabular}{|c|c|c|c|}
\multicolumn{4}{c}{$M_{\la\lap}^{SLb}(2, 2,1)$:}\\
\hline
\backslashbox{$\la$}{$\lap$} & $0$ & $+$ & $-$ \\
\hline
$0$& $0$ & $0$ & $0$ \\
\hline
$+$& $0$ & $-1$ & $1$ \\
\hline
$-$& $0$ & $1$ & $-1$\\
\hline
\end{tabular}\quad\quad
\begin{tabular}{|c|c|c|c|}
\multicolumn{4}{c}{$M_{\la\lap}^{SLb}(3, 2,1)$:}\\
\hline
\backslashbox{$\la$}{$\lap$} & $0$ & $+$ & $-$ \\
\hline
$0$& $0$ & $0$ & $0$ \\
\hline
$+$& $0$ & $2$ & $-2$ \\
\hline
$-$& $0$ & $-2$ & $2$\\
\hline
\end{tabular}\\
\end{center}
For the $V_{SLc}$-term, we define the corresponding angular matrix elements by
\begin{align}
M_{\la\lap}^{SLc}(j, l,s)&=\int d\Omega\, \Phi^{j m_j l s\,\dagger}_\la(\theta,\phi)[
(\bm{r}\cdot\hat{\bm{r}}_\lambda^\dagger)(\bm{p}\times\bm{S})\cdot \hat{\bm{r}}_{\lambda'}
+\hat{\bm{r}}_\lambda^\dagger\cdot(\bm{p}\times \bm{S})(\bm{r}\cdot\hat{\bm{r}}_{\lambda'})]
\Phi^{j m_j l s}_{\lap}(\theta,\phi)\,.
\end{align}
$M_{\la\lap}^{SLc}(j, l,s)$ is calculated using Eq.~(\ref{eq:matrix_pxS}).
For $s=0$, we have $M_{\lambda\lambda'}^{SLc}(j, j,0)=0$. For $s=1$ and $l=0,1,2$,
$M_{\la\lap}^{SLc}(j, l,1)$ are given by\\
\begin{center}
\begin{tabular}{|c|c|c|c|}
\multicolumn{4}{c}{$M_{\la\lap}^{SLc}(1, 0, 1)$:}\\
\hline
\backslashbox{$\la$}{$\lap$} & $0$ & $+$ & $-$ \\
\hline
$0$& $0$ & $0$ & $0$ \\
\hline
$+$& $0$ & $0$ & $0$ \\
\hline
$-$& $0$ & $0$ & $0$\\
\hline
\end{tabular}\quad\quad
\begin{tabular}{|c|c|c|c|}
\multicolumn{4}{c}{$M_{\la\lap}^{SLc}(0, 1, 1)$:}\\
\hline
\backslashbox{$\la$}{$\lap$} & $0$ & $+$ & $-$ \\
\hline
$0$& $-4$ & $1$ & $1$ \\
\hline
$+$& $1$ & $0$ & $0$ \\
\hline
$-$& $1$ & $0$ & $0$\\
\hline
\end{tabular}\quad\quad
\begin{tabular}{|c|c|c|c|}
\multicolumn{4}{c}{$M_{\la\lap}^{SLc}(1, 1,1)$:}\\
\hline
\backslashbox{$\la$}{$\lap$} & $0$ & $+$ & $-$ \\
\hline
$0$& $-2$ & $\frac{1}{2}$ & $\frac{1}{2}$ \\
\hline
$+$& $\frac{1}{2}$ & $0$ & $0$ \\
\hline
$-$& $\frac{1}{2}$ & $0$ & $0$\\
\hline
\end{tabular}\quad\quad
\begin{tabular}{|c|c|c|c|}
\multicolumn{4}{c}{$M_{\la\lap}^{SLc}(2, 1, 1)$:}\\
\hline
\backslashbox{$\la$}{$\lap$} & $0$ & $+$ & $-$ \\
\hline
$0$& $2$ & $-\frac{1}{2}$ & -$\frac{1}{2}$ \\
\hline
$+$& $-\frac{1}{2}$ & $0$ & $0$ \\
\hline
$-$& $-\frac{1}{2}$ & $0$ & $0$\\
\hline
\end{tabular}\\
\end{center}
\begin{center}
\begin{tabular}{|c|c|c|c|}
\multicolumn{4}{c}{$M_{\la\lap}^{SLc}(1, 2, 1)$:}\\
\hline
\backslashbox{$\la$}{$\lap$} & $0$ & $+$ & $-$ \\
\hline
$0$& $-6$ & $\frac{\sqrt{3}}{2}$ & $\frac{\sqrt{3}}{2}$ \\
\hline
$+$& $\frac{\sqrt{3}}{2}$ & $0$ & $0$ \\
\hline
$-$& $\frac{\sqrt{3}}{2}$ & $0$ & $0$\\
\hline
\end{tabular}\quad\quad
\begin{tabular}{|c|c|c|c|}
\multicolumn{4}{c}{$M_{\la\lap}^{SLc}(2, 2, 1)$:}\\
\hline
\backslashbox{$\la$}{$\lap$} & $0$ & $+$ & $-$ \\
\hline
$0$& $-2$ & $\frac{1}{2\sqrt{3}}$ & $\frac{1}{2\sqrt{3}}$ \\
\hline
$+$& $\frac{1}{2\sqrt{3}}$ & $0$ & $0$ \\
\hline
$-$& $\frac{1}{2\sqrt{3}}$ & $0$ & $0$\\
\hline
\end{tabular}\quad\quad
\begin{tabular}{|c|c|c|c|}
\multicolumn{4}{c}{$M_{\la\lap}^{SLc}(3, 2,1)$:}\\
\hline
\backslashbox{$\la$}{$\lap$} & $0$ & $+$ & $-$ \\
\hline
$0$& $4$ & $-\frac{1}{\sqrt{3}}$ & $-\frac{1}{\sqrt{3}}$ \\
\hline
$+$& $-\frac{1}{\sqrt{3}}$ & $0$ & $0$ \\
\hline
$-$& $-\frac{1}{\sqrt{3}}$ & $0$ & $0$\\
\hline
\end{tabular}
\end{center}
For the $V_{S_{12}a}$-term, we define the corresponding angular matrix elements by
\begin{align}
M_{\la\lap}^{S_{12}a}(j, l, s)&=\int d\Omega\, \Phi^{j m_j l s\,\dagger}_\la(\theta,\phi)
S_{12}\delta_{\lambda\lambda'}
\Phi^{j m_j l s}_{\lap}(\theta,\phi)\,.
\end{align}
For $s=0$, $M^{S_{12}a}_{\la\lap}(j, j, 0)=0$. For $s=1$ and $l=0,1,2$, $M^{S_{12}a}_{\la\lap}(j, l, 1)=0$ are given by\\
\begin{center}
\begin{tabular}{|c|c|c|c|}
\multicolumn{4}{c}{$M_{\la\lap}^{S_{12}a}(1,0,1)$:}\\
\hline
\backslashbox{$\la$}{$\lap$} & $0$ & $+$ & $-$ \\
\hline
$0$& $0$ & $0$ & $0$ \\
\hline
$+$& $0$ & $0$ & $0$ \\
\hline
$-$& $0$ & $0$ & $0$\\
\hline
\end{tabular}\quad\quad
\begin{tabular}{|c|c|c|c|}
\multicolumn{4}{c}{$M_{\la\lap}^{S_{12}a}(0,1,1)$:}\\
\hline
\backslashbox{$\la$}{$\lap$} & $0$ & $+$ & $-$ \\
\hline
$0$& $-4$ & $0$ & $0$ \\
\hline
$+$& $0$ & $2$ & $0$ \\
\hline
$-$& $0$ & $0$ & $2$\\
\hline
\end{tabular}\quad\quad
\begin{tabular}{|c|c|c|c|}
\multicolumn{4}{c}{$M_{\la\lap}^{S_{12}a}(1,1,1)$:}\\
\hline
\backslashbox{$\la$}{$\lap$} & $0$ & $+$ & $-$ \\
\hline
$0$& $2$ & $0$ & $0$ \\
\hline
$+$& $0$ & $-1$ & $0$ \\
\hline
$-$& $0$ & $0$ & $-1$\\
\hline
\end{tabular}\quad\quad
\begin{tabular}{|c|c|c|c|}
\multicolumn{4}{c}{$M_{\la\lap}^{S_{12}a}(2, 1, 1)$:}\\
\hline
\backslashbox{$\la$}{$\lap$} & $0$ & $+$ & $-$ \\
\hline
$0$& $-\frac{2}{5}$ & $0$ & $0$ \\
\hline
$+$& $0$ & $\frac{1}{5}$ & $0$ \\
\hline
$-$& $0$ & $0$ & $\frac{1}{5}$\\
\hline
\end{tabular}\\
\end{center}
\begin{center}
\begin{tabular}{|c|c|c|c|}
\multicolumn{4}{c}{$M_{\la\lap}^{S_{12}a}(1,2,1)$:}\\
\hline
\backslashbox{$\la$}{$\lap$} & $0$ & $+$ & $-$ \\
\hline
$0$& $-2$ & $0$ & $0$ \\
\hline
$+$& $0$ & $-1$ & $0$ \\
\hline
$-$& $0$ & $0$ & $-1$\\
\hline
\end{tabular}\quad\quad
\begin{tabular}{|c|c|c|c|}
\multicolumn{4}{c}{$M_{\la\lap}^{S_{12}a}(2,2,1)$:}\\
\hline
\backslashbox{$\la$}{$\lap$} & $0$ & $+$ & $-$ \\
\hline
$0$& $2$ & $0$ & $0$ \\
\hline
$+$& $0$ & $1$ & $0$ \\
\hline
$-$& $0$ & $0$ & $1$\\
\hline
\end{tabular}\quad\quad
\begin{tabular}{|c|c|c|c|}
\multicolumn{4}{c}{$M_{\la\lap}^{S_{12}a}(3,2,1)$:}\\
\hline
\backslashbox{$\la$}{$\lap$} & $0$ & $+$ & $-$ \\
\hline
$0$& $-\frac{4}{7}$ & $0$ & $0$ \\
\hline
$+$& $0$ & $-\frac{2}{7}$ & $0$ \\
\hline
$-$& $0$ & $0$ & $-\frac{2}{7}$\\
\hline
\end{tabular}\quad\quad
\end{center}
For the $V_{S_{12}b}$-term, we define the corresponding angular matrix elements by
\begin{align}
M_{\la\lap}^{S_{12}b}(j, l, s)&=\int d\Omega\, \Phi^{j m_j l s\,\dagger}_\la(\theta,\phi)
\hat{r}^{i\dag}_{\la}\hat{r}^k_{\lap}\left(S^i_1S^k_2+S^i_2S^k_1\right)
\Phi^{j m_j l s}_{\lap}(\theta,\phi)\,.
\end{align}
For $s=0$, $M^{S_{12}b}_{\la\lap}(j, j, 0)=-\delta_{\la\lap}/2$. For $s=1$ and $l=0,1,2$, $M^{S_{12}b}_{\la\lap}(j, l, 1)$ are given by\\
\begin{center}
\begin{tabular}{|c|c|c|c|}
\multicolumn{4}{c}{$M_{\la\lap}^{S_{12}b}(1,0,1)$:}\\
\hline
\backslashbox{$\la$}{$\lap$} & $0$ & $+$ & $-$ \\
\hline
$0$& $\frac{1}{6}$ & $0$ & $0$ \\
\hline
$+$& $0$ & $0$ & $0$ \\
\hline
$-$& $0$ & $0$ & $0$\\
\hline
\end{tabular}\quad\quad
\begin{tabular}{|c|c|c|c|}
\multicolumn{4}{c}{$M_{\la\lap}^{S_{12}b}(0,1,1)$:}\\
\hline
\backslashbox{$\la$}{$\lap$} & $0$ & $+$ & $-$ \\
\hline
$0$& $-\frac{1}{2}$ & $-\frac{1}{2}$ & $-\frac{1}{2}$ \\
\hline
$+$& $-\frac{1}{2}$ & $0$ & $-1$ \\
\hline
$-$& $-\frac{1}{2}$ & $-1$ & $0$\\
\hline
\end{tabular}\quad\quad
\begin{tabular}{|c|c|c|c|}
\multicolumn{4}{c}{$M_{\la\lap}^{S_{12}b}(1,1,1)$:}\\
\hline
\backslashbox{$\la$}{$\lap$} & $0$ & $+$ & $-$ \\
\hline
$0$& $\frac{1}{2}$ & $\frac{1}{4}$ & $\frac{1}{4}$ \\
\hline
$+$& $\frac{1}{4}$ & $\frac{1}{4}$ & $\frac{1}{2}$ \\
\hline
$-$& $\frac{1}{4}$ & $\frac{1}{2}$ & $\frac{1}{4}$\\
\hline
\end{tabular}\quad\quad
\begin{tabular}{|c|c|c|c|}
\multicolumn{4}{c}{$M_{\la\lap}^{S_{12}b}(2, 1, 1)$:}\\
\hline
\backslashbox{$\la$}{$\lap$} & $0$ & $+$ & $-$ \\
\hline
$0$& $\frac{1}{10}$ & $-\frac{1}{20}$ & $-\frac{1}{20}$ \\
\hline
$+$& $-\frac{1}{20}$ & $\frac{3}{20}$ & $-\frac{1}{10}$ \\
\hline
$-$& $-\frac{1}{20}$ & $-\frac{1}{10}$ & $\frac{3}{20}$\\
\hline
\end{tabular}\\
\end{center}
\begin{center}
\begin{tabular}{|c|c|c|c|}
\multicolumn{4}{c}{$M_{\la\lap}^{S_{12}b}(1,2,1)$:}\\
\hline
\backslashbox{$\la$}{$\lap$} & $0$ & $+$ & $-$ \\
\hline
$0$& $-\frac{1}{6}$ & $-\frac{1}{4\sqrt{3}}$ & $-\frac{1}{4\sqrt{3}}$ \\
\hline
$+$& $-\frac{1}{4\sqrt{3}}$ & $\frac{1}{4}$ & $-\frac{1}{2}$ \\
\hline
$-$& $-\frac{1}{4\sqrt{3}}$ & $-\frac{1}{2}$ & $\frac{1}{4}$\\
\hline
\end{tabular}\quad\quad
\begin{tabular}{|c|c|c|c|}
\multicolumn{4}{c}{$M_{\la\lap}^{S_{12}b}(2,2,1)$:}\\
\hline
\backslashbox{$\la$}{$\lap$} & $0$ & $+$ & $-$ \\
\hline
$0$& $\frac{1}{2}$ & $\frac{1}{4\sqrt{3}}$ & $\frac{1}{4\sqrt{3}}$ \\
\hline
$+$& $\frac{1}{4\sqrt{3}}$ & $\frac{1}{12}$ & $\frac{1}{2}$ \\
\hline
$-$& $\frac{1}{4\sqrt{3}}$ & $\frac{1}{2}$ & $\frac{1}{12}$\\
\hline
\end{tabular}\quad\quad
\begin{tabular}{|c|c|c|c|}
\multicolumn{4}{c}{$M_{\la\lap}^{S_{12}b}(3,2,1)$:}\\
\hline
\backslashbox{$\la$}{$\lap$} & $0$ & $+$ & $-$ \\
\hline
$0$& $\frac{1}{14}$ & $-\frac{1}{14\sqrt{3}}$ & $-\frac{1}{14\sqrt{3}}$ \\
\hline
$+$& $-\frac{1}{14\sqrt{3}}$ & $\frac{4}{21}$ & $-\frac{1}{7}$ \\
\hline
$-$& $-\frac{1}{14\sqrt{3}}$ & $-\frac{1}{7}$ & $\frac{4}{21}$\\
\hline
\end{tabular}\quad\quad
\end{center}


\bibliography{biblongspin}

\end{document}